\documentclass[pdflatex,sn-mathphys-num]{sn-jnl}

\usepackage{graphicx}%
\usepackage{multirow}%
\usepackage{amsmath,amssymb,amsfonts}%
\usepackage{amsthm}%
\usepackage{mathrsfs}%
\usepackage[title]{appendix}%
\usepackage{xcolor}%
\usepackage{textcomp}%
\usepackage{manyfoot}%
\usepackage{booktabs}%
\usepackage{algorithm}%
\usepackage{algorithmicx}%
\usepackage{algpseudocode}%
\usepackage{listings}%
\usepackage{tabularx}
\usepackage{caption}
\usepackage[many]{tcolorbox}
\usepackage{booktabs}
\usepackage{pifont}
\usepackage{array}
\usepackage{subcaption}
\usepackage{adjustbox}
\usepackage{ragged2e}
\usepackage{threeparttablex}
\usepackage{placeins}

\usepackage{booktabs}
\usepackage{fontawesome5} 
\usepackage{xcolor}       
\usepackage{xltabular}

\usepackage{tikz}
\usetikzlibrary{shapes.geometric, arrows.meta, positioning, fit}

\theoremstyle{thmstyleone}%
\theoremstyle{thmstyletwo}%

\theoremstyle{thmstylethree}%

\begin{document}

\title[Article Title]{Maintenance and Support in Community-Driven Scientific Pipeline Ecosystems: A Cross-Platform Empirical Study of nf-core}


\author*[1]{\fnm{Khairul} \sur{Alam}}\email{kha060@usask.ca}
\author[2]{\fnm{Kowsik} \sur{Roy}}\email{kowsik.roy@g.bracu.ac.bd}
\author[3]{\fnm{Md Shamimur} \sur{Rahman}}\email{shamimur.rahman@usask.ca}
\author[4]{\fnm{Banani} \sur{Roy}}\email{banani.roy@usask.ca}

\affil*[1,3,4]{\orgdiv{Department of Computer Science}, \orgname{University of Saskatchewan}, \orgaddress{\city{Saskatoon}, \postcode{S7N 5A2}, \ \country{Canada}}}

\affil[2]{\orgdiv{Department of Computer Science and Engineering}, \orgname{BRAC University}, \orgaddress{\city{Dhaka}, \postcode{1212}, \country{Bangladesh}}}


\abstract{Community-driven scientific pipeline ecosystems are increasingly important for reproducible data-intensive research, but their sustainability depends on more than workflow engines, templates, and testing infrastructure. It also depends on how communities maintain pipelines, integrate contributions, and support users across heterogeneous execution environments. This paper presents a cross-platform empirical study of maintenance and support in nf-core, a large ecosystem of standardized Nextflow pipelines. We analyze 15,760 GitHub issues, 35,411 GitHub pull requests, and 895 Seqera Community Forum discussions to examine what maintenance and support concerns arise, how they differ across artifact types, which factors are associated with resolution outcomes, and how problems and solutions flow between repository-centered and community-centered spaces. Using topic modeling, statistical outcome analysis, direct-link mining, semantic similarity analysis, technical-signal overlap analysis, and qualitative inspection, we show that issues primarily capture repository-level problem reporting and maintenance coordination; pull requests capture implementation, review, testing, dependency, and template-update work; and forum discussions capture user-facing support around execution failures, containers, cloud and HPC environments, MultiQC reporting, and Nextflow usage. Resolution outcomes are associated with actionability, coordination, and diagnostic evidence. Issue closure is linked to assignees, comments, milestones, bug labels, error mentions, and version information. Pull request integration varies by author role, automation type, draft status, checklists, linked issues, and review routing. Forum accepted answers are more likely when discussions include code blocks, sustained interaction, and concrete technical evidence, while cloud, HPC, and workflow-semantics questions are harder to resolve. Cross-platform analysis reveals strong repository-internal traceability within GitHub, but limited explicit linkage between forum discussions and repository artifacts. These findings show that sustaining scientific pipeline ecosystems requires not only technical standardization, but also actionable support processes, review-ready contribution workflows, infrastructure-specific guidance, and stronger traceability between user support and repository maintenance.}





\maketitle

\section{Introduction}
\label{sec:introduction}
Modern data-intensive science increasingly depends on computational pipelines to transform raw data into reliable scientific results \cite{ewels2020nf, liu2015survey}. In domains such as bioinformatics, astronomy and cosmology, and earth and climate science, these pipelines coordinate many interdependent steps, including quality control, alignment, quantification, statistical analysis, reporting, visualization, simulation, and large-scale data processing. They also bind together software tools, parameters, reference datasets, containers, execution profiles, and infrastructure-specific configurations \cite{alam2026analyzing,di2017nextflow, ewels2020nf}. As scientific analyses become more computationally intensive and distributed, reproducibility depends not only on sharing data and code, but also on making the full analytical process portable, executable, inspectable, and maintainable across changing computational environments \cite{sansone2019fairsharing, wilkinson2016fair,cohen2017scientific,wratten2021reproducible,barker2022introducing}. However, achieving these goals remains difficult because pipelines must operate across changing software dependencies, container images, operating systems, cloud services, high-performance computing (HPC) schedulers, file systems, and institutional configurations \cite{alam2025empirical}.

Scientific Workflow Systems (SWSs) have emerged as a central response to these challenges. Systems such as Nextflow \cite{di2017nextflow}, Snakemake \cite{koster2012snakemake}, Galaxy \cite{goecks2010galaxy}, Pegasus \cite{deelman2015pegasus}, Taverna \cite{oinn2004taverna}, and others \cite{suter2026terminology} provide abstractions for specifying, executing, scaling, and reusing multi-step computational analyses. These systems improve portability and reproducibility by separating workflow logic from execution infrastructure and by supporting execution across local machines, high-performance computing (HPC) clusters, cloud platforms, and containerized environments \cite{leipzig2017review,da2017characterization,wratten2021reproducible}. Among these systems, Nextflow \cite{di2017nextflow} has become particularly prominent because it combines a dataflow programming model with support for containers, package managers, cloud execution, HPC schedulers, and scalable deployment \cite{langer2025empowering}. However, the existence of a workflow engine alone does not guarantee that pipelines remain reusable, well-tested, documented, executable, or sustainably maintained. These qualities also depend on software engineering processes, review practices, documentation, testing infrastructure, dependency management, and community support.

Within this broader landscape, nf-core\footnote{https://nf-co.re/} has become one of the most visible community-driven ecosystems for standardized Nextflow pipelines. We use the term \emph{community-driven scientific pipeline ecosystem} to refer to a socio-technical system in which reusable scientific pipelines, shared components, development tools, infrastructure conventions, contributors, maintainers, and users jointly sustain computational analyses over time. Software ecosystem research emphasizes that such systems are shaped by both technical dependencies and social coordination among actors, projects, and shared infrastructure \cite{jansen2009sense,manikas2016revisiting,storey2016social}. nf-core exemplifies this form of ecosystem because it combines Nextflow with shared development guidelines, reusable modules and subworkflows, automated testing, container support, documentation requirements, stable releases, and community review \cite{ewels2020nf,langer2025empowering}. Over time, nf-core has expanded from a collection of pipelines into a broader ecosystem that includes reusable DSL2 modules, subworkflows, shared test datasets, helper tools, governance structures, training material, outreach activities, and user-support channels.

The sustainability of nf-core depends not only on technical standardization but also on continuous maintenance and user support. Pipelines depend on evolving tools, modules, reference datasets, container images, package versions, execution profiles, and infrastructure-specific settings. Users run these pipelines across heterogeneous environments, including local workstations, institutional HPC clusters, commercial cloud platforms, container runtimes, and managed execution services. As a result, they encounter problems related to input formatting, software versions, resource limits, executor behavior, storage systems, reference files, containers, profiles, and pipeline-specific parameters. These problems are not always conventional software defects. Addressing these problems requires more than bug fixing; it requires triage, documentation, review, dependency maintenance, template synchronization, infrastructure troubleshooting, and community-based knowledge sharing.

These activities are distributed across multiple socio-technical spaces. GitHub issues \cite{github-issue-tracker} capture repository-centered problem reporting, bug reports, documentation gaps, module behavior, configuration problems, and pipeline-specific failures. GitHub pull requests \cite{tomayko2010pullrequests} capture implementation and integration work, including fixes, tests, dependency updates, template changes, documentation edits, review discussion, and automated maintenance. Seqera Community Forum discussions \cite{seqera_community} capture user-facing support needs, including pipeline execution failures, container and dependency problems, cloud and HPC deployment questions, MultiQC reporting issues, Nextflow syntax questions, and troubleshooting practices. Prior empirical software engineering research has shown that GitHub artifacts and developer discussion spaces provide valuable evidence about collaboration, contribution evaluation, maintenance practices, and user support \cite{tsay2014influence,gousios2014exploratory,kalliamvakou2016depth,hata2022github,hellman2022characterizing}. For nf-core, mining these artifacts is appropriate because maintenance and support work is explicitly recorded through issue reports, pull request discussions, review metadata, forum questions, accepted answers, links, labels, and other observable traces. At the same time, these traces must be interpreted carefully because repository and forum data are incomplete records of broader community activity \cite{kalliamvakou2016depth}.

Studying only one of these spaces would provide an incomplete view of nf-core maintenance and support. Studying only GitHub issues and pull requests would emphasize formal repository maintenance while missing many practical user-facing execution problems. Studying only forum discussions would reveal support needs but miss how maintainers implement, review, and integrate durable changes. A cross-platform perspective is therefore necessary to understand how maintenance and support concerns arise, how they differ across artifact types, how they are resolved, and whether user-facing support problems are traceably connected to repository-level maintenance work.

Prior work has established the importance of SWSs for reproducible computational science and has described nf-core's role in standardizing and curating Nextflow pipelines \cite{cohen2017scientific,ewels2020nf,wratten2021reproducible,langer2025empowering,suter2026terminology}. Recent empirical studies provide useful context for this work. Alam and Roy~\cite{alam2026analyzing} analyzed issues and pull requests from nf-core pipeline repositories to characterize repository-level development and maintenance challenges. Alam et al.~\cite{alam2025empirical} examined broader SWS development challenges using Stack Overflow and GitHub data across multiple workflow-system contexts. However, these studies did not examine nf-core as a broader maintenance and support ecosystem spanning issues, pull requests, and community forum discussions, nor did they analyze how user-facing support problems connect to repository-level maintenance work. Related empirical software engineering studies have mined GitHub issues, pull requests, and forum discussions to understand open-source maintenance, contribution acceptance, collaboration, and user support \cite{rahman2025investigating, tsay2014influence,gousios2014exploratory,kalliamvakou2014promises,rahman2014insight,hata2022github,hellman2022characterizing}. However, little is known about how these dynamics unfold in community-driven scientific pipeline ecosystems, where maintenance problems often involve not only software defects but also workflow logic, containers, dependencies, reference data, HPC and cloud execution, reporting tools, and user configuration. In particular, we lack empirical evidence on what maintenance and support concerns dominate nf-core, how these concerns differ across repository-centered and community-centered spaces, which artifact-level characteristics are associated with resolution, and whether user-facing support problems are explicitly or implicitly connected to repository-level maintenance work.

To address this gap, we present a cross-platform empirical study of maintenance and support in the nf-core ecosystem. We analyze 15,760 GitHub issues, 35,411 GitHub pull requests, and 895 Seqera Community Forum discussions. Methodologically, we combine topic modeling, manual topic interpretation, statistical outcome analysis, direct-link mining, semantic similarity analysis, technical-signal overlap analysis, and qualitative inspection. This design allows us to characterize what kinds of maintenance and support concerns arise, where they appear, how they are resolved, and how problems and solutions flow between repository-centered development spaces and community-centered support spaces.
\newline
\newline
We investigate the following research questions (\textbf{RQs}):

\begin{itemize}
    \item \textbf{RQ1. What maintenance and support topics emerge across nf-core GitHub issues, pull requests, and Seqera Community Forum discussions?}

    \item \textbf{RQ2. How do maintenance and support topics differ across GitHub issues, pull requests, and Seqera Community Forum discussions?}

    \item \textbf{RQ3. What factors are associated with resolution outcomes across nf-core GitHub issues, pull requests, and  Seqera Community Forum discussions?}

    \item \textbf{RQ4. How do problems and solutions flow between repository-centered development spaces and community-centered support spaces?}
\end{itemize}

Our results show that nf-core maintenance and support are distributed across distinct but complementary spaces. GitHub issues primarily capture repository-level problem reporting and maintenance coordination, including execution failures, runtime diagnostics, container problems, reporting issues, testing gaps, linting behavior, and modernization requests. Pull requests capture implementation-oriented work, including module development, review, testing, dependency updates, template propagation, metadata management, input validation, and community infrastructure maintenance. Forum discussions capture user-facing support needs around pipeline execution, Nextflow dataflow, containers, cloud and HPC environments, MultiQC reporting, development tooling, and pipeline-specific troubleshooting.

Resolution outcomes are shaped by actionability, coordination, and diagnostic evidence. Issue closure is associated with assignees, comments, milestones, bug labels, error mentions, and version information. Pull request integration varies by author role, automation group, draft status, linked issues, requested reviewers, checklists, and contribution-readiness signals. Forum accepted answers are more likely when discussions include code blocks, sustained community interaction, and concrete technical evidence, while cloud, HPC, and workflow-semantics questions are harder to resolve. Cross-platform analysis shows strong issue--pull-request traceability within GitHub but limited explicit linkage between forum discussions and GitHub artifacts, suggesting that many user-facing support problems remain only implicitly connected to repository-level maintenance work.
\newline
\newline
This paper makes the following contributions:
\begin{itemize}
    \item We provide a large-scale cross-platform empirical analysis of maintenance and support in a community-driven scientific pipeline ecosystem, using GitHub issues, pull requests, and Seqera Community Forum discussions from nf-core.
    
    \item We develop a topic-based characterization of nf-core maintenance and support, showing how repository-centered and community-centered platforms expose different layers of ecosystem work.
    
    \item We analyze resolution outcomes across issues, pull requests, and forum discussions, identifying artifact-level features associated with issue closure, pull request integration, closed-without-merge outcomes, and accepted-answer status.
    
    \item We examine explicit and implicit problem--solution flow across platforms, showing strong repository-internal traceability within GitHub but weak explicit linkage between user-facing forum support and formal repository maintenance.
    
    \item We derive actionable recommendations for improving documentation, issue and forum templates, cross-platform traceability, contributor workflows, review readiness, and infrastructure-specific support for containers, cloud, and HPC execution.
\end{itemize}

By studying nf-core as a socio-technical scientific pipeline ecosystem, this work extends prior research on scientific workflow systems beyond workflow design and execution. It shows that the sustainability of reusable scientific pipelines also depends on how communities organize maintenance, support users, integrate contributions, preserve traceability, and translate recurring support problems into durable ecosystem improvements.

\section{Background and Related Work}
\label{sec:background}
This section situates our study in four areas of prior work: scientific workflow systems and pipeline sustainability, the nf-core ecosystem, repository mining for software maintenance, and community-support analysis. We then position our study by identifying the gap addressed by our cross-platform empirical design.
\subsection{Scientific Workflow Systems and Pipeline Sustainability}

Scientific Workflow Systems (SWSs) are specialized platforms for designing, executing, and managing complex computational analyses, often organized as scientific pipelines \cite{liu2015survey}. They allow researchers to specify analytical steps, data dependencies, software requirements, parameters, and execution logic in a structured workflow rather than through ad hoc scripts or manual command sequences \cite{cohen2017scientific,wratten2021reproducible,djaffardjy2023developing}. Examples include Galaxy \cite{goecks2010galaxy}, Nextflow \cite{di2017nextflow}, Snakemake \cite{koster2012snakemake}, Pegasus \cite{deelman2015pegasus}, and Taverna \cite{oinn2004taverna}. By making workflow structure and dependencies explicit, SWSs support reproducibility, scalability, provenance tracking, and reuse across computational settings \cite{alam2023reusability,goble2020fair,alam2022challenges, barker2022introducing}.

However, workflow specification alone does not ensure long-term pipeline sustainability. Scientific pipelines depend on evolving tools, reference datasets, container images, package versions, executor profiles, cloud services, and HPC schedulers \cite{leipzig2017review,wratten2021reproducible}. A pipeline may therefore be well specified but still fail when dependencies change, containers become unavailable, input schemas evolve, or infrastructure-specific configurations no longer work. Sustaining pipelines requires continued maintenance of code, documentation, tests, configuration profiles, containers, reference resources, and user-facing guidance.

This socio-technical view is central to our study. Users encounter workflows through concrete execution attempts, error messages, resource limits, installation problems, scheduler policies, cloud credentials, missing files, and unclear documentation. These problems become visible through maintenance and support artifacts: GitHub issues that report failures or request changes, pull requests that update and validate pipelines, and forum discussions that document user-facing troubleshooting. Analyzing these artifacts therefore helps explain how scientific pipeline sustainability is maintained after pipelines are released.

\subsection{Nextflow and the nf-core Ecosystem}

Nextflow is a widely used SWS that supports scalable and portable execution through a dataflow programming model \cite{di2017nextflow}. Its process--channel abstraction allows developers to define computational steps as independent processes and connect them through explicit data dependencies. This design enables the same pipeline logic to be executed across local machines, HPC schedulers, cloud platforms, and containerized environments. Nextflow also integrates with technologies such as Docker, Singularity/Apptainer, Conda, and cloud or batch execution backends, making it well suited for large-scale scientific data analysis \cite{langer2025empowering}.

nf-core builds on Nextflow by organizing pipelines, reusable modules, subworkflows, templates, test datasets, documentation, and development tools into a shared community ecosystem \cite{ewels2020nf}. Rather than functioning as a collection of isolated repositories, nf-core provides common conventions for pipeline structure, review, testing, release, documentation, and execution. These conventions help standardize pipeline development, but they also create continuous maintenance demands as tools, dependencies, templates, containers, and execution environments evolve. This makes nf-core an appropriate empirical setting for our study. Its maintenance work includes conventional software-engineering activities, such as fixing bugs, reviewing pull requests, updating documentation, and managing dependencies. At the same time, it includes scientific-pipeline-specific work, such as updating modules, validating sample sheets, maintaining container images, revising execution profiles, resolving reference-data problems, synchronizing templates, and supporting users across local, HPC, cloud, and managed execution environments. Because these activities are distributed across GitHub issues, pull requests, and community forum discussions, nf-core provides a rich setting for examining how community-driven scientific pipeline ecosystems are maintained and supported in practice.

\subsection{Mining GitHub Artifacts for Software Maintenance}

GitHub has become a major source of socio-technical evidence for studying software maintenance. Prior studies have analyzed issues, pull requests, commits, and review discussions to understand collaboration, contribution evaluation, pull-based development, bug fixing, feature evolution, and software quality \cite{braga2023not,tan2024crossfix,gousios2014exploratory,tsay2014influence,kalliamvakou2014promises,rahman2014insight,soares2015acceptance,lenarduzzi2021does}. These studies show that repository artifacts capture both technical and social aspects of maintenance, including how contributors propose changes, how maintainers evaluate them, and which factors influence acceptance or delay.

Repository mining is particularly useful in specialized technical domains, where maintenance challenges are shaped by domain-specific tools, assumptions, and execution contexts \cite{han2020programmers,li2021understanding,yang2023users,alam2025empirical}. In the nf-core context, Alam and Roy~\cite{alam2026analyzing} analyzed GitHub issues and pull requests from pipeline repositories to characterize repository-level development and maintenance challenges. However, their analysis was limited to pipeline repositories. The present study examines the broader nf-core ecosystem, including pipelines, modules, tooling, configurations, infrastructure, and community-facing repositories, together with Seqera Community Forum discussions. It also compares the roles of issues, pull requests, and forum discussions, analyzes their artifact-specific resolution outcomes, and examines cross-artifact traceability and correspondence. This broader perspective captures maintenance concerns involving not only source-code defects, but also reusable components, test data, templates, containers, execution environments, and user support that may not be visible through pipeline repositories alone.

\subsection{Community Support and Discussion Platforms}
Community support platforms play a central role in open-source ecosystems by capturing how users ask questions, report problems, share workarounds, clarify documentation, and learn from previous discussions. Prior studies have shown that user forums and discussion platforms support troubleshooting, knowledge sharing, onboarding, and collaboration beyond formal repository workflows \cite{hata2022github,hellman2022characterizing}. Such spaces are especially important when users encounter problems that are difficult to classify as software defects, such as environment-specific failures, installation issues, configuration errors, or uncertainty about expected behavior.

For nf-core, community support is particularly important because successful pipeline execution depends on the interaction between pipeline definitions and their execution contexts. Users may encounter failures caused by Singularity or Apptainer image retrieval, Docker layer errors, Wave authentication, Conda environment creation, AWS Batch or Google Cloud configuration, Slurm queue policies, file permissions, memory limits, profile settings, reference-data paths, or incomplete logs. These problems directly affect whether users can run and trust pipelines, even when the pipeline code itself is correct. The Seqera Community Forum therefore provides a complementary perspective on nf-core maintenance by revealing practical user-facing problems that may not be visible in GitHub repositories alone. At the same time, support discussions can contain maintenance-relevant knowledge. Repeated forum questions may reveal documentation gaps, missing examples, confusing error messages, usability barriers, or pipeline defects. If these discussions are not linked to GitHub issues or pull requests, maintainers may miss opportunities to translate recurring support problems into durable repository-level improvements. This motivates our cross-platform analysis of whether and how problems and solutions flow between community support spaces and repository-centered development spaces.

\subsection{Topic Modeling for Maintenance and Support Analysis}

Topic modeling is widely used to identify recurring themes in large collections of unstructured text, supporting document organization, information retrieval, recommendation, and software repository analysis \cite{jelodar2019latent,asuncion2010software,bagheri2014adm,gethers2010using}. Prior studies have applied topic modeling to understand developer and user challenges in domains such as quantum software engineering, cryptography APIs, open-source AI repositories, deep learning frameworks, desktop web applications, scientific workflow systems, Eclipse forums, and online discussion forums \cite{li2021understanding,nadi2016jumping,yang2023users,han2020programmers,scoccia2021challenges,alam2025empirical,kahani2016problems,barker2024viability,dhasade2020towards,jokhio2021mining,wang2019does}.

Traditional topic models such as Latent Dirichlet Allocation (LDA) and its variants (e.g., sLDA, CTM, RTM, DTM)  have been extensively used for software text analysis \cite{blei2003latent}. However, these models often require a predefined number of topics, are sensitive to hyperparameter choices, and may produce incoherent topics when applied to short, noisy, and technical texts. BERTopic   \cite{grootendorst2022bertopic} addresses several of these limitations by combining transformer-based embeddings with clustering to generate more contextual and interpretable topics. Because GitHub issues, pull requests, and community forum posts often contain domain-specific terminology, short descriptions, code fragments, logs, and noisy text, we utilize BERTopic for identifying maintenance and support topics in the nf-core ecosystem.

\subsection{Research Gap}

Prior work has established the importance of SWSs for reproducible research and has described nf-core's role in standardizing Nextflow pipelines \cite{cohen2017scientific,wratten2021reproducible,ewels2020nf,langer2024empowering}. Empirical software engineering research has also shown that GitHub artifacts and support forums can reveal maintenance, collaboration, and user-support practices \cite{tsay2014influence,gousios2014exploratory,hata2022github,hellman2022characterizing}. However, we still lack empirical evidence on how maintenance and support are organized in a community-driven scientific pipeline ecosystem where problems span workflow code, reusable modules, containers, dependencies, reference data, execution profiles, cloud services, HPC schedulers, reporting tools, and user configuration. In particular, existing work does not explain: (i) what maintenance and support topics dominate nf-core across repository and community spaces; (ii) how these topics differ between GitHub issues, pull requests, and forum discussions; (iii) which artifact-level characteristics are associated with closure, merge, or accepted-answer outcomes; and (iv) whether user-facing support problems are explicitly or implicitly connected to repository-level maintenance work.

Our study addresses this gap through a cross-platform empirical analysis of nf-core GitHub issues, pull requests, and Seqera Community Forum discussions. By treating these artifacts as complementary units of analysis, we examine maintenance and support as distributed ecosystem work rather than as isolated repository activity.

\section{Study Design}
\label{sec:study-design}
\begin{figure*}[htbp]
\centering
\resizebox{0.94\textwidth}{!}{
\begin{tikzpicture}[
    font=\scriptsize,
    node distance=0.55cm and 0.85cm,
    >=Latex,
    box/.style={
        rectangle,
        rounded corners,
        draw=black,
        align=center,
        minimum width=3.15cm,
        minimum height=0.72cm,
        inner sep=2.5pt
    },
    smallbox/.style={
        rectangle,
        rounded corners,
        draw=black,
        align=center,
        text width=3.15cm,
        minimum height=1.05cm,
        inner sep=2.5pt
    },
    line/.style={draw, -Latex, thick},
    groupbox/.style={
        rectangle,
        rounded corners,
        dashed,
        draw=black,
        inner sep=0.20cm
    }
]

\node[box] (issues) {nf-core GitHub \\Issues};
\node[box, right=0.95cm of issues] (prs) {nf-core GitHub \\Pull Requests};
\node[box, right=0.95cm of prs] (forum) {Seqera Community \\Forum Discussions};

\node[box, below=0.8cm of prs] (collection) {Data Collection\\and Filtering};

\node[box, below=0.5cm of collection] (preprocess) {
Data preprocessing,\\
metadata extraction,\\
technical-signal construction
};

\node[box, below=0.5cm of preprocess] (analysisprep) {
Shared Analytical Preparation\\
topic-modeling corpus, resolution features,\\
and cross-platform linkage features
};

\node[smallbox, below left=0.85cm and 2.1cm of analysisprep] (rq1) {
\textbf{RQ1}\\
Topic discovery\\
BERTopic modeling, coherence tuning,\\
manual topic labeling
};

\node[smallbox, below=0.85cm of analysisprep] (rq2) {
\textbf{RQ2}\\
Platform comparison\\
topic distributions, shared concerns,\\
platform-specific roles
};

\node[smallbox, below right=0.85cm and 2.1cm of analysisprep] (rq3) {
\textbf{RQ3}\\
Resolution analysis\\
outcome definitions, bivariate tests,\\
multivariable and survival models
};

\node[smallbox, below=0.72cm of rq1] (rq4) {
\textbf{RQ4}\\
Problem--solution flow\\
direct links, semantic similarity,\\
technical-signal overlap
};

\node[smallbox, right=1.45cm of rq4] (rq5) {
\textbf{Synthesis}\\
Actionable improvements\\
documentation, templates, traceability,\\
review and infrastructure support
};

\node[box, below=0.82cm of rq5, xshift=-2.15cm, minimum width=4.6cm] (findings) {
Cross-platform empirical insights\\
into maintenance and support\\
in the nf-core ecosystem
};

\draw[line] (issues.south) -- ++(0,-0.22) -| (collection.north);
\draw[line] (prs.south) -- (collection.north);
\draw[line] (forum.south) -- ++(0,-0.22) -| (collection.north);

\draw[line] (collection) -- (preprocess);
\draw[line] (preprocess) -- (analysisprep);

\draw[line] (analysisprep) -- (rq1);
\draw[line] (analysisprep) -- (rq2);
\draw[line] (analysisprep) -- (rq3);

\draw[line] (rq1) -- (rq4);
\draw[line] (rq2) -- (rq4);
\draw[line] (rq3) -- (rq5);
\draw[line] (rq4) -- (rq5);

\draw[line] (rq5) -- (findings);

\node[groupbox, fit=(issues)(prs)(forum),
      label=above:{\textbf{Data Sources}}] {};

\node[groupbox, fit=(collection)(preprocess)(analysisprep),
      label={[align=center]right:{\textbf{Shared}\\\textbf{Processing Pipeline}}}] {};

\node[groupbox, fit=(rq1)(rq2)(rq3)(rq4)(rq5),
      label={[align=center]right:{\textbf{RQ-Specific}\\\textbf{Analyses and Synthesis}}}] {};

\end{tikzpicture}
}
\vspace{1em}
\caption{Overview of the study design for analyzing maintenance and support in the nf-core ecosystem.}
\label{fig:study_design}
\end{figure*}
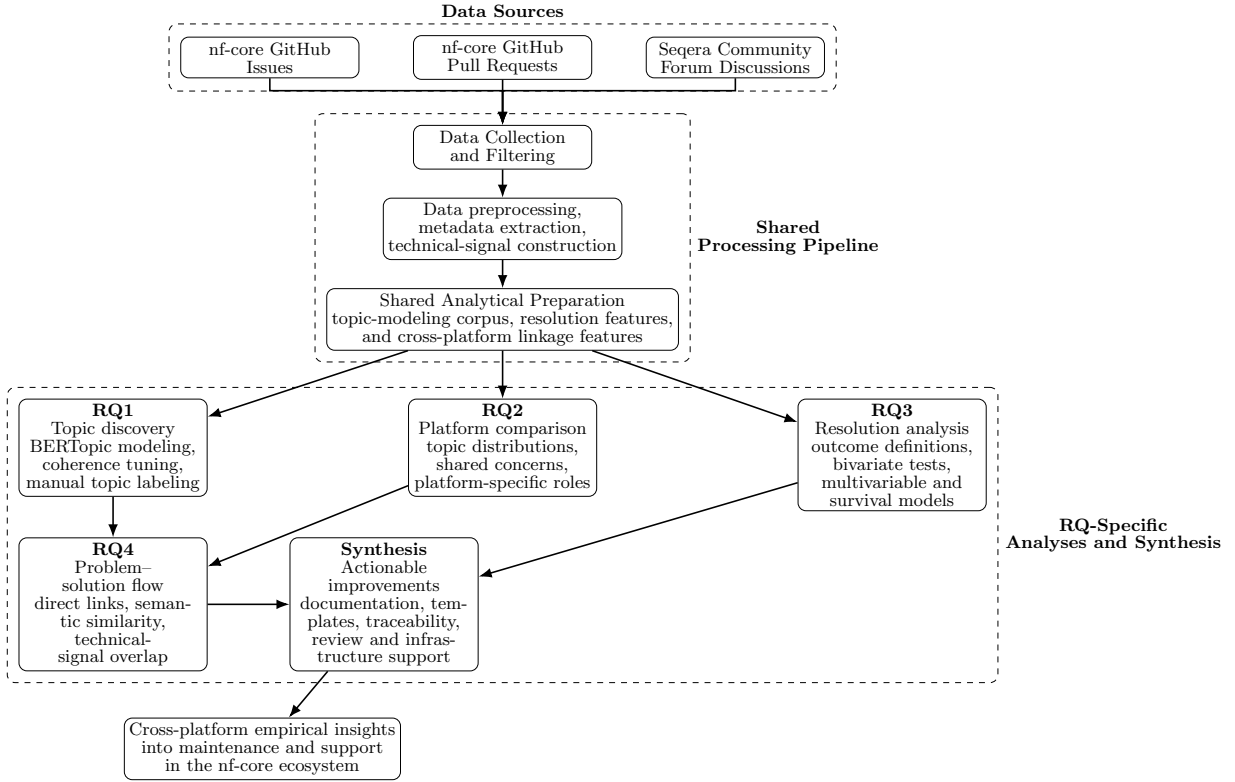

This study adopts a cross-platform empirical design to examine how maintenance and support are organized in the nf-core ecosystem. We analyze three complementary artifact types: GitHub issues, pull requests, and Seqera Community Forum discussions. These artifacts represent different layers of ecosystem activity. Figure~\ref{fig:study_design} summarizes the overall research process, including data collection, preprocessing, topic modeling, cross-platform comparison, resolution analysis, problem--solution flow analysis, and synthesis of actionable improvements.

\subsection{Data Collection}
We collected artifacts from three sources: nf-core GitHub issues, pull requests, and Seqera Community Forum discussions. At the time of data collection on May 8, 2026, the nf-core ecosystem contained 191 GitHub repositories. Of these, 19 repositories were archived. As archived repositories are no longer actively maintained and may not reflect ongoing maintenance and support activity, we excluded them from the analysis. This resulted in 172 active repositories for GitHub data collection.

To reduce right-censoring and allow artifacts sufficient time to receive responses or reach observable outcomes, we applied a one-month observation buffer. Accordingly, we included only GitHub issues, pull requests, and forum discussions created on or before April 8, 2026. This cutoff is important for the resolution-oriented analyses because artifacts created shortly before data collection may appear unresolved simply because maintainers, contributors, or community members have not yet had enough time to respond.

From the 172 active repositories, 149 repositories contained at least one issue created on or before the cutoff date, yielding \textit{15,763} GitHub issues for analysis. For pull requests, 164 repositories contained at least one pull request created on or before the cutoff date, yielding \textit{36,521} pull requests. We utilized the \textit{GitHub REST API} to collect issues and pull requests. However, some pull requests were directly associated with existing issues. Treating these linked pull requests as fully independent records could introduce redundancy, particularly when the same maintenance task was represented both as an issue and as an associated pull request. To reduce this duplication, following prior works \cite{alam2026analyzing, alam2025empirical}, we used the \textit{GitHub Search API} to identify pull requests explicitly associated with existing issues and excluded \textit{1,091} linked pull requests from the pull request dataset. After this filtering step, the final pull request dataset contained 35,430 pull requests. For community support data, we developed a custom Python script to collect discussions from the Seqera Community Forum. The initial forum dataset contained 957 discussions. Among them, 62 were categorized as events and announcements in the forum that were not directly related to maintenance or user support, including event-organizing posts, announcements, and general community updates. The remaining 895 forum discussions captured user facing problems and we utilized that for our analysis.

\begin{table*}[htbp]
\centering
\caption{Data-flow summary for artifact collection, filtering, and preprocessing}
\label{tab:data_flow_counts}

\begin{tabularx}{\textwidth}{Xrrr}
\toprule
\textbf{Stage} & \textbf{Issues} & \textbf{Pull requests} & \textbf{Forum discussions} \\
\midrule
Artifacts collected after repository/date filtering & 15,763 & 36,521 & 957 \\
After removing non-support forum discussions & 15,763 & 36,521 & 895 \\
After linked-pull-request de-duplication & 15,763 & 35,430 & 895 \\
Final topic-modeling corpus after text-quality filtering & 15,760 & 35,411 & 895 \\
\bottomrule
\end{tabularx}

\caption*{\footnotesize \textit{Note.} The linked-pull-request de-duplication step applies to the pull request corpus used for topic and outcome analyses. Explicit issue--pull-request relationships identified during this step are retained as link evidence for the problem--solution flow analysis.}

\end{table*}

Table~\ref{tab:data_flow_counts} reconciles the dataset counts used in the paper. We distinguish between collected artifacts, de-duplicated analysis corpora, and final topic-modeling corpora to avoid ambiguity. Unless otherwise stated, \textbf{RQ1--RQ3} use the final preprocessed corpora, while \textbf{RQ4} uses explicit link records and semantic-flow candidates derived during the linkage analysis. The exclusion of issue-linked pull requests from the pull request analysis corpus does not remove issue--pull request relationships from the study; rather, these relationships are retained as explicit traceability evidence for \textbf{RQ4}, where the goal is to examine problem--solution flow across artifacts. Thus, linked pull requests are excluded only to reduce redundancy in topic modeling and pull request outcome analyses, while their associations with issues remain part of the cross-artifact linkage analysis.

\subsection{Data Preprocessing}
We preprocessed textual content for topic modeling and separately retained metadata, structural indicators, and technical signals for outcome and flow analyses. This distinction is important because the requirements of topic modeling differ from the requirements of resolution analysis. Topic modeling benefits from reducing repeated syntax, markup, and boilerplate, while resolution analysis requires preserving whether artifacts contain code, URLs, errors, configuration details, or other diagnostic evidence.

For topic modeling, we constructed one textual document for each artifact. For GitHub issues and pull requests, we concatenated the title and body fields. For Seqera Community Forum discussions, we concatenated the discussion title and rendered discussion content. We removed or normalized HTML markup, URLs, code fragments, command-line snippets, stack traces, configuration blocks, and other structured technical fragments to prevent topic representations from being dominated by repeated paths, commands, environment variables, or syntax tokens. We then lowercased text, removed punctuation and non-alphabetic symbols, removed standard English stopwords \cite{hardeniya2016natural}, and removed high-frequency ecosystem terms such as \emph{nf-core}, \emph{Nextflow}, \emph{issue}, and \emph{pull request}. These terms identify the general study context but do not help distinguish one maintenance or support topic from another. Finally, we applied lemmatization using the \texttt{spaCy} \texttt{en\_core\_web\_sm} model \cite{vasiliev2020natural}.

Because technical fragments can contain meaningful diagnostic information in nf-core, we did not discard their presence from the overall analysis. Instead, we preserved structural and technical indicators as separate features, including code-block presence, URL presence, error mentions, version mentions, command mentions, system-information mentions, container terms, cloud terms, HPC terms, MultiQC terms, and reproduction-related terms. These features are used in RQ3 and RQ4 to analyze resolution outcomes and cross-platform technical-signal overlap.

After preprocessing, in the issue dataset, we found three non-informative records: two issues had only numeric titles, and one issue contained only a URL in the title, with an empty body. We removed these three records, resulting in 15,760 GitHub issues for subsequent analysis. Similarly, we identified and removed 19 non-informative pull request records, resulting in 35,411 pull requests. In contrast, all retained forum discussions contained meaningful title and discussion content after preprocessing. The final topic-modeling corpus therefore consisted of 15,760 issues, 35,411 pull requests, and 895 forum discussions.

It is important to note that this preprocessing was applied to the textual corpus used for topic modeling. For resolution-oriented analyses, we retained relevant metadata and structural indicators, such as labels, assignees, comments, requested reviewers, linked issues, draft status, reply counts, accepted-answer status, and the presence of code blocks or URLs. This separation allowed us to reduce textual noise for topic discovery while preserving artifact-level features needed to analyze maintenance and support outcomes.

\subsection{Identifying Maintenance and Support Topics}
\label{sec:topic-modeling}
To answer \textbf{RQ1}, we applied BERTopic \cite{grootendorst2022bertopic} separately to the issue, pull request, and forum corpora. We modeled three artifact types separately because each serves a different communicative role in the nf-core ecosystem. Issues are primarily used for problem reporting and maintenance coordination, pull requests for implementation and review, and forum discussions for user-facing troubleshooting. Separate modeling allowed platform-specific topic structures to emerge without forcing all artifacts into a single shared topic space.

BERTopic represents documents as semantic embeddings before clustering them into topics. For each corpus, we generated embeddings using the \texttt{all-mpnet-base-v2} sentence-transformer model, a fine-tuned \textit{MPNet} model available through Hugging Face that produces 768-dimensional dense vectors optimized for semantic similarity, clustering, and semantic search \cite{reimers2019sentence, sentence_transformers_huggingface, huggingface_models}. This model is well suited for short, technical, and domain-specific texts because it captures semantic similarity beyond exact keyword overlap. We then reduced the embedding dimensions using \textit{UMAP}, which preserves meaningful local structure in high-dimensional data \cite{mcinnes2018umap}, and clustered the reduced embeddings with \textit{HDBSCAN}, a hierarchical density-based method that can identify clusters of varying density while treating weakly related documents as outliers \cite{mcinnes2017hdbscan}. This embedding–reduction–clustering pipeline allowed us to identify semantically coherent topics without predefining the number of clusters.

We tuned BERTopic parameters separately for each artifact type to balance coherence, interpretability, and granularity. We explored UMAP settings for \emph{n\_neighbors} between 15 and 60 and \emph{n\_components} between 5 and 40, using cosine distance to align with sentence-transformer embeddings. For HDBSCAN, we varied \emph{min\_cluster\_size} between 30 and 300 and used Euclidean distance on the UMAP-reduced embeddings. We evaluated topic quality using topic coherence, which measures semantic similarity among top topic terms and supports interpretability assessment~\cite{roder2015exploring}. We used unigram and bigram features in CountVectorizer, following prior software-engineering topic-modeling studies~\cite{alam2025empirical,li2021understanding,abdellatif2020challenges,alam2026analyzing}.

The final topic models used in RQ1 produced 11 issue topics, 13 pull request topics, and 8 forum topics. BERTopic assigns the special label \emph{-1} to documents that do not fit clearly into any of the discovered topic clusters. This label is commonly used for outlier or mixed-topic documents whose semantic patterns are less coherent than those of the main clusters. Because the data had already been extensively preprocessed, we interpreted Topic \emph{-1} as a main topic category rather than excluding it as noise.

\paragraph{Manual topic validation.}
We validated topic interpretations through expert inspection. For each topic, the first author reviewed the top representative terms and at least 25 representative artifacts. This procedure was applied consistently across issues, pull requests, and forum discussions. Initial labels were assigned based on recurring technical concerns, artifact context, and representative examples. The proposed labels were then reviewed by co-authors with experience in SWSs, pipelines, and empirical software engineering. Borderline cases and ambiguous labels were discussed until consensus was reached. The detailed process is described in section \ref{maintences-process}. We documented representative examples and rationale for topic labels in the replication package to support transparency and reproducibility.

\subsection{Cross-Platform Topic Comparison}
To answer \textbf{RQ2}, we compared how maintenance and support concerns differ across GitHub issues, pull requests, and Community Forum discussions. Because topics were modeled separately for each artifact type, we did not force a one-to-one topic mapping. Instead, we compared topic names, representative terms, and representative artifacts to identify shared concerns and artifact-specific roles.

Two authors with over five and nine years of experience with scientific pipelines independently reviewed the topic labels and representative artifacts to identify recurring cross-platform concerns, including execution failures, containers, documentation, testing, dependencies, configuration, MultiQC/reporting, modules/subworkflows, and HPC/cloud execution. We then summarized how each concern appeared in each artifact type. For example, execution failures may appear in issues as runtime bug reports, in pull requests as fixes or tests, and in forum discussions as troubleshooting questions. Disagreements were resolved through discussion. This comparison allowed us to interpret the three artifact types as complementary views of nf-core maintenance and support rather than interchangeable text sources.

\subsection{Resolution Outcome Analysis}
\label{sec:resolution-analysis}

To answer \textbf{RQ3}, we defined artifact-specific resolution outcomes because issues, pull requests, and forum discussions follow different workflows. For GitHub issues, we measured closure status and time-to-close, computed from issue creation to closure, with open issues treated as censored in time-to-event analyses. For pull requests, we measured whether a pull request was merged, closed without merge, or remained open, along with time-to-merge and time-to-final-decision; open pull requests were treated as censored in lifecycle analyses. For forum discussions, we used accepted-answer status as the explicit support-resolution indicator. Because accepted-answer timestamps were unavailable, we used activity span, measured from discussion creation to last activity, as an exploratory lifecycle proxy rather than as exact time-to-answer.

We organized explanatory features into three groups to avoid causal or predictive overinterpretation. \emph{Submission-time features} are available from the initial artifact text or metadata, such as title/body text, author type, labels present at collection, code blocks, URLs, errors, versions, commands, container terms, cloud terms, HPC terms, and topic assignment. \emph{Process features} are created as the artifact is handled, such as comments, assignees, milestones, requested reviewers, replies, likes, and review-routing indicators. \emph{Outcome-adjacent features} are close to the resolution process itself, such as accepted-answer discussion activity or final review state. We therefore interpret all \textbf{RQ3} results as associations observed over the artifact lifecycle, not as causal effects or purely submission-time predictions.

We first summarized outcome rates and lifecycle distributions for each artifact type. We then conducted bivariate analyses to examine associations between artifact features and outcomes~\cite{agresti2013categorical}. For binary features, we used contingency-table analyses~\cite{fagerland2017statistical} and reported odds ratios where appropriate. For numeric features, we used nonparametric comparisons and effect sizes because comments, replies, views, and resolution times were highly skewed. We applied Benjamini--Hochberg \cite{benjamini2001control} false-discovery-rate correction for multiple comparisons.

We then fitted multivariable models to examine whether associations remained after accounting for other artifact characteristics. For issues, we used logistic regression for closure and Cox proportional hazards models for time-to-close~\cite{hosmer2013applied,kalbfleisch2023fifty}. For pull requests, we used logistic and regularized logistic models for merge and closed-without-merge outcomes~\cite{friedman2010regularization}. For forum discussions, we used regularized logistic regression for accepted-answer status. In addition, we used Kaplan--Meier curves \cite{d2021methods} for all three artifact types to visualize lifecycle differences across selected binary features. For issues, the event was issue closure; for pull requests, the event was a final decision, either merged or closed without merge; and for forum discussions, the event was receiving an accepted answer, using activity span as an exploratory proxy because exact accepted-answer timestamps were unavailable.

Continuous predictors were log-transformed where appropriate to reduce skew and standardized before multivariable modeling; binary predictors were encoded as 0/1. Sparse categorical levels were collapsed, and constant, duplicate, or highly correlated predictors were removed before fitting. For regularized logistic models, we used L1 regularization to handle sparse and correlated predictors and treated coefficients as adjusted directional associations rather than formal significance tests. We report odds ratios, hazard ratios, confidence intervals where available, adjusted $p$-values where applicable, and effect sizes to avoid relying only on statistical significance. For Cox models, we checked proportional-hazards assumptions and interpret results cautiously when assumptions were imperfect. Full model specifications and robustness checks are included in the replication package.

\subsection{Problem--Solution Flow Analysis}
\label{sec:problem_solution_flow}

To answer \textbf{RQ4}, we examined how problems and solutions are connected across issues, pull requests, and forum discussions. We analyzed both explicit traceability and weaker implicit relatedness. Explicit links provide direct evidence of cross-artifact relationships, while semantic similarity and shared technical signals identify candidate relationships that require cautious interpretation.

We first extracted explicit references from artifact text. These included GitHub issue URLs, pull request URLs, forum URLs, GitHub-style references such as \texttt{\#123}, and pull request closing-keyword references such as \emph{fixes}, \emph{closes}, and \emph{resolves}. Closing-keyword references were treated as stronger evidence of problem--solution flow because they indicate that a pull request is intended to address a specific issue. We resolved extracted references against the collected artifacts and classified the resulting relationships by source and target artifact type, such as pull request-to-issue, issue-to-pull request, forum-to-GitHub, and GitHub-to-forum.

Because many related artifacts may not be explicitly linked, we also examined implicit relatedness between issues and pull requests. We represented artifact text using TF--IDF features and computed cosine similarity between candidate cross-artifact pairs. We removed directly linked pairs before semantic analysis so that the semantic-flow results would capture additional candidate relationships rather than rediscover known links. We then applied temporal constraints to identify plausible issue-to-pull-request sequences and pull-request-before-issue sequences. These pairs are interpreted as \emph{candidate semantic flows}, not definitive causal links.

To reduce overinterpretation, we validated a sample of semantic-flow candidates through manual inspection. Candidate pairs were classified as exact or strongly related, broadly related, weakly related, or unrelated. Exact or strongly related pairs discuss the same problem and likely implementation response; broadly related pairs share a technical concern but do not show clear problem--solution continuity; weak or unrelated pairs share vocabulary but do not provide meaningful evidence of flow. We report these validation results in RQ4 and use them to qualify the strength of the semantic-flow interpretation.

We also extracted shared technical signals across artifact types, including execution errors, workflow execution, containers, cloud, HPC, MultiQC/reporting, testing/linting, documentation/training, dependency updates, template synchronization, modules/subworkflows, configuration, and channel/dataflow concerns. Signal overlap helps identify recurring problem domains that appear across issues, pull requests, and forum discussions even when artifacts are not directly linked. Together, direct links, semantic-flow candidates, manual validation, and signal overlap allow us to characterize both visible traceability and weaker thematic connections between community support and repository maintenance.

\section{Results}
\subsection{RQ1: Maintenance and Support Topics}
\label{maintences-process}
\subsubsection{Motivation}
Maintenance and support for nf-core span multiple artifact types, but the recurring concerns discussed in these spaces have not been systematically characterized. GitHub issues, pull requests, and Seqera Community Forum discussions capture different forms of ecosystem activity: users and maintainers report problems and coordinate maintenance tasks, contributors implement and review changes, and community members seek help with execution, configuration, containers, cloud/HPC environments, reporting, and Nextflow behavior. Without first identifying the topics that appear in these artifacts, it is difficult to understand what kinds of problems users encounter, what kinds of maintenance work contributors perform, and which support needs emerge outside repository-centered development. \textbf{RQ1} therefore establishes the empirical foundation of the study by identifying the main maintenance and support topics across nf-core issues, pull requests, and forum discussions.

\subsubsection{Approach}
To identify maintenance and support topics, we applied BERTopic separately to the preprocessed issue, pull request, and forum corpora, as described in Section~\ref{sec:topic-modeling}. The modeling process yields 11 topics for issues, 13 topics for pull requests, and 8 topics for forum discussions, along with representative keywords and documents for each topic.

We then manually assigned meaningful and contextually accurate labels to the discovered topics. Following established practices in prior studies \cite{alam2025empirical, alam2026analyzing, li2021understanding, openja2020analysis, yang2016security, bagherzadeh2019going, scoccia2021challenges}, the first author proposed initial topic labels based on the top representative keywords and a manual review of at least 25 representative artifacts per topic. The first author has over five years of experience working with nf-core pipelines and more than a decade of professional software development experience. The proposed labels were then reviewed and refined through discussions with another author with over nine years of experience in scientific pipeline research and an additional expert with more than two decades of experience in empirical software engineering. Ambiguous labels and borderline cases were discussed until consensus was reached. This manual validation helped ensure that the final labels reflected both the topic-model output and the domain-specific context of nf-core maintenance and support.

\subsubsection{Results of RQ1}
\textbf{RQ1} identifies distinct but complementary maintenance and support topics across the three artifact types. As shown in Tables~\ref{tab:issue_topics}, \ref{tab:pr_topics}, and \ref{tab:forum_topics}, GitHub issues, pull requests, and forum discussions expose different layers of nf-core ecosystem work.
\begin{ThreePartTable}

\begin{TableNotes}
\footnotesize
\item \textit{Note.} Topic names were assigned based on BERTopic representations and inspection of representative issue discussions.
\end{TableNotes}

\footnotesize
\setlength{\tabcolsep}{2.5pt}
\renewcommand{\arraystretch}{0.88}

\begin{xltabular}{\textwidth}{
@{}
>{\centering\arraybackslash}p{0.025\textwidth}
>{\raggedright\arraybackslash}p{0.135\textwidth}
>{\raggedright\arraybackslash}p{0.17\textwidth}
>{\raggedright\arraybackslash}X
@{}}

\caption{Maintenance and support topics identified from nf-core GitHub issue discussions.}
\label{tab:issue_topics}\\

\toprule
\textbf{SL} & \textbf{Topic Name} & \textbf{Representation} & \textbf{Concise Description with Examples} \\
\midrule
\endfirsthead

\caption[]{Maintenance and support topics identified from nf-core GitHub issue discussions.}\\

\toprule
\textbf{SL} & \textbf{Topic Name} & \textbf{Representation} & \textbf{Concise Description with Examples} \\
\midrule
\endhead

\midrule
\multicolumn{4}{r}{\textit{Continued on next page}}\\
\endfoot

\bottomrule
\insertTableNotes
\endlastfoot

1 & \textbf{Pipeline Execution and Debugging Support}
& pipeline, output, version, command use, add, relevant file, description feature, module, file response, check
& This topic captures running and debugging issues, including command-line errors, execution failures, unexpected outputs, version warnings, and relevant-file inspection.  Examples include \texttt{Run cannot start due to "Unknown option"}, which reports a pipeline failure caused by an unrecognized option, and \texttt{Pipeline line warns igenomes-related files missing, despite specifying skip\_features}, which describes an unexpected iGenomes-related warning in \texttt{nf-core/taxprofiler}.\\
\addlinespace[1pt]

2 & \textbf{Module Development, Updates, and Migration}
& work module, exist module, open search, module search, exist open, planning, track work, add facilitate, issue information
& This topic represents coordination issues around module-level work, including new module requests, searches for existing implementations, module updates, and migration to newer nf-core standards. For example, issues to propose new modules such as \texttt{new module: FASTQDL}, track implementation progress, improve existing module behavior such as \texttt{Improve module specific resource requests}, or coordinate standardization work such as \texttt{migrate phispy to nf-test}.\\
\addlinespace[1pt]

3 & \textbf{Sequencing, Alignment, and Genome Reference Support}
& genome, sequence, read, add, description feature, variant, module, propose pipeline, support, alignment
& This topic focuses on sequencing and genome-resource issues, including reference genomes, sequencing reads, alignment steps, variant processing, and database inputs.  Examples include \texttt{Reference genome downloaded for each sample}, which concerns repeated reference-genome handling in \texttt{nf-core/methylseq}; \texttt{bwameth index creation fail}, which reports an indexing failure while using the \texttt{bwameth aligner}; and \texttt{No variant calling jobs if bait\_padding is set to zero}, where \texttt{nf-core/raredisease} skips DeepVariant jobs and then runs GLnexus without VCF inputs, causing the pipeline to crash.\\
\addlinespace[1pt]

4 & \textbf{Pipeline Tooling, Documentation, and Feature Support}
& update, documentation, feature, pipeline, make, module, browser, community, template, rfc
& This topic captures tooling, documentation, and feature-support issues, including pipeline-download problems, missing documentation, tooling behavior, and usability-focused requests.  For example, \texttt{nf-core pipeline downloads issues when downloading pipelines for --platform} reports bugs in \texttt{nf-core/tools} when downloading pipelines for Seqera Platform use; \texttt{Missing instruction on how to update nf-core tools in Tools documentation} identifies a documentation gap; and \texttt{Add support for giving NCBI acc2tax files to MALT databases} requests support for providing NCBI accession-to-taxonomy files during MALT database construction.\\
\addlinespace[1pt]

5 & \textbf{Subworkflow Testing and nf-test Migration}
& subworkflow, nftest, module, test, feature relate, additional context, problem, response alternative, create, update
& This topic highlights subworkflow testing, validation, installation, and migration to \texttt{nf-test}. Examples include \texttt{Subworkflow install is not properly finding modules with "\_"}, which reports missing module downloads during subworkflow installation; \texttt{Migrate all subworkflows to nf-test}, which tracks the transition to \texttt{nf-test}; and \texttt{Add ability to install a subworkflow composed nf-core and non-nf-core modules/subworkflows}, which requests more flexible subworkflow installation support.\\
\addlinespace[1pt]

6 & \textbf{Runtime Bugs and Diagnostics Reports}
& description bug, version, run, terminal output, pipeline, relevant file, information, check, step
& This topic focuses on runtime bugs and diagnostic reports supported by version details, commands, terminal output, relevant files, and system information. Examples include \texttt{Sarek possibly supplying the wrong number of args for CONTROLFREEC\_ASSESSSIGNIFICANCE?} in \texttt{nf-core/sarek}, which reports a Control-FREEC runtime failure involving R-script arguments, Slurm, and Singularity.\\
\addlinespace[1pt]

7 & \textbf{Container Images and Runtime Environment Failures}
& singularity, image, container, version, pipeline, description bug, docker image, use terminal, output, response information
& This topic covers containerized runtime environments, including Singularity/Apptainer, Docker images, image downloads, offline execution, and executor-related failures. Examples include \emph{Issue with pulling singularity images} in \texttt{nf-core/rnafusion}, \emph{docker: failed to register layer: Error processing tar file} in \texttt{nf-core/rnaseq}, and \emph{Change devcontainer to install tools at current commit} in \texttt{nf-core/tools}.\\
\addlinespace[1pt]

8 & \textbf{MultiQC Reporting and Quality-Control}
& multiqc report, plot, add, description feature, table, summary, qc, stat, generate, pipeline
& This topic centers on generating, extending, and correcting quality-control reports, especially MultiQC summaries, plots, statistics, and pipeline-specific reporting behavior.  Examples include \texttt{Increase the number of decimal points in MultiQC reporting of Endogenous DNA} in \texttt{nf-core/eager}, which requests more precise reporting of endogenous DNA values, and \texttt{SortMeRNA stats in MultiQC report are reported for read2} in \texttt{nf-core/rnaseq}, which reports unexpected read-specific SortMeRNA statistics in the MultiQC output.\\
\addlinespace[1pt]

9 & \textbf{Pipeline Test Infrastructure and CI Validation}
& datum, description feature, test dataset, add test, pipeline, ci test, test profile, create, aws, stub
& This topic captures test infrastructure for validating nf-core pipelines, modules, and subworkflows, including test datasets, CI profiles, stub runs, runtime optimization, and test-configuration tooling. Examples include \texttt{Move the test data into a new repository and set up clones}, which proposes moving bundled test data into a separate repository for easier testing; \texttt{Reduce run time of CI tests}, which discusses reducing long continuous-integration runtimes through test-profile or parallelization changes.\\
\addlinespace[1pt]

10 & \textbf{Pipeline Modernization and Template Migration}
& migrate, version, pipeline, template, local module, nftest, replace, rfc, update nfcoremodule, feature add
& This topic highlights migration to newer nf-core conventions, including template migration, local-module replacement, \texttt{nf-test} adoption, version compatibility, schema changes, and pipeline modernization.  Examples include \texttt{Update lint checks for modules\_testdata\_base\_path and pipelines\_testdata\_base\_path}, which discusses improving parameter-related lint checks, and \texttt{Make utils\_nfcore\_<pipeline>\_pipeline subworkflow nf-core linting compliant or do not lint file}, which reports linting problems in local subworkflow templates.\\
\addlinespace[1pt]

11 & \textbf{Automated Linting and Template Compliance}
& lint, pipeline, module, check, description bug, sync, terminal output, information, run, template branch
& This topic focuses on nf-core linting and automated template checks across pipelines, modules, and subworkflows. Examples include \emph{Update lint checks for modules\_testdata\_base\_path and pipelines\_testdata\_base\_path} and \emph{Make utils\_nfcore\_<pipeline>\_pipeline subworkflow nf-core linting compliant or do not lint file}.\\

\end{xltabular}
\end{ThreePartTable}

\paragraph{GitHub issues capture problem reporting and maintenance coordination.}
The issue topics show that GitHub issues are the primary repository-centered space for surfacing pipeline problems, requesting changes, and coordinating maintenance work. Several topics concern user-visible execution and runtime problems, including pipeline execution failures, debugging support, runtime bug reports, container-image failures, MultiQC reporting problems, and genome-reference handling. These topics suggest that issues often translate observed pipeline failures into repository-level concerns that can be discussed, triaged, and potentially converted into maintenance tasks.

Issues also capture coordination around reusable nf-core components and shared ecosystem standards. Topics related to module development, subworkflow testing, nf-test migration, CI validation, template migration, and automated linting show that issues are used to plan and track work that affects pipelines, modules, subworkflows, templates, and testing infrastructure. Thus, issue discussions are not limited to conventional bug reports; they also support planning, standardization, migration, validation, and documentation-oriented maintenance across the nf-core ecosystem.
\begin{ThreePartTable}

\begin{TableNotes}
\footnotesize
\item \textit{Note.} Topic names were assigned based on BERTopic representations and inspection of representative pull request discussions.
\end{TableNotes}

\footnotesize
\setlength{\tabcolsep}{2pt}
\renewcommand{\arraystretch}{0.95}

\begin{xltabular}{\textwidth}{
@{}
>{\centering\arraybackslash}p{0.025\textwidth}
>{\RaggedRight\arraybackslash}p{0.125\textwidth}
>{\RaggedRight\arraybackslash}p{0.17\textwidth}
>{\RaggedRight\arraybackslash}X
@{}}

\caption{Maintenance and support topics identified from nf-core GitHub pull request discussions.}
\label{tab:pr_topics}\\

\toprule
\textbf{SL} & \textbf{Topic Name} & \textbf{Representation} & \textbf{Concise Description with Example} \\
\midrule
\endfirsthead

\caption[]{Maintenance and support topics identified from nf-core GitHub pull request discussions (continued).}\\

\toprule
\textbf{SL} & \textbf{Topic Name} & \textbf{Representation} & \textbf{Concise Description with Example} \\
\midrule
\endhead

\midrule
\multicolumn{4}{r}{\textit{Continued on next page}}\\
\endfoot

\bottomrule
\insertTableNotes
\endlastfoot

1 & \textbf{Module Development and Quality Assurance}
& add test, module, new tool, convention, make, comment, documentation update, change reason, contain description, fix bug
& This topic captures pull requests that extend and validate the nf-core module ecosystem. It includes adding new modules, updating existing modules, integrating test data, standardizing module inputs and outputs, and validating modules through container-based CI. Examples include pull requests adding modules such as \texttt{BUSCO} and \texttt{gatk4/splitcram}, as well as updating \texttt{antismashlite} outputs with nf-core checklist requirements for tests, documentation, naming conventions, and parameters.\\

2 & \textbf{Pipeline Documentation, Testing, and Template Alignment}
& documentation update, new tool, add test, nfcoretestdataset repository, make lint, repository make, follow pipeline, output documentation, update include, update output
& This topic highlights pipeline- and ecosystem-level pull requests that align nf-core pipelines with community expectations for documentation, tests, outputs, templates, and usability. Examples include \texttt{epitopeprediction} pull requests for template updates, Zenodo link fixes, and empty prediction-result fixes; \texttt{airrflow} pull requests adding \texttt{nf-test}, merging template updates, and addressing minor bugs; and \texttt{website} pull requests updating ByteSize links, pipeline buttons, and interface spacing.\\

3 & \textbf{Module Review, Linting, and Interface Validation}
& requirement, module, test add, option guideline, follow parameter, input-output option, naming convention, docker singularity, quite flaky
& This topic captures pull request work on validating nf-core modules against review requirements, including linting, input/output definitions, parameter conventions, test data, documentation, and container compatibility. Examples include pull requests, \texttt{nf-core/modules} \texttt{update arriba}, \texttt{nf-core/sarek} \texttt{Fix linte}, and \texttt{nf-core/tools} \texttt{add module linting condition}, which address module conventions, deprecated parameters, tests, linting, and tooling support for module validation.\\

4 & \textbf{Routine Maintenance and Reliability Improvements}
& add test, comment contain, change reason, documentation update, fix bug, module, lint, file, make
& This topic captures routine pull requests that improve nf-core reliability across tools, pipelines, websites, configurations, and test-data repositories. It includes bug fixes, documentation updates, lint adjustments, parameter handling, test-data support, and usability improvements. Examples include \texttt{nf-core/tools} pull requests such as \texttt{Dump pipeline parameters into a json file}, \texttt{Add singularity note for offline use}, and \texttt{Remove params.enable\_conda}, which support JSON parameter export, offline Singularity guidance, and updated Conda configuration practices. It also includes smaller pipeline or website fixes, such as correcting module specification text.\\

5 & \textbf{Template Update Propagation Across Pipelines}
& nfcoretool template, automate attempt, apply relevant, make resolve, branch fork, resolve conflict, instruction information, complete make, new minor, update pipeline
& This topic covers pull requests that propagate shared nf-core template and tooling updates across existing pipelines. Such changes often require maintainers to apply template updates, resolve merge conflicts, update pipeline files, refresh documentation, and keep pipelines aligned with the current \texttt{nf-core/tools} release. Examples include repeated \texttt{Template update nfcoretool} pull requests in pipelines such as \texttt{eager}, \texttt{crisprseq}, and \texttt{proteomicslfq}; \texttt{nf-core/tools} pull requests such as \texttt{[FIX] Missing r-markdown dependency}, which fixes a missing template dependency; and follow-up template-merge changes such as \texttt{Some more tweaks from the final rnaseq template merge}.\\

6 & \textbf{Contribution Checklist and Review Preparation}
& contribute, update learn, appropriate delete, relevant common, thing comment, lint documentation, add test, necessary make, branch nfcoretestdataset, fix
& This topic represents pull requests structured around the nf-core contribution checklist for review preparation. These include documenting change rationale, identifying bug fixes or tool additions, adding tests, updating usage/output documentation, linking \texttt{nf-core/test-datasets} branches, and running checks. Examples include \texttt{nf-core/rnaseq} pull requests \texttt{rnaseqconfig} and \texttt{dupradar fusion}, \texttt{nf-core/scrnaseq} \texttt{Fix time limits}, and \texttt{nf-core/tools} \texttt{Don't pin nf-validation plugin version} whose bodies contain these review-preparation checklist items.\\

7 & \textbf{Execution Profile and Infrastructure Configuration}
& config, directory add, custom profile, file toplevel, step, cluster, target branch, relevant issue, include link, submit work
& This topic highlights pull requests that add or update nf-core execution profiles and infrastructure-specific configurations, including custom profiles, cluster settings, cloud/batch definitions, GPU behavior, and Singularity or executor options. Examples include new configuration-profile submissions such as \texttt{googlebatch} and \texttt{pawsey\_nimbus}; Azure Batch pool configuration in \texttt{nf-core/test-datasets}; GPU profile fixes across Docker, Singularity/Apptainer, Slurm, AWS Batch, Google Cloud, and Kubernetes; and pipeline-level configuration changes in \texttt{nf-core/methylseq} and \texttt{nf-core/rnafusion}.\\

8 & \textbf{Community Infrastructure and Website Maintenance}
& netlify, project, local site, hackathon, group, update, registration, time
& This topic captures pull requests that maintain nf-core's community-facing infrastructure, including website content, documentation pages, event and hackathon resources, local hub information, registration workflows, ByteSize materials, and Netlify-backed deployment behavior. Examples include \texttt{nf-core/website} pull requests such as \texttt{Add high-level section in style-guide.md}, \texttt{yt and figshare link for bytesize precommit}, and \texttt{identify busy netlify functions}.\\

9 & \textbf{Automated Dependency and Repository Workflow Maintenance}
& rebase retry, compare source, dependabot, update, define automerge, time schedule, branch creation, repository job, generate mend, renovate view
& This topic focuses on pull requests that maintain nf-core's automated repository workflows and development dependencies, including Renovate/Dependabot updates, GitHub Actions updates, pre-commit hook updates, Docker/devcontainer digests, automation schedules, linting support, and template-related workflow checks. Examples include \texttt{nf-core/tools} pull requests such as \texttt{Update GitHub Actions}, \texttt{Update astral-sh/setup-uv action to v8}, and \texttt{Update actions/checkout action to v6}.\\

10 & \textbf{Reference Data and Alignment Test Resources}
& file, add, genome, sequence, alignment, test datum, bam, dataset, include
& This topic highlights pull requests that add, update, or fix reference and alignment-related data resources used to validate nf-core pipelines and modules, including genome files, sequence inputs, BAM/alignment files, database files, and test datasets. Examples include \texttt{nf-core/test-datasets} pull requests such as \texttt{Replace corrupted sample url file} and \texttt{Salmon results set for tximport testing}, as well as module and pipeline updates such as \texttt{nf-core/modules} pull requests for \texttt{update datavzrd} and \texttt{fix bismark \& bwameth align fasta symlink for multiple samples}.\\

11 & \textbf{Runtime Environment and Container Dependency}
& conda, singularity, dependency, test, schedule, docker image, automerge, tag, update fix
& This topic represents pull requests that maintain software environments used by nf-core modules and pipelines, especially Conda references, Docker/Singularity container images, BioContainers tags, dependency pinning, and container-related test behavior. Examples include \texttt{nf-core/tools} pull requests such as \texttt{Fix conda environment reference} and \texttt{Download: Seqera container support - Patch 1}; \texttt{nf-core/methylseq} pull requests updating the \texttt{bwameth} container image version; and \texttt{nf-core/modules} updates involving BioContainers Docker tags and module container versions.\\

12 & \textbf{Subworkflow Integration and Pipeline Metadata Management}
& update json, documentation update, tool, automate, test add, subworkflow, repository make, pipeline convention, nfcoretestdataset repository, include new
& This topic focuses on pull requests that integrate or refine subworkflows and update pipeline metadata, documentation, JSON/schema files, test data links, and output descriptions needed to make new pipeline functionality reviewable and reproducible. Examples include the \texttt{nf-core/viralrecon} pull request \texttt{Add Minimap2, seqwish and vg variant calling for all assemblers} and the \texttt{nf-core/taxprofiler} pull request \texttt{Fix MultiQC mixing}, where changes involve subworkflow-level functionality, output documentation, test additions, and pipeline-convention updates.\\

13 & \textbf{Sample Sheet and Input Validation}
& test dataset, module, file, samplesheet, add datum, pipeline, read, testdataset, new, need
& This topic reflects pull requests focused on validating pipeline inputs through sample sheets, FASTA references, read files, and pipeline-specific test cases. Examples include \texttt{nf-core/methylseq} pull requests such as \texttt{Create a test for samplesheet with technical replicates}, which adds test coverage for sample sheet handling, and \texttt{Add fastqc outputs for testing MultiQC}, which supports downstream report testing. It also includes \texttt{nf-core/test-datasets} updates such as \texttt{Added lima and refine samplesheets to test injection points individually}, which extend test resources for validating pipeline input behavior.\\

\end{xltabular}
\end{ThreePartTable}

\begin{table*}[htbp]
\centering
\caption{Maintenance and support topics identified from Seqera Community Forum discussions.}
\label{tab:forum_topics}

\footnotesize
\setlength{\tabcolsep}{2.5pt}
\renewcommand{\arraystretch}{0.86}

\begin{tabularx}{\textwidth}{
@{}
>{\centering\arraybackslash}p{0.025\textwidth}
>{\raggedright\arraybackslash}p{0.13\textwidth}
>{\raggedright\arraybackslash}p{0.13\textwidth}
>{\raggedright\arraybackslash}X
@{}}
\toprule
\textbf{SL} & \textbf{Topic Name} & \textbf{Representation} & \textbf{Concise Description with Example} \\
\midrule

1 & \textbf{Pipeline Execution and Debugging Support}
& run, error, try, workflow, script, path, batch, make, version
& This topic centers on forum discussions where users seek help running, configuring, and debugging nf-core pipelines. These discussions often involve runtime errors, scripts, file paths, version-specific behavior, empty or incomplete logs, and uncertainty about failed tasks. Examples include \texttt{Workflow execution completed unsuccessfully?}, where a pipeline fails on Seqera Platform despite running from the command line, and \texttt{How to debug a task in the context of Fusion}, which asks how to inspect a failed task under Fusion-enabled execution.\\
\addlinespace[1pt]

2 & \textbf{Dataflow, Channels, and Output Propagation}
& channel, output, workflow, tuple, operator, collect, input file, resume
& This topic focuses on how data move through Nextflow and nf-core workflows, including channels, tuples, operators, process inputs and outputs, file collection, workflow-level outputs, and publishing behavior. Examples include \texttt{Using workflow outputs to publish files from a nested map channel}, \texttt{Stage single input for multiple instances of one process}, and \texttt{Multi-channel output cannot be applied to operator combine for which argument is already provided}.\\
\addlinespace[1pt]

3 & \textbf{Quality-Control Reporting and Result Visualization}
& table, plot, MultiQC report, custom content, generate, HTML, Picard, include
& This topic represents discussions about generating, extending, and troubleshooting MultiQC-based reporting in nf-core and Nextflow pipelines, including custom tables, plots, HTML content, images, Picard metrics, workflow-status summaries, and report-layout behavior. Examples include \texttt{Include nextflow workflow status information in MultiQC report}, \texttt{Picard QualityYieldMetrics does not display all data in QC report}, and \texttt{Pandoc fails to convert html to pdf in the presence of a unicode $\geq$ sign}.\\
\addlinespace[1pt]

4 & \textbf{Containerized Runtime Environments and Dependency Resolution}
& Singularity, image, Wave, pipeline, Docker image, Conda, directory, module, container, platform
& This topic centers on configuring and troubleshooting software environments used to run nf-core and Nextflow pipelines, especially Singularity/Apptainer images, Docker images, Wave containers, Conda environments, private registries, and container-related execution behavior. Examples include \texttt{Single Seqera container failing to build}, \texttt{Wave cannot connect with my Seqera access token}, and \texttt{Are container pulls and conda environment creations counted in realtime in a nextflow trace?}.\\
\addlinespace[1pt]

5 & \textbf{Cloud and Batch Execution Support}
& AWS Batch, run, pipeline, compute environment, error, job, log, Google, cloud
& This topic highlights discussions about running nf-core and Nextflow pipelines on cloud and managed batch infrastructures, especially AWS Batch, Google Cloud, Seqera Platform compute environments, job queues, storage permissions, Fusion, Wave, and cloud-execution logs. Examples include \texttt{Job failing in GCP due to no CredentialsProvider}, \texttt{Unable to run the submitted job through AWS batch with Seqera listed firewall configuration IP}, and \texttt{Issue with nf-core/taxprofiler Pipeline Stuck in Runnable Status on AWS Batch}.\\
\addlinespace[1pt]

6 & \textbf{Nextflow Development Tooling and Syntax Support}
& VSCode, extension, language server, config, linter, strict syntax, block, pipeline, define
& This topic centers on tools and configurations used to develop, validate, and debug Nextflow or nf-core workflows, including VSCode extension behavior, language-server initialization, linting, strict syntax, Gradle library support, \texttt{nf-test} installation, and development-environment configuration. Examples include \texttt{Importing Nextflow as a library in Gradle no longer supported?}, \texttt{Nftest installation}, and \texttt{Codespace running in recovery mode due to a configuration error}.\\
\addlinespace[1pt]

7 & \textbf{Pipeline-Specific Runtime Errors and Version Issues}
& pipeline, run, error, version, sequence, rnaseq, job, data, GRCh, step
& This topic centers on runtime problems in specific nf-core pipeline contexts, often involving version-dependent behavior, failed jobs, data handling, pipeline steps, and execution differences across environments. Examples include \texttt{Own variant calling nextflow pipeline - error with Variant Recalibrator. Argument resource is missing}, \texttt{MultiQC does not find picard rnaseqmetrics files}, and \texttt{Final step of my pipeline halts successfully after 1 sample}.\\
\addlinespace[1pt]

8 & \textbf{HPC Execution, Process Scheduling, and Resource Allocation}
& run, process, submit, number, GB, cluster, HPC, limit, executor, Slurm job
& This topic focuses on running Nextflow and nf-core pipelines on HPC and cluster environments, especially Slurm job submission, executor behavior, process parallelism, CPU and memory directives, queue limits, GPU use, and resource-allocation failures. Examples include \texttt{nf-core/rnaseq v3.18.0 -- Minimum-55-CPU rule, account suspensions \& debug-queue limits}, \texttt{Assistance Needed for Configuring SLURM to Utilize GPUs in VM for AlphaFold}, and \texttt{Slurm requiring multiple resumes for pipeline advancement}.\\

\bottomrule
\end{tabularx}
\end{table*}
\paragraph{GitHub pull requests capture implementation, review, and integration work.}
Pull request topics represent the implementation side of nf-core maintenance. Table~\ref{tab:pr_topics} shows how maintenance needs become reviewable changes to modules, pipelines, documentation, tests, templates, dependencies, execution profiles, reference data, metadata, and community infrastructure. Topics such as module development and quality assurance, module review and linting, sample-sheet validation, runtime-environment dependency updates, and subworkflow integration show that pull requests are the main mechanism through which nf-core changes are tested, reviewed, standardized, and integrated.

The pull request topics also reveal ecosystem-level maintenance mechanisms that are less visible in issue or forum discussions. Template propagation, automated dependency updates, contribution checklists, repository workflow maintenance, and CI-related changes show how nf-core keeps many repositories aligned with shared conventions. This indicates that pull requests are not only responses to individual problems; they also sustain the infrastructure and standards that make nf-core pipelines reusable and maintainable across the ecosystem.

\paragraph{Forum discussions capture user-facing support and execution contexts.}
Forum topics (\ref{tab:forum_topics}) differ from GitHub topics by foregrounding practical user-support needs. Many discussions concern running, debugging, and configuring pipelines in real execution environments. Users ask about failed runs, incomplete logs, version-specific behavior, dataflow and channel behavior, output propagation, container setup, dependency resolution, cloud and batch execution, HPC scheduling, and resource allocation. These topics capture operational problems that may not immediately appear as repository issues or pull requests, especially when the problem depends on the user's execution environment.

The forum also captures support needs around Nextflow development and reporting. Discussions about development tooling, strict syntax, nf-test installation, MultiQC reports, custom visualizations, and pipeline-specific runtime errors show that users need help both in running standardized nf-core pipelines and in developing or adapting workflow logic. Compared with GitHub issues and pull requests, the forum provides a more direct view of the barriers users encounter when applying nf-core and Nextflow pipelines across heterogeneous computational environments.

\begin{center}
\fbox{
\begin{minipage}{0.96\linewidth}
\textbf{Summary of RQ1.}
\textbf{RQ1} shows that maintenance and support in nf-core extend beyond correcting source-code defects. GitHub issues expose repository-level problems and coordination needs; pull requests implement, validate, and integrate maintenance changes; and forum discussions surface user-facing execution and infrastructure problems. Together, the topics show that sustaining nf-core requires continuous work on modules, tests, templates, documentation, containers, dependencies, execution profiles, reporting outputs, and support for users running pipelines across heterogeneous environments.
\end{minipage}
}
\end{center}

\subsection{RQ2: Cross-Platform Differences in Maintenance and Support Topics}
\subsubsection{Motivation}
\textbf{RQ1} identifies the maintenance and support topics that emerge within each artifact type. However, identifying topics separately does not show how maintenance work is distributed across the nf-core ecosystem. GitHub issues, pull requests, and Community Forum discussions play different but complementary roles: issues surface problems and maintenance needs, pull requests implement and validate changes, and forum discussions capture user-facing execution support. Comparing topics across these spaces is therefore necessary to understand which concerns are handled through repository-centered development and which concerns emerge primarily through community-centered troubleshooting. \textbf{RQ2} examines these differences to characterize the division of maintenance and support work across nf-core GitHub artifacts and community forum discussions.

\subsubsection{Approach}

To compare topics across artifact types, we used the topic labels and representative artifacts identified in RQ1. Because the topic models were trained separately for issues, pull requests, and forum discussions, we did not impose a one-to-one topic mapping. Instead, two authors with over five and nine years of experience with scientific pipelines research compared topic names, representative keywords, and representative artifacts to identify shared maintenance and support concerns that appeared across artifact types. These concerns included execution failures, containers and runtime environments, documentation and training, testing and validation, dependency updates, templates, modules and subworkflows, reporting outputs, configuration, and HPC/cloud execution.

We then examined how each shared concern appeared in each artifact type. For example, execution failures may appear in issues as runtime bug reports, in pull requests as fixes or tests, and in forum discussions as troubleshooting questions. Similarly, container concerns may appear in issues as image-pull failures, in pull requests as container or dependency updates, and in forum discussions as runtime setup questions. This comparison treats issues, pull requests, and forum discussions as complementary evidence sources rather than interchangeable text records.

\subsubsection{Results of RQ2}
RQ2 shows that similar maintenance and support concerns take different forms across the nf-core ecosystem. Table~\ref{tab:rq2_cross_artifact_matrix} summarizes these differences. GitHub issues primarily capture reported problems and coordination needs; pull requests capture implementation and integration work; and forum discussions capture user-facing execution and troubleshooting contexts.

\begin{table*}[htbp]
\centering
\caption{Cross-artifact mapping of shared maintenance and support concerns in nf-core.}
\label{tab:rq2_cross_artifact_matrix}
\scriptsize
\setlength{\tabcolsep}{4pt}
\renewcommand{\arraystretch}{1.12}
\begin{tabularx}{\textwidth}{
@{}
>{\raggedright\arraybackslash}p{0.16\textwidth}
>{\raggedright\arraybackslash}X
>{\raggedright\arraybackslash}X
>{\raggedright\arraybackslash}X
@{}}
\toprule
\textbf{Concern} & \textbf{GitHub issues} & \textbf{GitHub pull requests} & \textbf{Forum discussions} \\
\midrule
Execution failures & Runtime bug reports, failed runs, command errors, unexpected outputs, and diagnostic logs. & Fixes, tests, validation changes, and pipeline behavior updates. & Troubleshooting failed runs, interpreting logs, debugging tasks, and understanding runtime behavior. \\

Containers and runtime environments & Docker, Singularity/Apptainer, image-pull failures, offline execution, and runtime-environment problems. & Container image updates, Conda references, BioContainers tags, dependency pinning, and container-related tests. & Runtime setup questions involving Docker, Singularity/Apptainer, Wave, Conda, private registries, and platform execution. \\

Documentation and usability & Missing or outdated documentation, tooling behavior, feature requests, and user-facing guidance gaps. & Documentation updates, output documentation, template-aligned docs, and website/community content changes. & Usage confusion, requests for explanation, training-related questions, and guidance for applying nf-core or Nextflow features. \\

Testing and validation & CI failures, test datasets, stub runs, nf-test migration, linting, and validation gaps. & New tests, nf-test adoption, module validation, CI updates, checklists, and review-readiness evidence. & Questions about nf-test, output correctness, development tooling, and workflow validation behavior. \\

Modules and subworkflows & Requests for new modules, module updates, subworkflow testing, migration, and reusable-component coordination. & Module implementation, subworkflow integration, interface validation, metadata updates, and quality assurance. & Questions about adapting workflow logic, process behavior, channels, tuples, and outputs. \\

Templates and standards & Template migration, lint failures, schema changes, and compliance with nf-core conventions. & Template propagation, repository synchronization, checklist compliance, and standardization changes. & Less prominent; appears indirectly through user confusion about expected behavior or development setup. \\

Reporting and outputs & MultiQC report problems, plots, tables, statistics, and quality-control output issues. & Output documentation, MultiQC-related fixes, report test data, and reporting integration changes. & MultiQC customization, report generation, visualization, HTML/PDF output, and reporting interpretation. \\

HPC and cloud execution & Configuration problems, executor behavior, resource limits, and infrastructure-specific issue reports. & Execution profiles, cluster/cloud configuration, GPU settings, and infrastructure-specific pull request changes. & User support for Slurm, AWS Batch, Google Cloud, Seqera Platform, queues, storage, credentials, and resource allocation. \\
\bottomrule
\end{tabularx}
\end{table*}

\paragraph{GitHub issues emphasize problem reporting and maintenance coordination.}
GitHub issue topics show where nf-core users and maintainers identify problems, request changes, and coordinate maintenance work. Issues contain runtime failures, debugging reports, container-image problems, genome-reference issues, MultiQC reporting problems, linting failures, template migration requests, and test-infrastructure concerns. These topics indicate that issues often function as a repository-centered entry point where user-visible problems are translated into maintainable tasks. Issues also support coordination around reusable components, including module development, subworkflow testing, nf-test migration, and ecosystem-wide standardization.

\paragraph{GitHub pull requests emphasize implementation, review, and integration.}
Pull request topics are more implementation-oriented. They capture the concrete changes through which nf-core maintenance work is performed, including module development, test additions, documentation updates, dependency updates, template propagation, execution-profile changes, sample-sheet validation, metadata updates, and community-infrastructure maintenance. Pull requests therefore represent the integration layer of the ecosystem: they convert maintenance needs into reviewable changes that can be tested, standardized, and merged. Compared with issues, pull requests more directly expose nf-core's review and quality-assurance practices, including contribution checklists, linting, interface validation, CI updates, and automation.

\paragraph{Forum discussions emphasize user-facing support and execution contexts.}
Forum discussions differ from GitHub artifacts by foregrounding practical support problems encountered during pipeline use. Forum topics include pipeline execution failures, Nextflow dataflow and channel behavior, container and dependency resolution, MultiQC reporting, cloud and batch execution, HPC scheduling, resource allocation, and development-tooling questions. These discussions often involve local or infrastructure-specific context that may not immediately correspond to a repository defect. The forum therefore exposes operational barriers that arise when users run nf-core or Nextflow workflows across local, cloud, HPC, containerized, and managed execution environments.

\paragraph{Cross-artifact interpretation.}
The comparison shows that the same broad concern often appears differently depending on artifact type. Execution problems appear in issues as bug reports, in pull requests as fixes or tests, and in forum discussions as troubleshooting requests. Container concerns appear in issues as runtime failures, in pull requests as dependency or image updates, and in forum discussions as setup and execution questions. Documentation appears in issues as gaps or requests, in pull requests as concrete documentation changes, and in forum discussions as usage confusion. HPC and cloud concerns appear in issues as configuration or executor problems, in pull requests as profile and infrastructure updates, and in forum discussions as user-support questions about queues, credentials, storage, and resources.

\begin{center}
\fbox{
\begin{minipage}{0.96\linewidth}
\textbf{Summary of RQ2.}
\textbf{RQ2} shows that no single artifact type provides a complete view of nf-core maintenance and support. GitHub issues reveal what needs attention, pull requests show how changes are implemented and integrated, and forum discussions reveal where users encounter practical barriers during execution and development. This division of work supports the cross-platform design of the study: understanding maintenance and support in nf-core requires analyzing repository-centered and community-centered spaces together.
\end{minipage}
}
\end{center}

\subsection{RQ3: Factors Associated with Resolution Outcomes}

\subsubsection{Motivation}

\textbf{RQ1} and \textbf{RQ2} show what maintenance and support concerns appear in nf-core and how these concerns differ across GitHub issues, pull requests, and Seqera Community Forum discussions. However, topic distributions alone do not show how the ecosystem responds to reported problems, proposed changes, and user-support requests. Some artifacts are closed, merged, or answered quickly, while others remain open, are closed without merge, or receive no accepted answer. Understanding these differences is important because resolution outcomes provide evidence about how maintenance and support work is handled across the ecosystem.

\textbf{RQ3} therefore examines artifact-level factors associated with resolution outcomes. Because this is an observational repository- and forum-mining study, we interpret the results as lifecycle associations rather than causal effects or early prediction models. In particular, some features are available when an artifact is submitted, such as text, author type, code blocks, URLs, error mentions, version mentions, and topic assignment. Other features emerge during handling, such as comments, assignees, milestones, requested reviewers, replies, and likes. We therefore treat these later features as process signals that describe how artifacts are handled, not as independent causes of resolution.

\subsubsection{Approach}
We analyzed resolution outcomes separately for the three artifact types, following the procedure described in Section~\ref{sec:resolution-analysis}. For issues, we examined closure and time-to-close. For pull requests, we examined merge, closed-without-merge, and time-to-final-decision. For forum discussions, we examined accepted-answer status and engagement; because accepted-answer timestamps were unavailable, forum lifecycle curves are interpreted as exploratory activity-span summaries rather than exact time-to-answer estimates.

For each artifact type, we combine descriptive outcome summaries, bivariate associations, multivariable models, and lifecycle curves. The descriptive analysis reports closure, merge, closed-without-merge, accepted-answer, and timing distributions. The bivariate analysis compares outcome rates across individual features. The multivariable analysis examines adjusted associations while accounting for other artifact characteristics. Kaplan--Meier curves are used descriptively to compare lifecycle trajectories across selected binary features.  This artifact-specific design allows us to compare how resolution differs across problem reports, implementation work, and user-support discussions while respecting the distinct workflows of GitHub and the community forum.

\subsubsection{Results of RQ3}
Overall, resolution outcomes in nf-core are associated with three recurring conditions: actionability, coordination, and diagnostic evidence. Issues were more likely to close when they showed maintainer attention or contained concrete technical details. Pull requests were more likely to merge when they were review-ready, linked to issues, supported by tests or checklists, or authored by contributors close to the project. Forum discussions were more likely to receive accepted answers when they contained concrete technical evidence and sustained interaction. In contrast, infrastructure-specific and coordination-intensive artifacts, especially those involving cloud, HPC, templates, or workflow semantics, were less consistently resolved or took longer to reach a final state.

\paragraph{GitHub issue resolution.}
Among 15,760 GitHub issues, 12,471 were closed, corresponding to a closure rate of 79.13\%, while 3,289 remained open. Closure time was strongly right-skewed. Among closed issues, the median time-to-close was 27.91 days, compared with a mean of 131.25 days. The percentile distribution further shows this long tail: 25\% of closed issues were resolved within 4.05 days, 50\% within 27.91 days, 75\% within 140.01 days, and 95\% within 655.03 days. Most issues with an explicit state reason were closed as \texttt{completed} ($n=11{,}954$), while fewer were marked \texttt{not\_planned} ($n=483$). Duplicate ($n=34$) and reopened ($n=53$) issues were rare. This indicates that issue closure in nf-core is dominated by completed maintenance or support actions rather than duplicate removal or rejection.

\begin{figure}[htbp]
    \centering
    \includegraphics[width=0.75\textwidth]{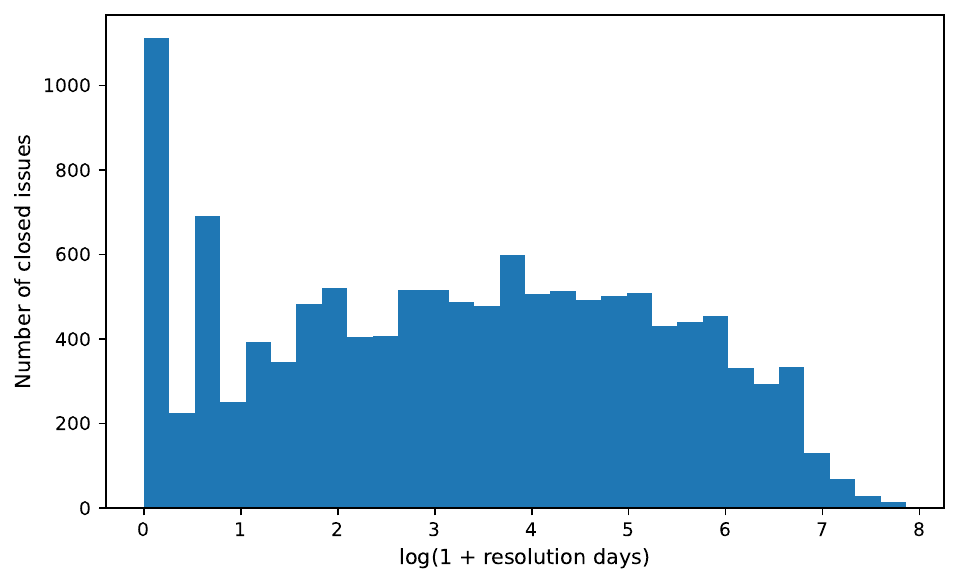}
    \caption{Log-transformed distribution of time-to-close for closed issues.}
    \label{fig:issues_time_to_close_distribution}
\end{figure}

Figure~\ref{fig:issues_time_to_close_distribution} shows the log-transformed distribution of time-to-close for closed nf-core issues. The distribution confirms that issue resolution is highly right-skewed: many issues are closed within a short period, while a substantial tail of issues remains open for much longer before closure. This pattern suggests that nf-core handles many issues quickly, but some issues require extended discussion, delayed prioritization, complex fixes, or long-term tracking.

\begin{table}[htbp]
\centering
\caption{Selected bivariate associations between issue features and closure}
\label{tab:issue_bivariate_closure}

\footnotesize
\setlength{\tabcolsep}{3.2pt}
\renewcommand{\arraystretch}{1.05}

\begin{tabularx}{.75\columnwidth}{Xrrrr}
\toprule
\textbf{Feature} & \textbf{Absent} & \textbf{Present} & \textbf{OR} & \textbf{Adj. $p$} \\
\midrule
Has comments           & 66.28\% & 86.22\% & 3.18 & $<.001$ \\
Has assignee           & 73.42\% & 88.11\% & 2.68 & $<.001$ \\
Good first issue label & 78.69\% & 91.60\% & 2.93 & $<.001$ \\
Has milestone          & 77.27\% & 86.34\% & 1.86 & $<.001$ \\
Bug label              & 76.83\% & 84.79\% & 1.68 & $<.001$ \\
Mentions error         & 76.84\% & 83.21\% & 1.49 & $<.001$ \\
Mentions version       & 77.85\% & 83.22\% & 1.41 & $<.001$ \\
Has checklist          & 78.32\% & 83.73\% & 1.42 & $<.001$ \\
Has URL                & 80.97\% & 76.66\% & 0.77 & $<.001$ \\
Enhancement label      & 83.56\% & 68.62\% & 0.43 & $<.001$ \\
\bottomrule
\end{tabularx}

\caption*{\footnotesize \textit{Note.} Absent and Present indicate issue-closure rates when the feature is absent or present. OR denotes the odds ratio for closure when the feature is present relative to absent. Adjusted $p$-values use Benjamini--Hochberg correction.}

\end{table}

\textbf{Bivariate associations with issue closure.} We examined a broader set of binary issue features, including triage metadata, author-role indicators, textual evidence, interaction signals, and common labels. For each feature, we compared closure rates when the feature was absent versus present, estimated odds ratios, and applied Benjamini--Hochberg correction for multiple comparisons. Table~\ref{tab:issue_bivariate_closure} reports the most influential and interpretable associations.

Bivariate results show that issue closure was strongly associated with interaction and triage signals. Issues with comments had a much higher closure rate than issues without comments (86.22\% vs. 66.28\%; OR=3.18), and assigned issues were also more likely to close than unassigned issues (88.11\% vs. 73.42\%; OR=2.68). Milestone-linked issues followed the same pattern (86.34\% vs. 77.27\%; OR=1.86), suggesting that discussion, ownership, and planning are closely related to issue resolution. Label-based patterns show that \texttt{good first issue} labels were associated with the highest closure rate among the selected features (91.60\%; OR=2.93), while \texttt{bug} labels were also positively associated with closure (84.79\% vs. 76.83\%; OR=1.68). In contrast, \texttt{enhancement} issues were much less likely to close (68.62\% vs. 83.56\%; OR=0.43), indicating that feature-oriented requests may require longer deliberation or remain open as future work. Textual evidence also mattered: issues mentioning errors, versions, or checklists had higher closure rates, whereas issues containing URLs had lower closure rates (76.66\% vs. 80.97\%; OR=0.77). These results suggest that closure is linked to visibility, ownership, and diagnostic clarity, while feature-oriented or context-heavy issues often require longer discussion or remain open as future work. The complete set of tested features is available in the replication package.

\begin{figure}[htbp]
    \centering
    \includegraphics[width=0.75\textwidth]{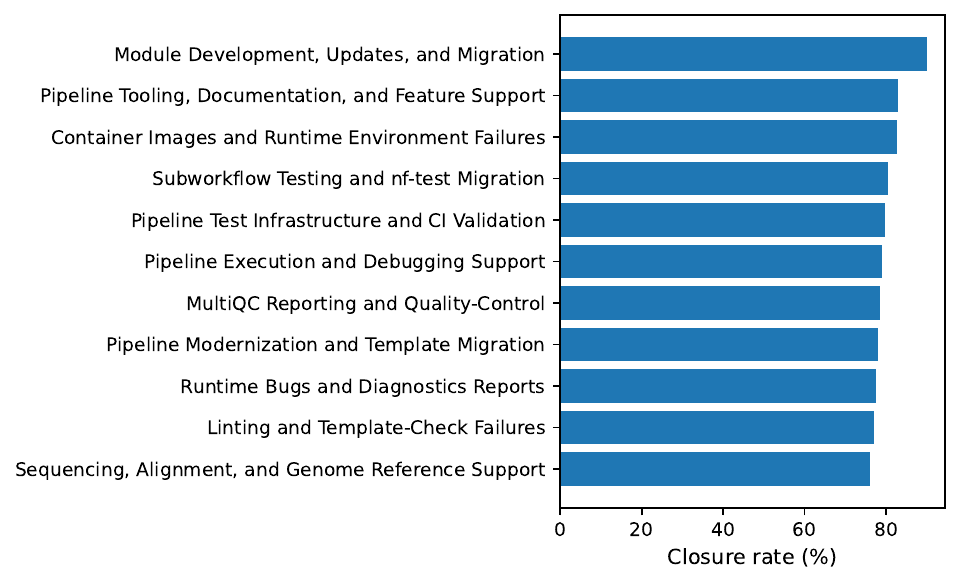}
    \caption{Issue closure rate by topic.}
    \label{fig:issue_closure_rate_by_topic}
\end{figure}

\textbf{Topic-level variation in issue resolution.} Issue closure rates varied across maintenance and support topics, although most topics showed relatively high resolution levels. As shown in Figure~\ref{fig:issue_closure_rate_by_topic}, \textit{Module Development, Updates, and Migration} had the highest closure rate, indicating that module-related coordination and migration tasks are often well scoped and actionable. \textit{Pipeline Tooling, Documentation, and Feature Support} and \textit{Container Images and Runtime Environment Failures} also showed comparatively high closure rates. In contrast, \textit{Sequencing, Alignment, and Genome Reference Support}, \textit{Automated Linting and Template Compliance}, and \textit{Runtime Bugs and Diagnostics Reports} had lower closure rates, suggesting that issues involving domain-specific resources, template/linting behavior, or runtime diagnosis often require more investigation and coordination before resolution. Overall, topic-level variation indicates that issue resolution depends not only on whether an issue receives attention, but also on the technical scope and maintainability of the reported concern.

\begin{table}[htbp]
\centering
\caption{Selected predictors from the multivariable logistic regression model for issue closure}
\label{tab:issue_logistic_regression}

\scriptsize
\setlength{\tabcolsep}{3pt}
\renewcommand{\arraystretch}{1.08}

\begin{tabularx}{0.85\columnwidth}{@{}
>{\raggedright\arraybackslash}X
>{\centering\arraybackslash}p{0.10\columnwidth}
>{\centering\arraybackslash}p{0.150\columnwidth}
>{\centering\arraybackslash}p{0.15\columnwidth}
>{\centering\arraybackslash}p{0.10\columnwidth}
@{}}
\toprule
\textbf{Predictor} & \textbf{OR} & \textbf{95\% CI low} & \textbf{95\% CI high} & \textbf{Adj. $p$} \\
\midrule
Number of assignees & 2.45 & 2.22 & 2.71 & $<.001$ \\
Comments & 1.14 & 1.11 & 1.16 & $<.001$ \\
Enhancement label & 0.49 & 0.43 & 0.56 & $<.001$ \\
Has milestone & 1.61 & 1.40 & 1.85 & $<.001$ \\
Has URL & 0.74 & 0.67 & 0.82 & $<.001$ \\
Mentions version & 1.46 & 1.28 & 1.67 & $<.001$ \\
Bug label & 1.62 & 1.37 & 1.92 & $<.001$ \\
Mentions system information & 0.75 & 0.67 & 0.84 & $<.001$ \\
Has parent issue & 0.61 & 0.50 & 0.75 & $<.001$ \\
Number of labels & 0.82 & 0.75 & 0.89 & $<.001$ \\
Author is member/owner & 1.35 & 1.18 & 1.53 & $<.001$ \\
Has code block & 0.74 & 0.64 & 0.85 & $<.001$ \\
Mentions error & 1.40 & 1.19 & 1.65 & $<.001$ \\
Good first issue label & 1.76 & 1.23 & 2.53 & $<.01$ \\
Text length & 0.91 & 0.85 & 0.97 & $<.01$ \\
Author is contributor & 1.17 & 1.03 & 1.34 & $<.05$ \\
\bottomrule
\end{tabularx}

\caption*{\footnotesize \textit{Note.} OR denotes the odds ratio for issue closure after adjusting for other predictors in the model. Values above 1 indicate higher odds of closure; values below 1 indicate lower odds. Adjusted $p$-values use Benjamini--Hochberg correction.}

\end{table}

\textbf{Multivariable model of issue closure.} To examine which issue-level factors were independently associated with closure, we fitted a robust multivariable logistic regression model with \texttt{is\_closed} as the outcome. The model included issue metadata, triage signals, textual evidence, selected labels, and categorical controls for repository and issue type. To avoid estimation problems caused by sparse categories and collinearity, rare categorical levels were collapsed, categorical controls were one-hot encoded, and constant, duplicate, and near-perfectly correlated predictors were removed before fitting the model. The final model used 15,760 issues, contained 113 predictors after encoding and cleaning, and was full rank. The standard logistic regression model converged successfully, with McFadden's pseudo-$R^2$ of 0.203.

Table~\ref{tab:issue_logistic_regression} reports the most interpretable predictors. Triage and coordination signals remained strongly associated with closure: additional assignees (OR=2.45), comments (OR=1.14), and milestones (OR=1.61) all increased closure odds. Label and textual signals also mattered. \texttt{Bug} issues (OR=1.62), \texttt{good first issue} issues (OR=1.76), version mentions (OR=1.46), and error mentions (OR=1.40) were positively associated with closure, while \texttt{enhancement} issues were less likely to close (OR=0.49). More complex or context-heavy reports showed lower closure odds, including issues with URLs (OR=0.74), system-information mentions (OR=0.75), code blocks (OR=0.74), parent-issue links (OR=0.61), more labels (OR=0.82), and longer text (OR=0.91). Overall, issue closure in nf-core is most strongly associated with ownership, discussion, planning, and concrete diagnostic evidence, while feature-oriented or more complex reports are less likely to reach closure within the observation window.

\begin{table}[htbp]
\centering
\caption{Selected Cox proportional hazards results for issue time-to-closure}
\label{tab:issue_cox_time_to_close}

\scriptsize
\setlength{\tabcolsep}{3pt}
\renewcommand{\arraystretch}{1.08}

\begin{tabularx}{0.85\columnwidth}{@{}
>{\raggedright\arraybackslash}X
>{\centering\arraybackslash}p{0.10\columnwidth}
>{\centering\arraybackslash}p{0.150\columnwidth}
>{\centering\arraybackslash}p{0.15\columnwidth}
>{\centering\arraybackslash}p{0.10\columnwidth}
@{}}
\toprule
\textbf{Predictor} & \textbf{HR} & \textbf{95\% CI low} & \textbf{95\% CI high} & \textbf{Adj. $p$} \\
\midrule
Enhancement label      & 0.71 & 0.68 & 0.74 & $<.001$ \\
Has assignee           & 1.41 & 1.33 & 1.49 & $<.001$ \\
Number of labels       & 0.86 & 0.83 & 0.89 & $<.001$ \\
Good first issue label & 1.43 & 1.30 & 1.58 & $<.001$ \\
Bug label              & 1.21 & 1.15 & 1.28 & $<.001$ \\
Has URL                & 0.89 & 0.85 & 0.92 & $<.001$ \\
Mentions error         & 1.18 & 1.12 & 1.24 & $<.001$ \\
Mentions version       & 1.14 & 1.09 & 1.20 & $<.001$ \\
Has milestone          & 1.11 & 1.06 & 1.16 & $<.001$ \\
Author is contributor  & 1.10 & 1.05 & 1.15 & $<.001$ \\
Mentions system info   & 0.91 & 0.88 & 0.95 & $<.001$ \\
Text length            & 0.97 & 0.95 & 0.99 & $<.01$ \\
\bottomrule
\end{tabularx}

\caption*{\footnotesize \textit{Note.} HR denotes the hazard ratio from the Cox proportional hazards model. HR values above 1 indicate faster issue closure; values below 1 indicate slower closure. Adjusted $p$-values use Benjamini--Hochberg correction. The full model output is available in the replication package.}

\end{table}

\textbf{Issue Time-to-closure analysis.} We further analyzed issue resolution speed using a Cox proportional hazards model, treating open issues as censored observations. The model included 15,760 issues, with 12,471 observed closures and 3,289 censored open issues, and achieved a concordance index of 0.601. In this model, hazard ratios above 1 indicate faster closure, while values below 1 indicate slower closure.

Table~\ref{tab:issue_cox_time_to_close} shows that triage and actionable diagnostic signals were associated with faster closure. Issues with assignees (HR=1.41), \texttt{good first issue} labels (HR=1.43), \texttt{bug} labels (HR=1.21), error mentions (HR=1.18), version mentions (HR=1.14), and milestones (HR=1.11) moved more quickly toward closure. In contrast, \texttt{enhancement} issues closed more slowly (HR=0.71), as did issues with more labels (HR=0.86), URLs (HR=0.89), system-information mentions (HR=0.91), and longer text (HR=0.97). These results reinforce the logistic-regression findings: issue resolution is faster when ownership and concrete diagnostic evidence are present, while feature-oriented or more complex environment-specific reports tend to progress more slowly.

\begin{figure*}[htbp]
    \centering
    \begin{subfigure}{0.32\textwidth}
        \centering
        \includegraphics[width=\linewidth]{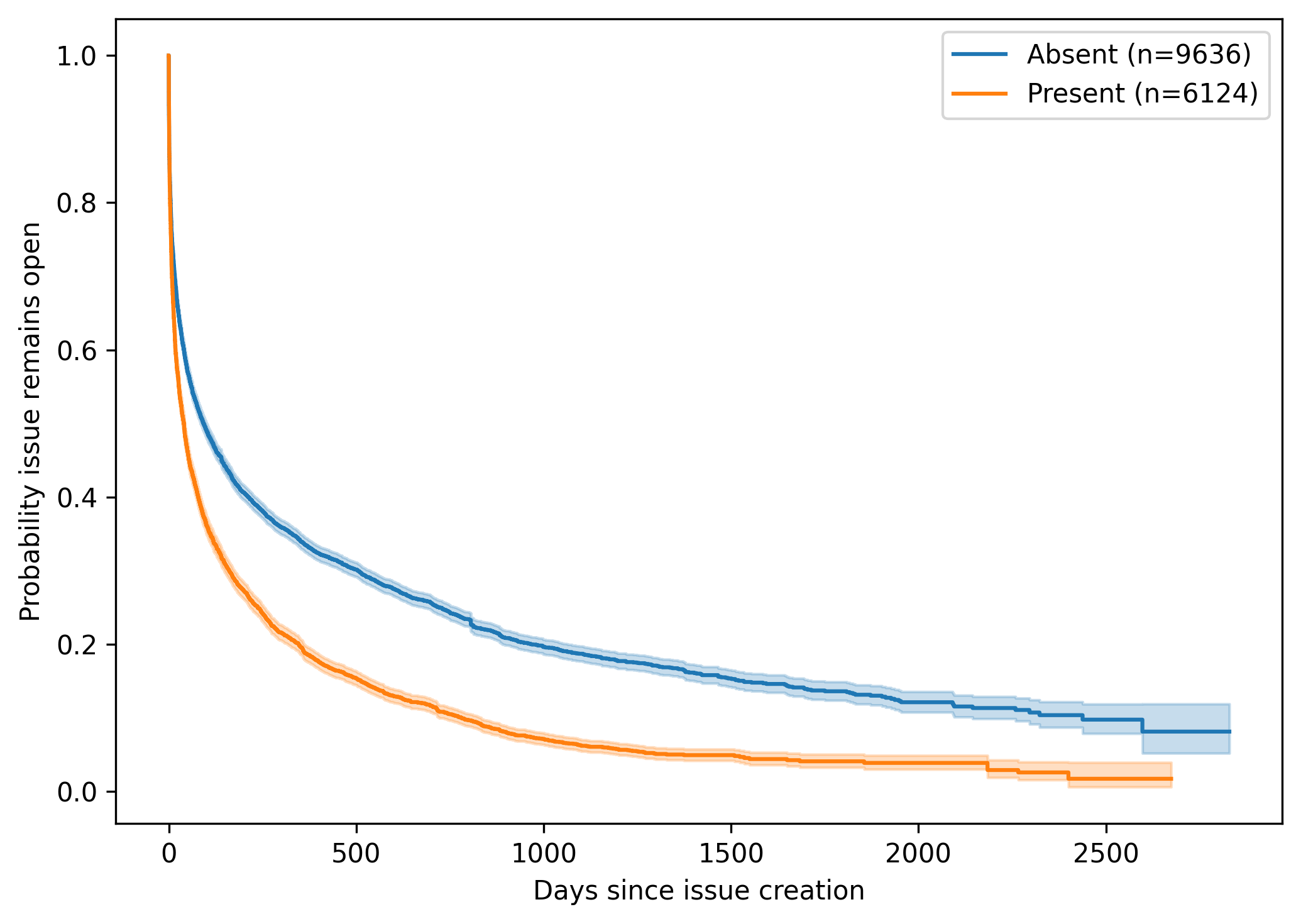}
        \caption{Assignee presence}
        \label{fig:km_has_assignee}
    \end{subfigure}
    \hfill
    \begin{subfigure}{0.32\textwidth}
        \centering
        \includegraphics[width=\linewidth]{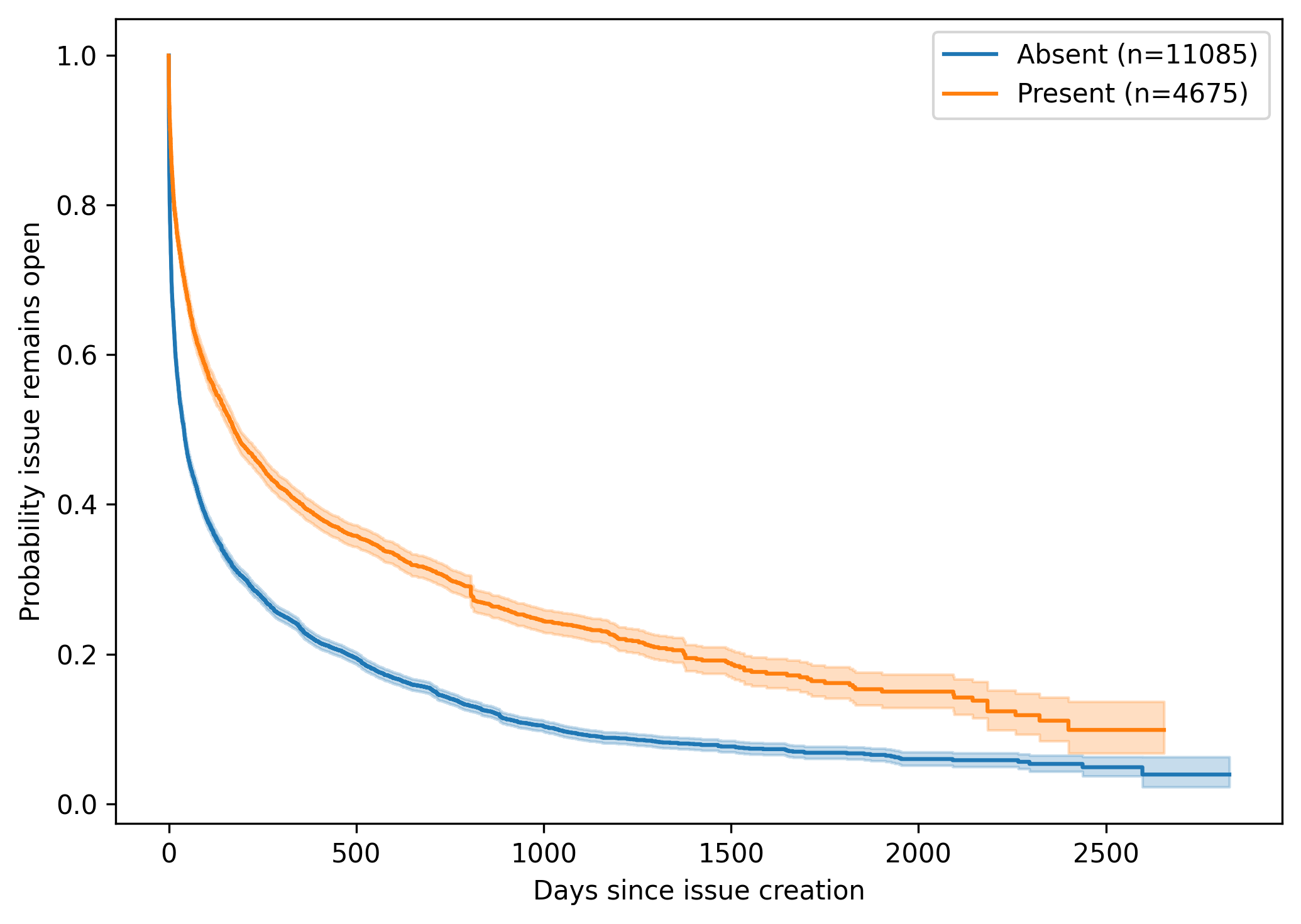}
        \caption{Enhancement label}
        \label{fig:km_label_enhancement}
    \end{subfigure}
    \hfill
    \begin{subfigure}{0.32\textwidth}
        \centering
        \includegraphics[width=\linewidth]{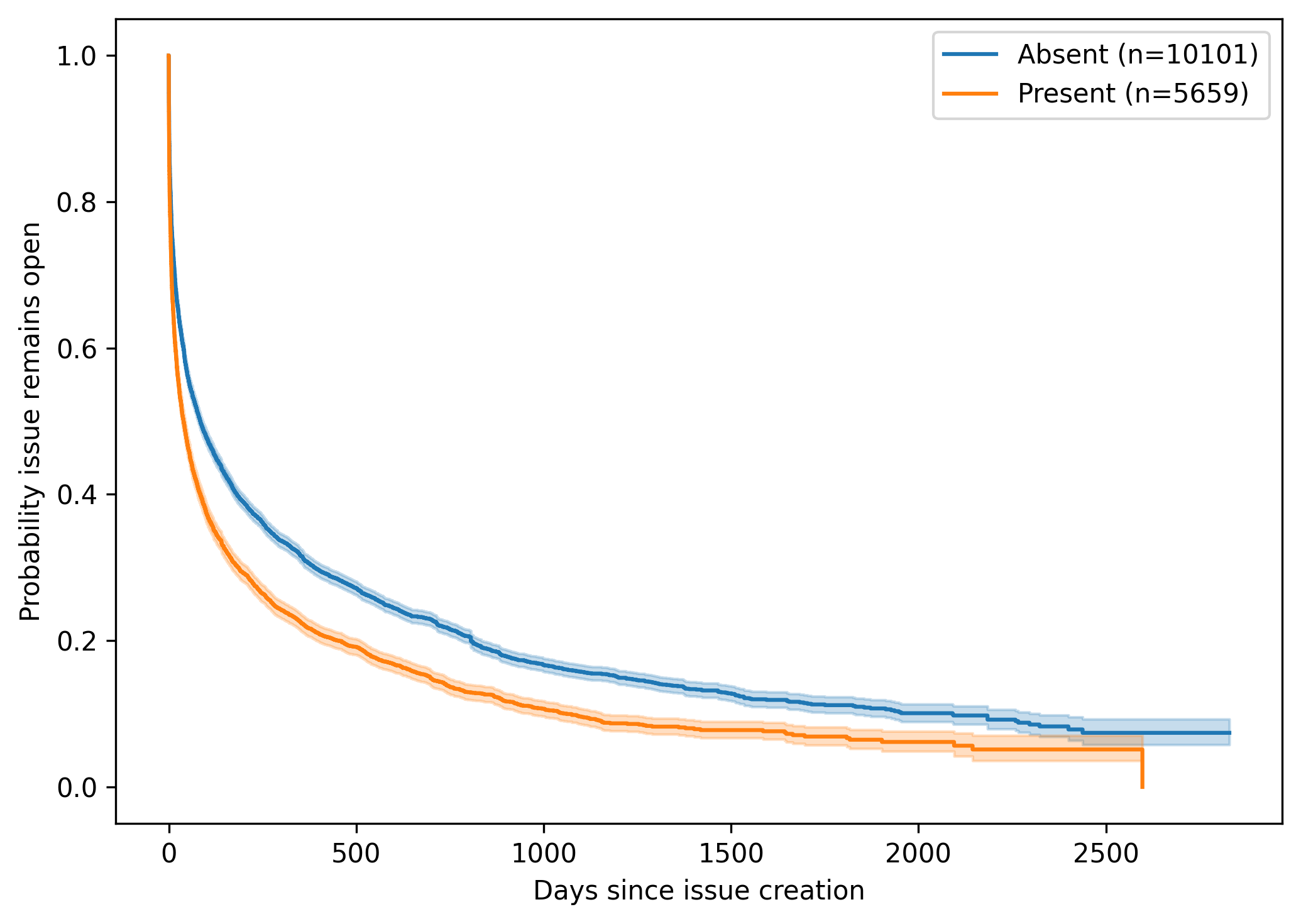}
        \caption{Error mention}
        \label{fig:km_mentions_error}
    \end{subfigure}
    \caption{Kaplan--Meier curves for selected issue features. The y-axis shows the probability that an issue remains open over time.}
    \label{fig:issue_km_selected}
\end{figure*}
\textbf{Kaplan--Meier analysis of issue-resolution trajectories.} We generated Kaplan--Meier curves for several binary issue features to examine how resolution trajectories differed over time. For space, Figure~\ref{fig:issue_km_selected} reports three representative factors that summarize the main patterns observed in the survival analysis: assignee presence, enhancement label, and error mention. The y-axis shows the probability that an issue remains open; therefore, curves that decline more quickly indicate faster closure.

Issues with assignees moved toward closure faster than unassigned issues, consistent with the Cox model result that assignee presence was associated with faster closure (HR=1.41). Issues mentioning errors also showed faster closure trajectories than issues without error mentions, supporting the interpretation that concrete diagnostic evidence makes issues more actionable (HR=1.18). In contrast, enhancement-labeled issues remained open for longer than non-enhancement issues, consistent with their lower closure hazard (HR=0.71). Overall, the Kaplan--Meier curves reinforce the regression results: issues close faster when they have clear ownership or actionable diagnostic information, whereas feature-oriented issues tend to remain open longer because they often require prioritization, design discussion, or longer-term planning.

\begin{figure}[htbp]
    \centering
    \includegraphics[width=.75\textwidth]{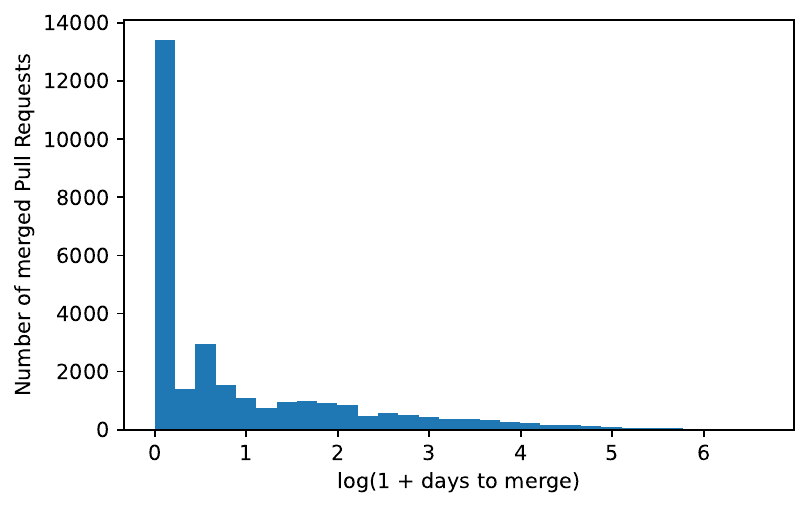}
    \caption{Log-transformed distribution of time to merge for merged pull requests.}
    \label{fig:pr_log_time_merge}
\end{figure}

\paragraph{GitHub pull request resolution.}
Among 35,411 pull requests, 32,518 were merged (91.83\%), 2,290 were closed without merge (6.47\%), and 603 remained open (1.70\%). Pull request integration was generally fast but right-skewed. The median time-to-merge was 0.50 days, while the mean was 8.04 days. Similarly, the median time-to-final-decision was 0.71 days, compared with a mean of 17.54 days. As shown in Figure~\ref{fig:pr_log_time_merge}, many pull requests were merged shortly after submission, while a smaller subset required substantially longer review, validation, or coordination.

Automation status further differentiated pull request outcomes. Human-authored pull requests formed the largest group and had a high merge rate (96.93\%; median time to merge = 0.39 days). Renovate pull requests had the highest merge rate (98.46\%; median = 0.43 days), suggesting that these automated updates align well with nf-core's review and validation workflows. In contrast, Dependabot pull requests were merged less consistently (72.13\%; median = 0.54 days), and other bot-authored pull requests had the lowest merge rate (51.94\%) and longest median time to merge (4.95 days). These results show that automated maintenance is heterogeneous: some automation supports efficient integration, whereas other bot-generated changes are more likely to require manual intervention, become obsolete, or be closed without merge.

\textbf{Review routing and issue linkage. }Pull requests with linked issues were more likely to be merged than pull requests without linked issues (96.71\% vs. 91.35\%), but they also took longer to merge (median 0.87 vs. 0.26 days). A similar pattern appears for requested reviewers: pull requests with requested reviewers had a higher merge rate than those without requested reviewers (95.70\% vs. 91.81\%), while also having a longer median time-to-merge (0.88 vs. 0.30 days). These results suggest that issue linkage and reviewer assignment support eventual integration by providing context and review routing, but they also mark pull requests that require additional coordination, verification, or maintainer attention before merging.

\begin{figure}[htbp]
    \centering
    \includegraphics[width=0.75\textwidth]{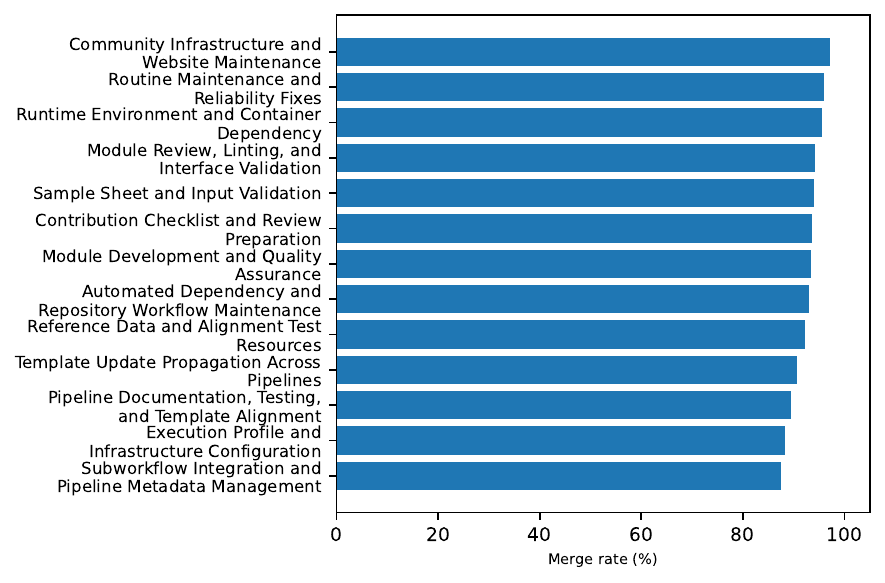}
    \caption{Merge rate by pull request topic.}
    \label{fig:pr_topic_merge_rate}
\end{figure}

\textbf{Topic-level variation in pull request integration.} Pull request merge rates varied by topic, although all topics showed high integration levels overall. As shown in Figure~\ref{fig:pr_topic_merge_rate}, \textit{Community Infrastructure and Website Maintenance} had the highest merge rate, followed by \textit{Routine Maintenance and Reliability Fixes} and \textit{Runtime Environment and Container Dependency}. These topics often involve localized website updates, reliability fixes, or concrete environment-maintenance changes that can be reviewed and integrated relatively efficiently. In contrast, \textit{Subworkflow Integration and Pipeline Metadata Management}, \textit{Execution Profile and Infrastructure Configuration}, and \textit{Pipeline Documentation, Testing, and Template Alignment} had comparatively lower merge rates, although they still remained above 87\%. These lower rates suggest that changes involving subworkflow integration, metadata, execution profiles, documentation, and template alignment may require more coordination, repository-specific review, or validation before integration. Overall, the topic-level pattern indicates that nf-core pull request integration is broadly successful, but coordination-intensive and infrastructure-oriented changes are somewhat less consistently merged than localized maintenance or community-infrastructure updates.

\begin{table}[htbp]
\centering
\caption{Topic-level pull request resolution summary}
\label{tab:pr_topic_resolution}

\scriptsize
\setlength{\tabcolsep}{3pt}
\renewcommand{\arraystretch}{1.08}

\begin{tabularx}{0.9\columnwidth}{@{}
>{\raggedright\arraybackslash}X
>{\centering\arraybackslash}p{0.1\columnwidth}
>{\centering\arraybackslash}p{0.1\columnwidth}
>{\centering\arraybackslash}p{0.1\columnwidth}
@{}}
\toprule
\textbf{Topic} & \textbf{Merged} & \textbf{Closed} & \textbf{Med. days} \\
\midrule
Module Development and Quality Assurance & 93.41\% & 5.87\% & 0.44 \\
Pipeline Documentation, Testing, and Template Alignment & 89.57\% & 9.60\% & 0.69 \\
Module Review, Linting, and Interface Validation & 94.17\% & 4.82\% & 0.66 \\
Routine Maintenance and Reliability Fixes & 95.92\% & 3.53\% & 0.41 \\
Template Update Propagation Across Pipelines & 90.71\% & 8.10\% & 0.61 \\
Contribution Checklist and Review Preparation & 93.54\% & 5.91\% & 0.69 \\
Execution Profile and Infrastructure Configuration & 88.30\% & 11.06\% & 0.73 \\
Community Infrastructure and Website Maintenance & 97.06\% & 2.59\% & 0.11 \\
Automated Dependency and Repository Workflow Maintenance & 92.98\% & 5.61\% & 0.28 \\
Reference Data and Alignment Test Resources & 92.28\% & 6.37\% & 0.19 \\
Runtime Environment and Container Dependency & 95.58\% & 3.01\% & 0.51 \\
Subworkflow Integration and Pipeline Metadata Management & 87.53\% & 11.22\% & 0.62 \\
Sample Sheet and Input Validation & 94.01\% & 5.72\% & 0.05 \\
\bottomrule
\end{tabularx}

\caption*{\footnotesize \textit{Note.} Closed denotes pull requests closed without merge. Median days are calculated for merged pull requests only.}

\end{table}

Table~\ref{tab:pr_topic_resolution} adds the closed-without-merge and median-time perspective. \textit{Community Infrastructure and Website Maintenance} had the highest merge rate (97.06\%) and a short median time-to-merge (0.11 days), followed by \textit{Routine Maintenance and Reliability Improvements} (95.92\%) and \textit{Runtime Environment and Container Dependency} (95.58\%). By contrast, \textit{Subworkflow Integration and Pipeline Metadata Management} (87.53\%), \textit{Execution Profile and Infrastructure Configuration} (88.30\%), and \textit{Pipeline Documentation, Testing, and Template Alignment} (89.57\%) had comparatively lower merge rates and higher closed-without-merge rates. Overall, topic-level variation indicates that nf-core pull request integration is broadly successful, but coordination-intensive and infrastructure-oriented changes are somewhat less consistently merged than localized maintenance or community-infrastructure updates.

\begin{table}[htbp]
\centering
\caption{Selected bivariate associations between pull request features and merge outcome}
\label{tab:pr_bivariate_merge}

\scriptsize
\setlength{\tabcolsep}{3pt}
\renewcommand{\arraystretch}{1.08}

\begin{tabularx}{0.85\columnwidth}{@{}
>{\raggedright\arraybackslash}X
>{\centering\arraybackslash}p{0.13\columnwidth}
>{\centering\arraybackslash}p{0.13\columnwidth}
>{\centering\arraybackslash}p{0.10\columnwidth}
>{\centering\arraybackslash}p{0.10\columnwidth}
@{}}
\toprule
\textbf{Feature} & \textbf{Absent} & \textbf{Present} & \textbf{OR} & \textbf{Adj. $p$} \\
\midrule
Mentions test              & 84.33\% & 96.80\% & 5.62 & $<.001$ \\
Mentions fix               & 85.73\% & 96.99\% & 5.36 & $<.001$ \\
Has checklist              & 86.91\% & 96.99\% & 4.85 & $<.001$ \\
Has linked issue           & 91.35\% & 96.71\% & 2.78 & $<.001$ \\
Requested reviewer         & 91.81\% & 95.70\% & 1.98 & $<.001$ \\
Renovate pull request      & 92.49\% & 98.46\% & 5.01 & $<.001$ \\
Auto-merge enabled         & 92.53\% & 99.51\% & 13.93 & $<.001$ \\
Bot-authored               & 96.93\% & 61.96\% & 0.05 & $<.001$ \\
Draft pull request         & 93.17\% & 72.25\% & 0.19 & $<.001$ \\
Mentions template          & 97.34\% & 87.31\% & 0.19 & $<.001$ \\
Targets development branch & 97.16\% & 88.22\% & 0.22 & $<.001$ \\
Targets main/master branch & 89.50\% & 97.17\% & 4.02 & $<.001$ \\
\bottomrule
\end{tabularx}

\caption*{\footnotesize \textit{Note.} Absent and Present indicate merge rates when the feature is absent or present. OR denotes the odds ratio for merging when the feature is present relative to absent. Adjusted $p$-values use Benjamini--Hochberg correction.}

\end{table}

\textbf{Bivariate associations with merge outcome.} We examined a broad set of binary pull request features, including review signals, author and automation indicators, target-branch information, textual cues, and labels. Table~\ref{tab:pr_bivariate_merge} reports the most influential and interpretable associations. Bivariate merge results show that contribution-readiness and review-routing signals were associated with higher merge rates. Pull requests mentioning tests (96.80\% vs. 84.33\%; OR=5.62), fixes (96.99\% vs. 85.73\%; OR=5.36), or checklists (96.99\% vs. 86.91\%; OR=4.85) had substantially higher merge rates. Pull requests with linked issues (96.71\% vs. 91.35\%; OR=2.78) and requested reviewers (95.70\% vs. 91.81\%; OR=1.98) were also more likely to merge. However, these same review-routing features were associated with longer median time-to-merge: linked-issue pull requests took 0.87 days versus 0.26 days, and pull requests with requested reviewers took 0.88 days versus 0.30 days. Thus, issue linkage and reviewer assignment are associated with eventual integration, but they also appear to mark changes requiring more coordination or verification.

\begin{table}[htbp]
\centering
\caption{Selected bivariate associations between pull request features and closed-without-merge outcome}
\label{tab:pr_bivariate_nonmerge}

\scriptsize
\setlength{\tabcolsep}{3pt}
\renewcommand{\arraystretch}{1.08}

\begin{tabularx}{0.95\columnwidth}{@{}
>{\raggedright\arraybackslash}X
>{\centering\arraybackslash}p{0.13\columnwidth}
>{\centering\arraybackslash}p{0.13\columnwidth}
>{\centering\arraybackslash}p{0.10\columnwidth}
>{\centering\arraybackslash}p{0.10\columnwidth}
@{}}
\toprule
\textbf{Feature} & \textbf{Absent} & \textbf{Present} & \textbf{OR} & \textbf{Adj. $p$} \\
\midrule
Bot-authored               & 2.34\%  & 37.13\% & 24.68 & $<.001$ \\
Mentions template          & 2.03\%  & 11.75\% & 6.41  & $<.001$ \\
Targets development branch & 2.26\%  & 10.81\% & 5.23  & $<.001$ \\
Has URL                    & 1.90\%  & 8.76\%  & 4.94  & $<.001$ \\
Has body                   & 1.53\%  & 7.18\%  & 4.94  & $<.001$ \\
Draft pull request         & 6.22\%  & 22.70\% & 4.44  & $<.001$ \\
Dependabot pull request    & 6.55\%  & 22.81\% & 4.33  & $<.001$ \\
Mentions fix               & 13.52\% & 2.22\%  & 0.15  & $<.001$ \\
Mentions test              & 14.94\% & 2.39\%  & 0.14  & $<.001$ \\
Has checklist              & 12.35\% & 2.19\%  & 0.16  & $<.001$ \\
Has linked issue           & 7.98\%  & 2.17\%  & 0.26  & $<.001$ \\
Requested reviewer         & 7.46\%  & 3.38\%  & 0.43  & $<.001$ \\
Renovate pull request      & 6.71\%  & 1.56\%  & 0.23  & $<.001$ \\
\bottomrule
\end{tabularx}

\caption*{\footnotesize \textit{Note.} Absent and Present indicate closed-without-merge rates among pull requests that reached a final decision. OR denotes the odds ratio for being closed without merge when the feature is present relative to absent. Open pull requests are excluded. Adjusted $p$-values use Benjamini--Hochberg correction.}

\end{table}

\textbf{Bivariate associations with closed-without-merge outcomes.} Closed-without-merge outcomes show the complementary pattern. Table~\ref{tab:pr_bivariate_nonmerge} reports the most influential and interpretable associations. Bot-authored pull requests had a much higher closed-without-merge rate than non-bot pull requests (37.13\% vs. 2.34\%; OR=24.68). Template-related pull requests, development-branch targets, pull requests containing URLs or bodies, draft pull requests, and Dependabot updates were also more likely to be closed without merge. In contrast, pull requests mentioning fixes or tests, containing checklists, linking to issues, requesting reviewers, or coming from Renovate were less likely to be closed without merge. These results suggest that non-merged closure is concentrated in automation-heavy, draft, template-related, and branch-specific work, while review-ready and traceable pull requests are more consistently integrated.




\begin{table}[htbp]
\centering
\caption{Selected bivariate associations between pull request features and time-to-merge}
\label{tab:pr_bivariate_time_merge}

\scriptsize
\setlength{\tabcolsep}{3pt}
\renewcommand{\arraystretch}{1.08}

\begin{tabularx}{0.85\columnwidth}{@{}
>{\raggedright\arraybackslash}X
>{\centering\arraybackslash}p{0.13\columnwidth}
>{\centering\arraybackslash}p{0.13\columnwidth}
>{\centering\arraybackslash}p{0.10\columnwidth}
>{\centering\arraybackslash}p{0.13\columnwidth}
@{}}
\toprule
\textbf{Feature} & \textbf{Absent} & \textbf{Present} & \textbf{Diff.} & \textbf{Cliff's $\delta$} \\
\midrule
Has milestone          & 0.46 & 3.77 & 3.32 & 0.39 \\
Enhancement label      & 0.48 & 3.25 & 2.77 & 0.35 \\
Bot-authored           & 0.39 & 2.16 & 1.77 & 0.32 \\
Has label              & 0.33 & 1.80 & 1.47 & 0.31 \\
Has assignee           & 0.36 & 1.34 & 0.98 & 0.26 \\
Has body               & 0.03 & 0.68 & 0.65 & 0.40 \\
Has URL                & 0.12 & 0.76 & 0.65 & 0.24 \\
Mentions documentation & 0.15 & 0.76 & 0.61 & 0.21 \\
Has linked issue       & 0.26 & 0.87 & 0.62 & 0.19 \\
Requested reviewer     & 0.30 & 0.88 & 0.59 & 0.19 \\
Mentions template      & 0.22 & 0.81 & 0.59 & 0.17 \\
Has checklist          & 0.18 & 0.73 & 0.55 & 0.16 \\
\bottomrule
\end{tabularx}

\caption*{\footnotesize \textit{Note.} Absent and Present indicate median time-to-merge in days among merged pull requests. Diff. denotes the median difference when the feature is present relative to absent. All selected associations are significant after Benjamini--Hochberg correction ($p_{\mathrm{adj}}<.001$).}

\end{table}

\textbf{Bivariate associations with time-to-merge.} We analyzed time-to-merge among merged pull requests using Mann--Whitney U tests and Cliff's $\delta$ effect sizes. Table~\ref{tab:pr_bivariate_time_merge} reports the most influential and interpretable associations. Several features associated with successful integration were also associated with longer merge times. Pull requests linked to issues, requesting reviewers, containing checklists, or mentioning templates took longer to merge, suggesting that traceability and review-readiness signals often appear in contributions requiring additional verification or coordination. The largest delays were observed for pull requests linked to milestones, enhancement-labeled pull requests, bot-authored pull requests, and labeled pull requests, indicating that planned, feature-oriented, automated, or explicitly classified changes often require more review effort. Overall, time-to-merge results show that fast integration is common in nf-core, but pull requests involving coordination, metadata, documentation, templates, or automation tend to remain under review longer before successful integration.

\begin{table}[htbp]
\centering
\caption{Selected predictors from the regularized logistic regression model for pull request merging}
\label{tab:pr_logistic_regression}

\footnotesize
\setlength{\tabcolsep}{4pt}
\renewcommand{\arraystretch}{1.05}

\begin{tabularx}{0.85\columnwidth}{Xrr}
\toprule
\textbf{Predictor} & \textbf{Coefficient} & \textbf{OR} \\
\midrule
Author is member/owner     & 2.94  & 18.89 \\
Author is contributor      & 2.04  & 7.70 \\
Renovate pull request      & 1.89  & 6.59 \\
Mentions security          & 1.71  & 5.50 \\
Auto-merge enabled         & 1.70  & 5.49 \\
Has checklist              & 1.16  & 3.20 \\
Mentions fix               & 0.90  & 2.45 \\
Mentions test              & 0.59  & 1.80 \\
Mentions dependency        & 0.52  & 1.68 \\
Has linked issue           & 0.19  & 1.22 \\
Has milestone              & 0.17  & 1.19 \\
\midrule
Number of labels           & -0.22 & 0.80 \\
Mentions template          & -0.54 & 0.58 \\
Text length                & -0.78 & 0.46 \\
Dependabot pull request    & -1.30 & 0.27 \\
Bot-authored               & -1.41 & 0.24 \\
Draft pull request         & -2.58 & 0.08 \\
Targets development branch & -2.93 & 0.05 \\
\bottomrule
\end{tabularx}

\caption*{\footnotesize \textit{Note.} OR denotes the odds ratio from the L1-regularized logistic regression model. Values above 1 indicate higher adjusted odds of merging; values below 1 indicate lower adjusted odds. Repository and branch controls were included but omitted for readability. Because this is a regularized model, results are interpreted as adjusted directional associations rather than formal significance tests.}

\end{table}

\textbf{Multivariable model of pull request merging.} The regularized merge model shows that contributor familiarity and contribution-readiness signals remained associated with integration after adjustment. As shown in Table~\ref{tab:pr_logistic_regression}, pull requests authored by members or owners (OR=18.89) and contributors (OR=7.70) had substantially higher adjusted odds of merging. Renovate pull requests (OR=6.59), security mentions (OR=5.50), auto-merge (OR=5.49), checklists (OR=3.20), fix mentions (OR=2.45), test mentions (OR=1.80), dependency mentions (OR=1.68), linked issues (OR=1.22), and milestones (OR=1.19) were also associated with higher merge odds. Lower merge odds were associated with development-branch targets (OR=0.05), draft status (OR=0.08), bot authorship (OR=0.24), Dependabot updates (OR=0.27), longer text (OR=0.46), template mentions (OR=0.58), and more labels (OR=0.80). Because this model is regularized, these estimates are interpreted as adjusted directional associations rather than formal significance tests.

\begin{table}[htbp]
\centering
\caption{Selected adjusted odds ratios for pull requests closed without merge}
\label{tab:pr_logit_nonmerge}

\footnotesize
\setlength{\tabcolsep}{4pt}
\renewcommand{\arraystretch}{1.05}

\begin{tabularx}{0.85\columnwidth}{Xr}
\toprule
\textbf{Predictor} & \textbf{OR} \\
\midrule
Draft pull request         & 13.81 \\
Targets development branch & 4.43 \\
Bot-authored               & 4.10 \\
Dependabot pull request    & 3.07 \\
Longer text                & 2.19 \\
Mentions template          & 2.14 \\
Number of labels           & 1.27 \\
Has linked issue           & 0.64 \\
Mentions test              & 0.55 \\
Mentions dependency        & 0.55 \\
Mentions fix               & 0.40 \\
Mentions security          & 0.30 \\
Renovate pull request      & 0.29 \\
Has checklist              & 0.27 \\
Author is contributor      & 0.16 \\
Author is member/owner     & 0.06 \\
Auto-merge enabled         & 0.01 \\
\bottomrule
\end{tabularx}

\caption*{\footnotesize \textit{Note.} Odds ratios are from an L1-regularized logistic regression model comparing pull requests closed without merge against merged pull requests. Open pull requests are excluded. Values above 1 indicate higher adjusted odds of closed-without-merge outcomes; values below 1 indicate lower adjusted odds. Repository, topic, and author-association controls were included but omitted for readability.}

\end{table}

\paragraph{Multivariable model for closed-without-merge outcomes.}
We fitted a regularized logistic regression model to identify factors associated with pull requests being closed without merge rather than merged, excluding open pull requests. Table~\ref{tab:pr_logit_nonmerge} reports the most interpretable adjusted odds ratios. Draft status showed the strongest association with non-merged closure (OR=13.81), indicating that pull requests not ready for review were much more likely to be closed without integration. Development-branch targets (OR=4.43), bot authorship (OR=4.10), Dependabot updates (OR=3.07), longer text (OR=2.19), template mentions (OR=2.14), and larger numbers of labels (OR=1.27) were also associated with higher odds of closed-without-merge outcomes.

In contrast, several features were associated with lower odds of non-merged closure. Pull requests linked to issues (OR=0.64), mentioning tests (OR=0.55), dependencies (OR=0.55), fixes (OR=0.40), or security (OR=0.30), and those containing checklists (OR=0.27) were less likely to be closed without merge. Renovate pull requests also had lower odds of non-merged closure (OR=0.29), while contributor and member/owner authorship were strongly protective against non-merged closure. Overall, the model suggests that non-merged closure in nf-core is concentrated among draft, bot-generated, template-related, and branch-specific pull requests, whereas issue linkage, validation evidence, checklist use, and maintainer familiarity support successful integration.

\begin{figure*}[htbp]
    \centering

    \begin{subfigure}[t]{0.32\textwidth}
        \centering
        \includegraphics[width=\linewidth]{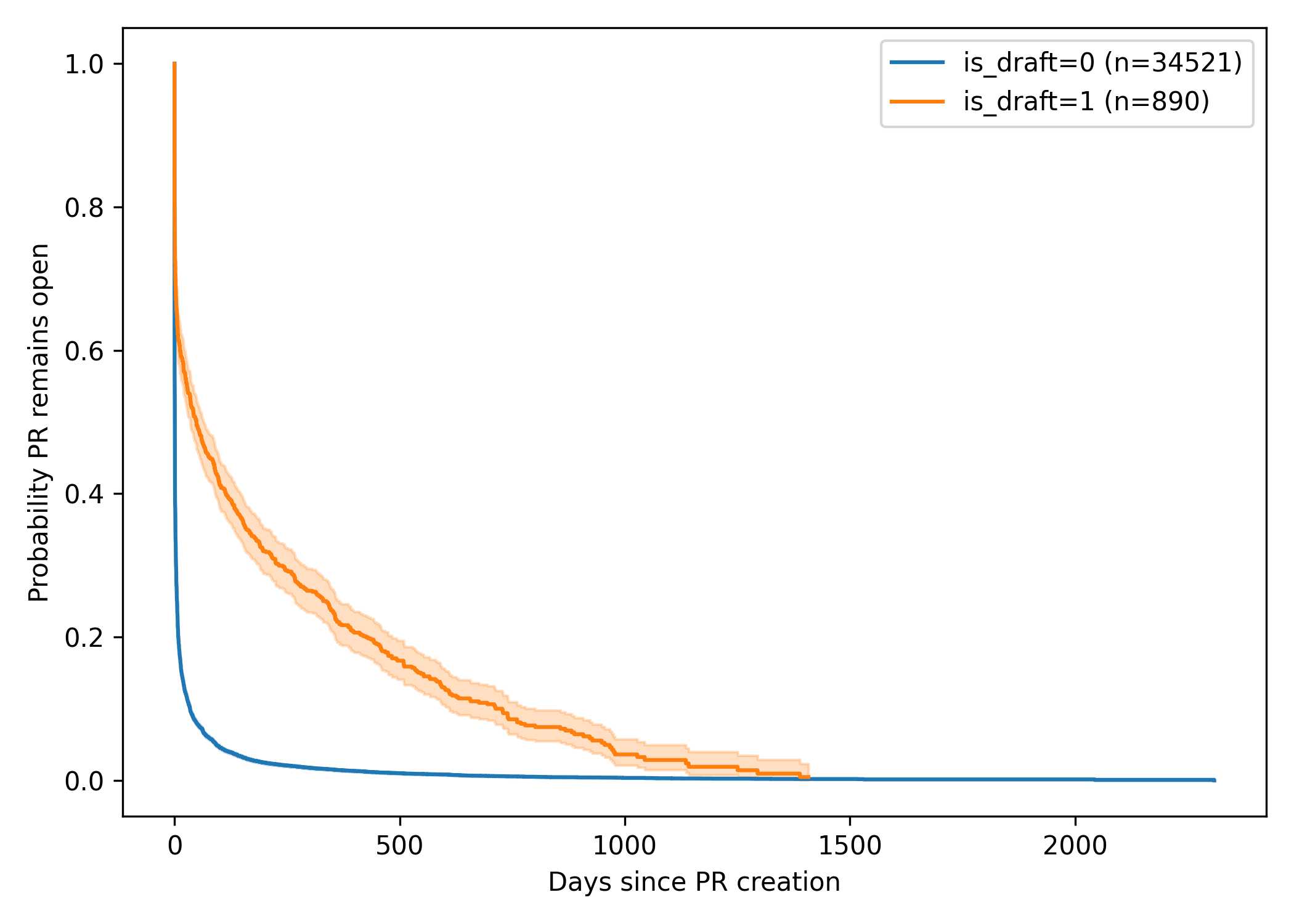}
        \caption{Draft status}
        \label{fig:pr_km_draft}
    \end{subfigure}
    \hfill
    \begin{subfigure}[t]{0.32\textwidth}
        \centering
        \includegraphics[width=\linewidth]{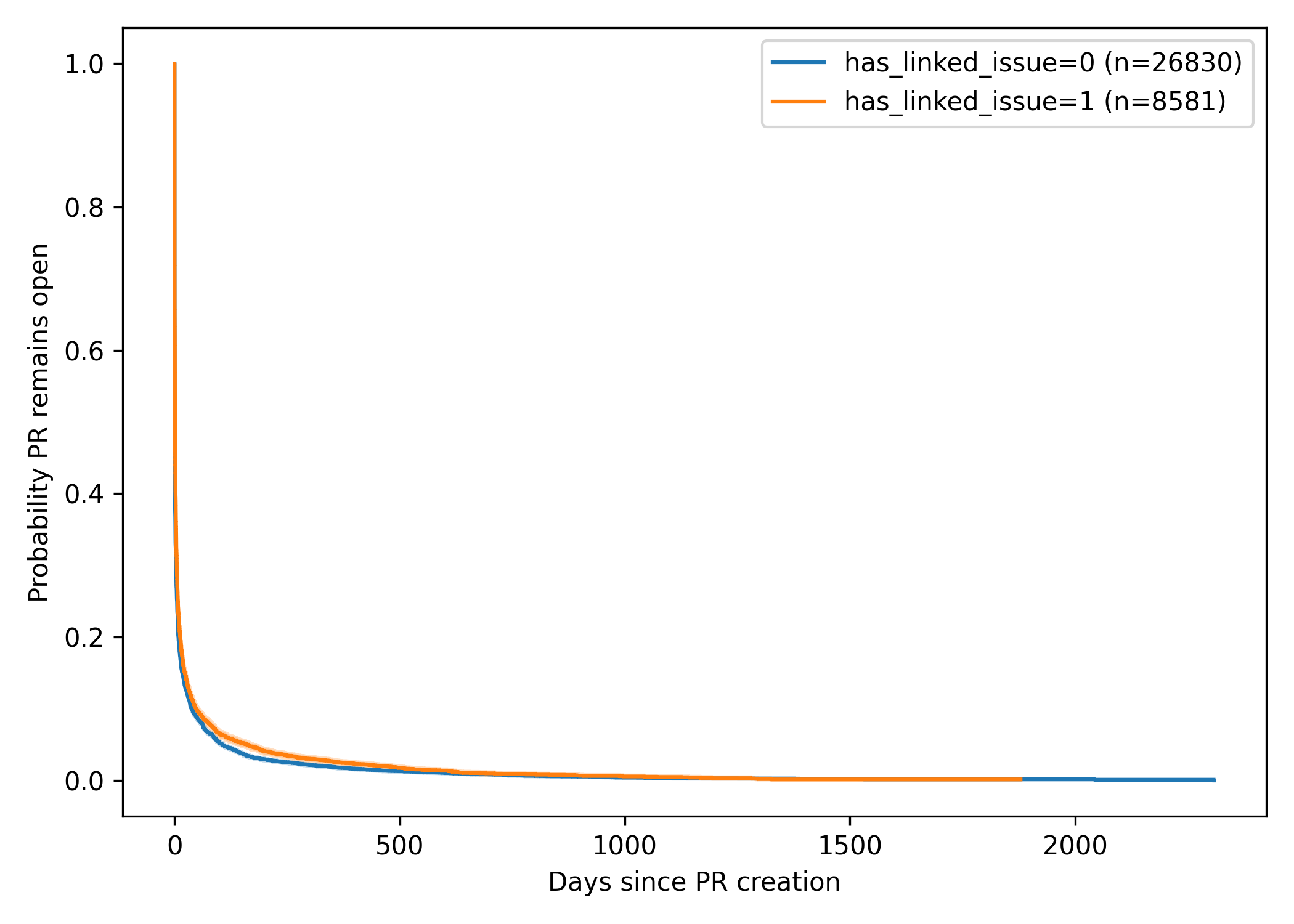}
        \caption{Linked issue}
        \label{fig:pr_km_linked_issue}
    \end{subfigure}
    \hfill
    \begin{subfigure}[t]{0.32\textwidth}
        \centering
        \includegraphics[width=\linewidth]{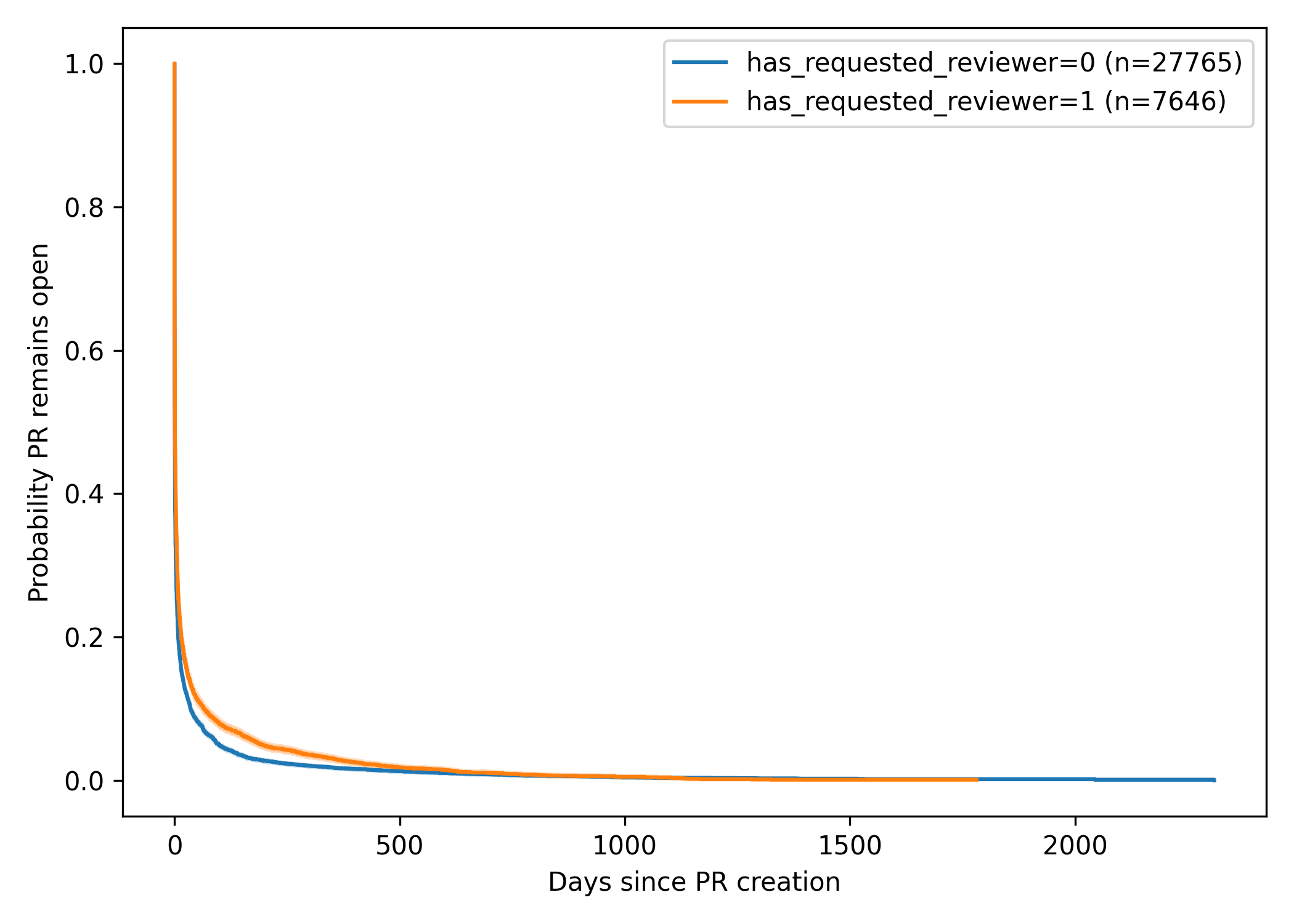}
        \caption{Requested reviewer}
        \label{fig:pr_km_requested_reviewer}
    \end{subfigure}

    \caption{Kaplan--Meier curves for time to final pull request decision across selected factors. The y-axis shows the probability that a pull request remains open. Draft pull requests remain open substantially longer, while pull requests with linked issues or requested reviewers also show longer decision trajectories.}
    \label{fig:pr_km_selected_factors}
\end{figure*}

\paragraph{Kaplan--Meier analysis of Pull Request-resolution trajectories}
We further examined time to final pull request decision using Kaplan--Meier curves. The event was a final decision, either merge or closed without merge, while pull requests that remained open were treated as censored. We explored several binary features, including checklist presence, linked issue, requested reviewer, bot authorship, automation status, draft status, dependency mentions, documentation mentions, template mentions, and test mentions. For space, Figure~\ref{fig:pr_km_selected_factors} reports three representative factors with clear interpretation: draft status, linked issue, and requested reviewer.

Draft status shows the strongest separation. Draft pull requests remained open substantially longer than non-draft pull requests, indicating that contributions marked as not ready for review move more slowly toward a final decision. Pull requests linked to issues and pull requests with requested reviewers also showed longer decision trajectories than their counterparts. This does not necessarily indicate weaker integration; earlier results show that linked issues and requested reviewers are associated with higher merge rates. Rather, these features appear to mark pull requests that require additional coordination, verification, or domain-specific review before resolution. Overall, the lifecycle analysis shows that most nf-core pull requests are resolved quickly, but draft work and coordination-intensive pull requests remain open longer before reaching a final decision.

\subsubsection{Factors Associated with Forum Discussion Resolution Outcomes:}

\paragraph{Forum support outcomes and engagement}
We analyzed 895 Seqera Community Forum discussions to examine how user-facing support requests reach resolution. Overall, 412 discussions received an accepted answer, corresponding to an accepted-answer rate of 46.03\%, while 483 discussions did not receive an accepted answer. A similar proportion received at least one reply (446 discussions; 49.83\%), indicating that support engagement is unevenly distributed across forum posts.

\begin{figure}[htbp]
    \centering
    \includegraphics[width=0.75\textwidth]{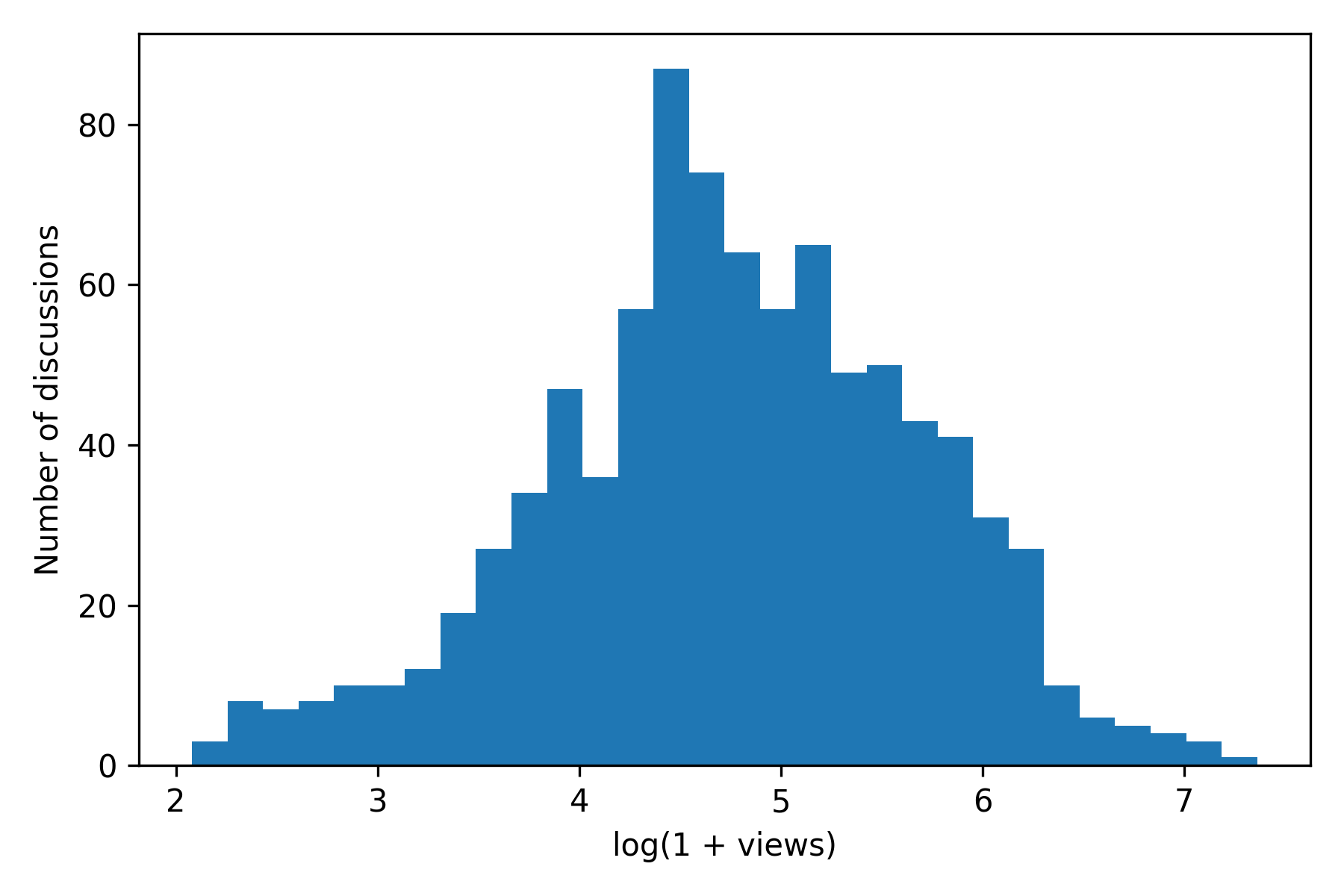}
    \caption{Log-transformed distribution of forum discussion views.}
    \label{fig:forum_log_views}
\end{figure}

The median number of replies was 0, although the mean was 1.15, showing that many discussions received no reply while a smaller subset generated more active exchanges. Discussions nevertheless attracted substantial visibility, with a median of 113 views and a mean of 172.45 views. Because raw view counts were skewed, Figure~\ref{fig:forum_log_views} presents the log-transformed distribution, which shows that most discussions received moderate visibility while a smaller number attracted substantially closer attention. The median activity span was 1.75 days, compared with a mean of 22.58 days, indicating that many discussions became inactive quickly, while a smaller subset remained active for much longer. These results suggest that forum resolution is less formalized than GitHub issue closure or pull request merging: accepted-answer outcomes depend on sustained community interaction, while even unresolved discussions may remain useful as visible support resources for later users facing similar nf-core or Nextflow execution problems.

\begin{figure}[htbp]
    \centering
    \includegraphics[width=0.75\textwidth]{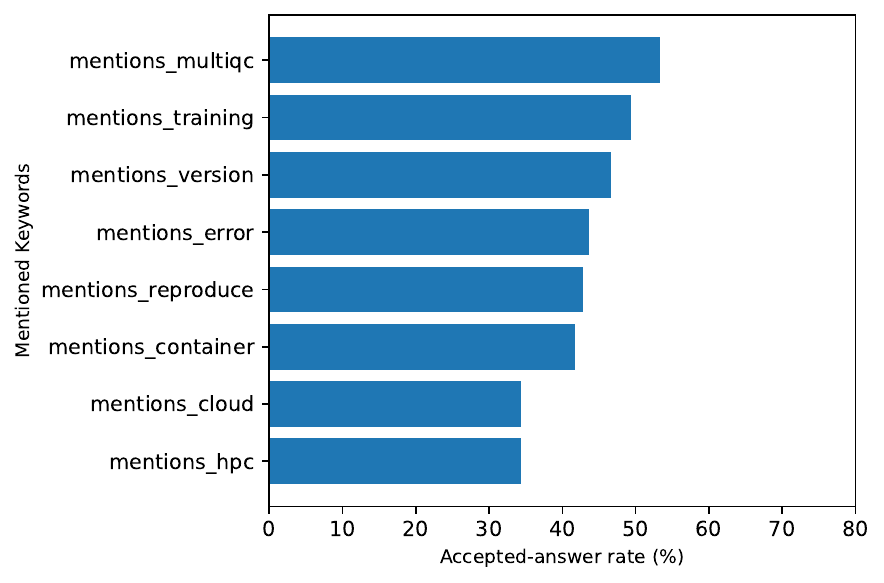}
    \caption{Accepted-answer rate by technical signal in forum discussions.}
    \label{fig:forum_signal_acceptance}
\end{figure}

\textbf{Technical evidence and accepted answers. }Accepted-answer rates varied by the technical evidence included in forum discussions. Discussions containing code blocks had a higher accepted-answer rate than those without code blocks (49.92\% vs. 38.41\%) and also showed greater engagement, with a median of one reply and 122 views compared with zero replies and 98 views for discussions without code blocks. This suggests that concrete technical evidence, such as commands, configuration snippets, workflow fragments, or error logs, makes support requests more diagnosable.

Accepted-answer rates also differed across technical signals. As shown in Figure~\ref{fig:forum_signal_acceptance}, discussions mentioning \textit{MultiQC} had the highest accepted-answer rate (53.28\%), followed by training-related discussions (49.38\%) and version-related discussions (46.60\%). Discussions mentioning errors, reproduction information, or containers had moderate accepted-answer rates, ranging from 41.67\% to 43.57\%. In contrast, cloud- and HPC-related discussions had the lowest accepted-answer rates, at 34.34\% and 34.31\%, respectively. These results suggest that reporting, training, or version-specific questions are more readily resolved, whereas cloud and HPC questions are harder to answer conclusively because they often depend on local infrastructure, credentials, schedulers, storage systems, or execution policies.

\begin{figure}[htbp]
    \centering
    \includegraphics[width=0.75\textwidth]{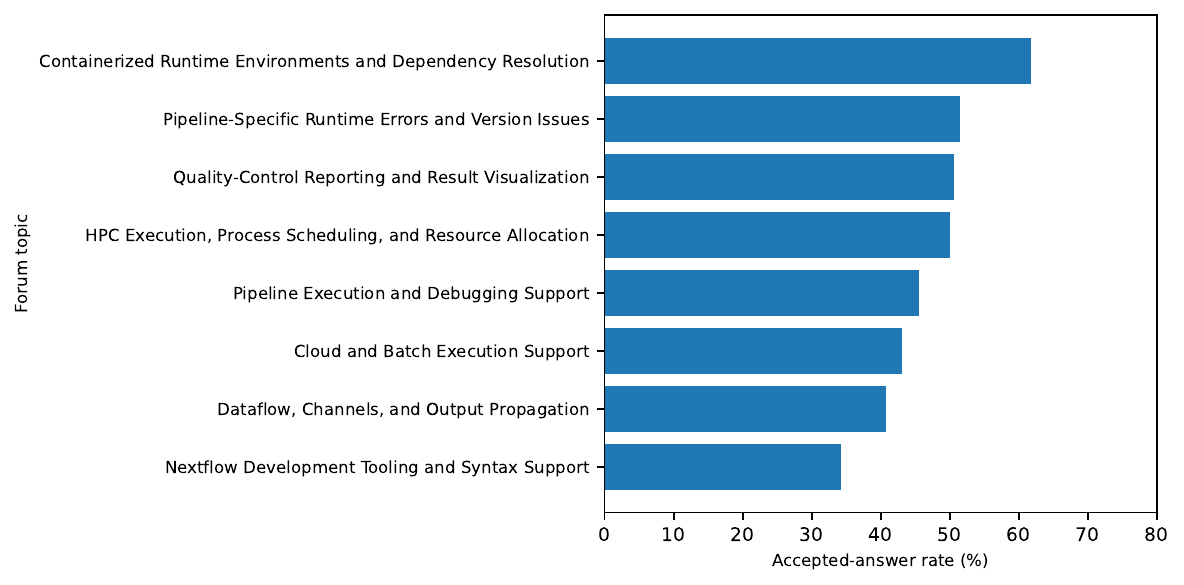}
    \caption{Accepted-answer rate by forum topic.}
    \label{fig:forum_topic_accepted_rate}
\end{figure}

\textbf{Topic-level variation in forum resolution. }Accepted-answer rates varied across forum topics, indicating that some support needs were easier to resolve than others. As shown in Figure~\ref{fig:forum_topic_accepted_rate}, \textit{Containerized Runtime Environments and Dependency Resolution} had the highest accepted-answer rate (61.84\%), followed by \textit{Pipeline-Specific Runtime Errors and Version Issues} (51.52\%), \textit{Quality-Control Reporting and Result Visualization} (50.62\%), and \textit{HPC Execution, Process Scheduling, and Resource Allocation} (50.00\%). These topics often involve concrete runtime symptoms, reporting outputs, or identifiable environment behavior. In contrast, \textit{Nextflow Development Tooling and Syntax Support} had the lowest accepted-answer rate (34.21\%), followed by \textit{Dataflow, Channels, and Output Propagation} (40.78\%) and \textit{Cloud and Batch Execution Support} (43.14\%). These lower rates suggest that questions involving workflow semantics, development tooling, or cloud execution are harder to resolve conclusively, likely because they require more context about user code, platform configuration, or infrastructure-specific constraints.

\begin{table}[htbp]
\centering
\caption{Selected bivariate associations between forum features and accepted-answer outcomes}
\label{tab:forum_bivariate_accepted}

\footnotesize
\setlength{\tabcolsep}{3.2pt}
\renewcommand{\arraystretch}{1.05}

\begin{tabularx}{0.85\columnwidth}{Xrrrr}
\toprule
\textbf{Feature} & \textbf{Absent} & \textbf{Present} & \textbf{OR} & \textbf{Adj. $p$} \\
\midrule
Has multiple comments & 0.00\%  & 51.95\% & 221.66 & $<.001$ \\
Has reply             & 29.62\% & 62.56\% & 3.96   & $<.001$ \\
Has likes             & 30.57\% & 62.24\% & 3.73   & $<.001$ \\
MultiQC tag           & 42.05\% & 76.70\% & 4.47   & $<.001$ \\
Has code block        & 38.41\% & 49.92\% & 1.60   & $.005$ \\
Mentions MultiQC      & 43.54\% & 53.28\% & 1.48   & $.037$ \\
Mentions cloud        & 49.35\% & 34.34\% & 0.54   & $.001$ \\
Mentions HPC          & 48.15\% & 34.31\% & 0.57   & $.012$ \\
SLURM tag             & 46.66\% & 0.00\%  & 0.05   & $.004$ \\
HPC tag               & 46.78\% & 20.00\% & 0.31   & $.029$ \\
\bottomrule
\end{tabularx}

\caption*{\footnotesize \textit{Note.} Absent and Present indicate accepted-answer rates when the feature is absent or present. OR denotes the odds ratio for receiving an accepted answer when the feature is present relative to absent. Adjusted $p$-values use Benjamini--Hochberg correction.}

\end{table}

\textbf{Bivariate associations with accepted answers. }We examined a broad set of binary forum features, including engagement signals, technical evidence, infrastructure mentions, and tag indicators. For each feature, we compared accepted-answer rates when the feature was absent versus present, estimated odds ratios using 2$\times$2 contingency tables, and applied Benjamini--Hochberg correction for multiple comparisons. Table~\ref{tab:forum_bivariate_accepted} reports the most influential and interpretable associations.

The strongest associations involved conversational engagement. Discussions with replies had a much higher accepted-answer rate than discussions without replies (62.56\% vs. 29.62\%; OR=3.96), and discussions with likes showed a similar pattern (62.24\% vs. 30.57\%; OR=3.73). Discussions with multiple posts also had a higher accepted-answer rate, although this should be interpreted as an engagement condition rather than an independent causal factor. Technical specificity was also associated with resolution: discussions containing code blocks were more likely to receive accepted answers (49.92\% vs. 38.41\%; OR=1.60), and MultiQC-related discussions showed higher acceptance, especially those tagged with \texttt{multiqc} (76.70\% vs. 42.05\%; OR=4.47). In contrast, infrastructure-related discussions were less likely to receive accepted answers. Cloud mentions (34.34\% vs. 49.35\%; OR=0.54), HPC mentions (34.31\% vs. 48.15\%; OR=0.57), and HPC/SLURM tags were associated with lower accepted-answer rates. Overall, forum resolution is most strongly associated with active community interaction and concrete diagnostic evidence, while infrastructure-dependent questions are harder to resolve conclusively.

\begin{table}[htbp]
\centering
\caption{Selected numeric predictors associated with accepted-answer outcomes in forum discussions}
\label{tab:forum_numeric_accepted}

\footnotesize
\setlength{\tabcolsep}{2.4pt}
\renewcommand{\arraystretch}{1.05}

\begin{tabularx}{0.90\columnwidth}{Xrrrrr}
\toprule
\textbf{Feature} & \textbf{No accepted} & \textbf{Accepted} & \textbf{Diff.} & \textbf{Cliff's $\delta$} & \textbf{Adj. $p$} \\
\midrule
Posts              & 2.00   & 5.00   & 3.00  & 0.56 & $<.001$ \\
Replies            & 0.00   & 1.00   & 1.00  & 0.33 & $<.001$ \\
Likes              & 0.00   & 1.00   & 1.00  & 0.32 & $<.001$ \\
Activity span days & 1.07   & 3.45   & 2.38  & 0.21 & $<.001$ \\
Code blocks        & 2.00   & 3.00   & 1.00  & 0.10 & $<.05$ \\
Views              & 107.00 & 122.00 & 15.00 & 0.10 & $<.05$ \\
\bottomrule
\end{tabularx}

\caption*{\footnotesize \textit{Note.} Values show medians for discussions without and with accepted answers. Diff. denotes the median difference for accepted-answer discussions. Adjusted $p$-values use Benjamini--Hochberg correction.}

\end{table}

\textbf{Numeric predictors of accepted answers.} We compared numeric discussion features between forum posts with and without accepted answers using Mann--Whitney U tests and Cliff's $\delta$ effect sizes. Numeric comparisons reinforce the engagement pattern. As shown in Table~\ref{tab:forum_numeric_accepted}, accepted-answer discussions had more posts (median 5 vs. 2; Cliff's $\delta=0.56$), more replies (1 vs. 0; $\delta=0.33$), more likes (1 vs. 0; $\delta=0.32$), and longer activity spans (3.45 vs. 1.07 days; $\delta=0.21$). Accepted-answer discussions also contained slightly more code blocks and received slightly more views, although these effects were small. In contrast, text-length measures showed negligible differences, suggesting that accepted answers are associated less with longer questions and more with conversational engagement, diagnostic evidence, and sustained community interaction.





\begin{table}[htbp]
\centering
\caption{Selected predictors from the regularized logistic regression model for accepted-answer outcomes in forum discussions}
\label{tab:forum_logit_accepted}

\footnotesize
\setlength{\tabcolsep}{3.5pt}
\renewcommand{\arraystretch}{1.05}

\begin{tabularx}{0.85\columnwidth}{Xrr}
\toprule
\textbf{Predictor} & \textbf{Coefficient} & \textbf{OR} \\
\midrule
Posts count, log-transformed   & 5.01  & 149.33 \\
Likes count, log-transformed   & 0.78  & 2.18 \\
Has code block                 & 0.69  & 2.00 \\
Has attachment                 & 0.54  & 1.72 \\
Mentions MultiQC               & 0.36  & 1.43 \\
Text length, log-transformed   & 0.05  & 1.05 \\
Mentions version               & 0.03  & 1.03 \\
\midrule
Mentions reproduce             & -0.25 & 0.78 \\
Mentions Seqera Platform       & -0.27 & 0.77 \\
Has image                      & -0.28 & 0.76 \\
Views, log-transformed         & -0.29 & 0.75 \\
Activity span, log-transformed & -0.35 & 0.71 \\
Mentions HPC                   & -0.56 & 0.57 \\
Mentions Nextflow              & -0.72 & 0.49 \\
Replies count, log-transformed & -1.73 & 0.18 \\
\bottomrule
\end{tabularx}

\caption*{\footnotesize \textit{Note.} OR denotes the odds ratio from the L1-regularized logistic regression model. Values above 1 indicate higher adjusted odds of receiving an accepted answer; values below 1 indicate lower adjusted odds. Topic/category controls were included but omitted for readability. Because this is a regularized model, results are interpreted as adjusted directional associations rather than formal significance tests.}

\end{table}

\textbf{Multivariable model of accepted-answer outcomes. }We fitted a regularized logistic regression model to examine which forum features were associated with receiving an accepted answer after accounting for engagement, technical evidence, infrastructure mentions, text length, and topic/category controls. The regularized logistic model for accepted-answer status shows similar adjusted associations. As shown in Table~\ref{tab:forum_logit_accepted}, accepted answers were positively associated with posts count (OR=149.33), likes count (OR=2.18), code blocks (OR=2.00), attachments (OR=1.72), MultiQC mentions (OR=1.43), text length (OR=1.05), and version mentions (OR=1.03). Lower adjusted odds were associated with reproduction mentions (OR=0.78), Seqera Platform mentions (OR=0.77), images (OR=0.76), views (OR=0.75), activity span (OR=0.71), HPC mentions (OR=0.57), Nextflow mentions (OR=0.49), and reply count (OR=0.18).

The negative coefficient for reply count should be interpreted cautiously because posts and replies are partially overlapping engagement measures. In combination, the model suggests that broader conversational development and concrete technical evidence are more informative than reply count alone. Because this is a regularized model, these coefficients are interpreted as adjusted directional associations rather than formal significance tests.

\begin{figure*}[htbp]
    \centering

    \begin{subfigure}[t]{0.32\textwidth}
        \centering
        \includegraphics[width=\linewidth]{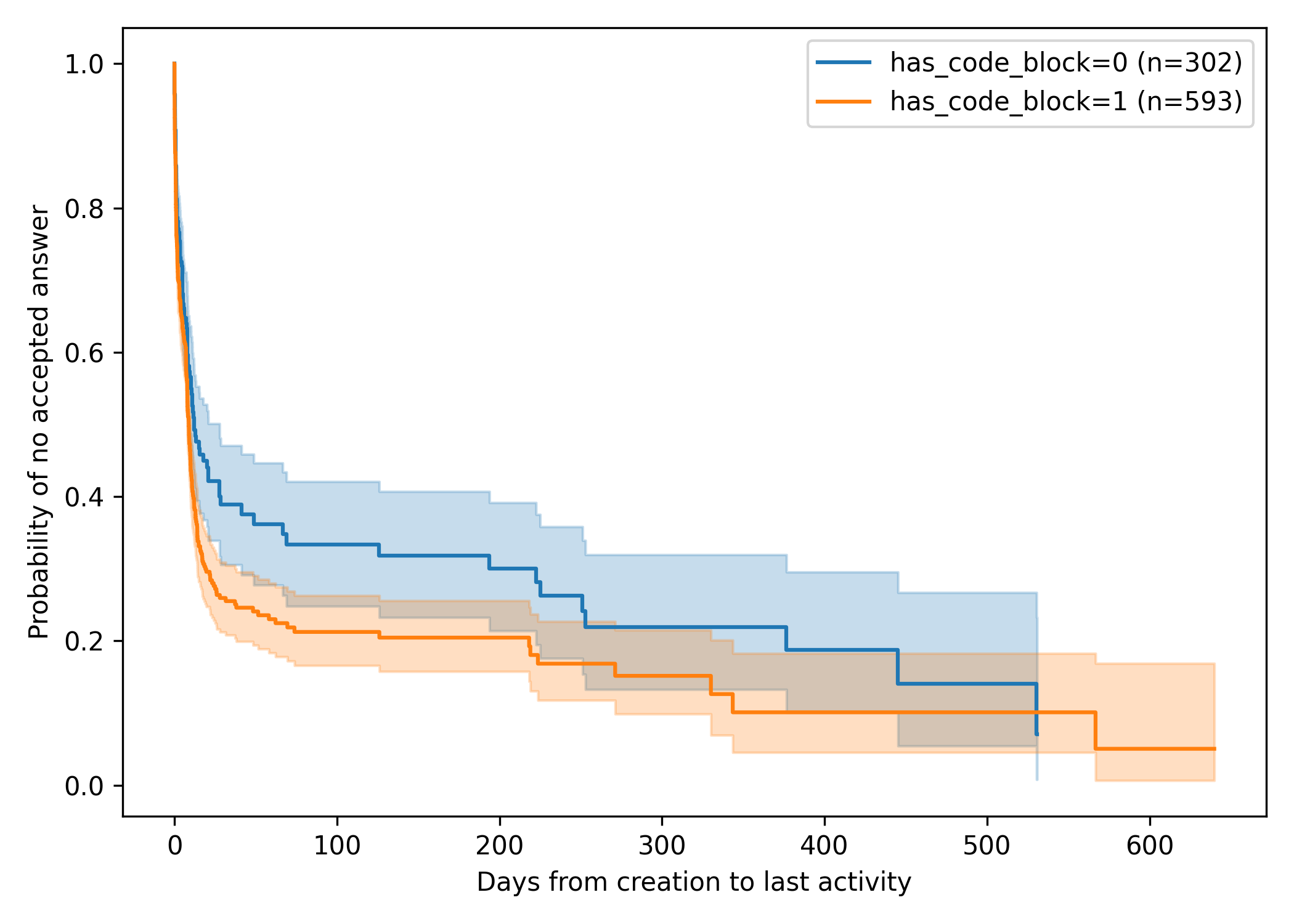}
        \caption{Code block}
        \label{fig:forum_km_code}
    \end{subfigure}
    \hfill
    \begin{subfigure}[t]{0.32\textwidth}
        \centering
        \includegraphics[width=\linewidth]{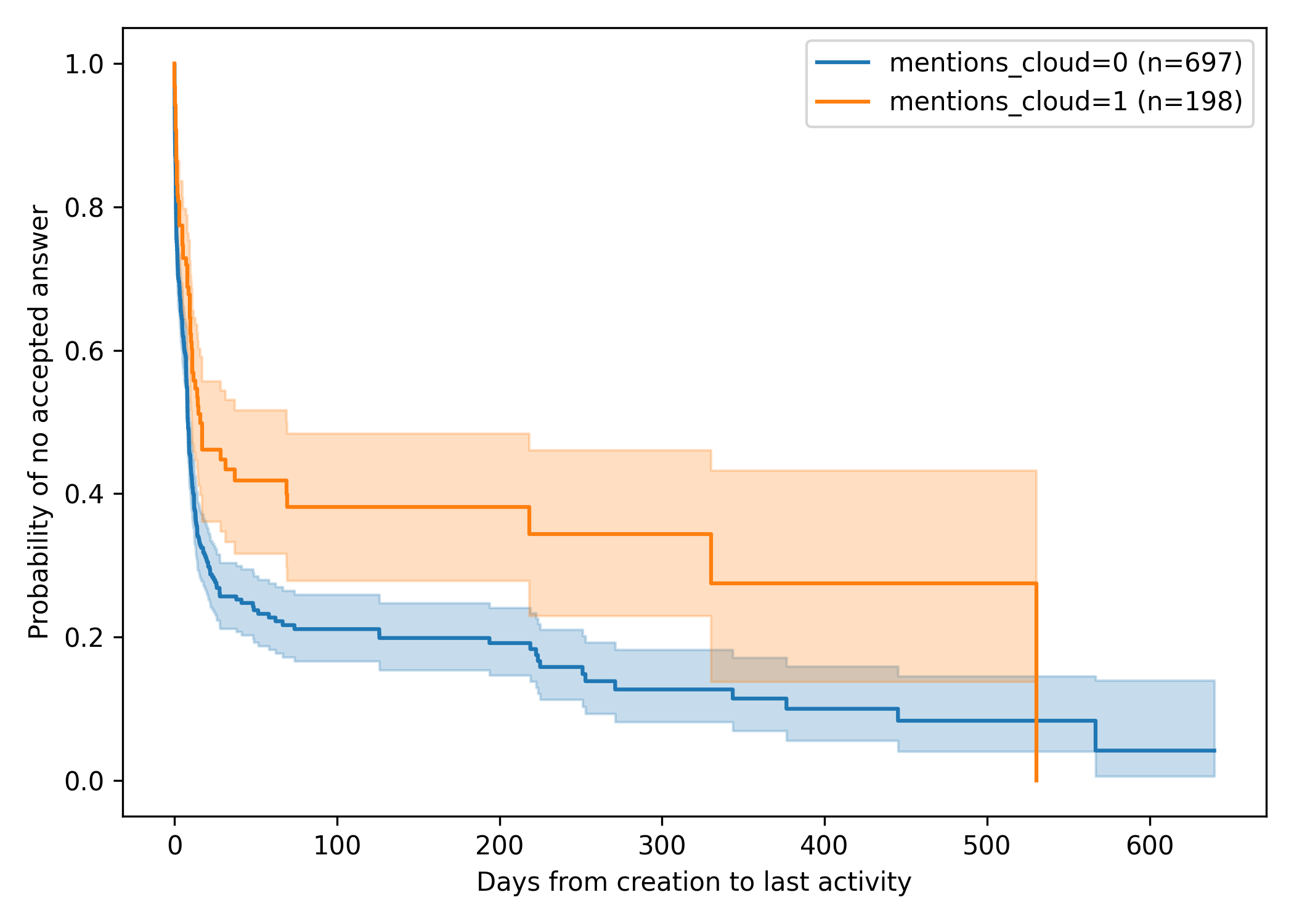}
        \caption{Cloud mention}
        \label{fig:forum_km_cloud}
    \end{subfigure}
    \hfill
    \begin{subfigure}[t]{0.32\textwidth}
        \centering
        \includegraphics[width=\linewidth]{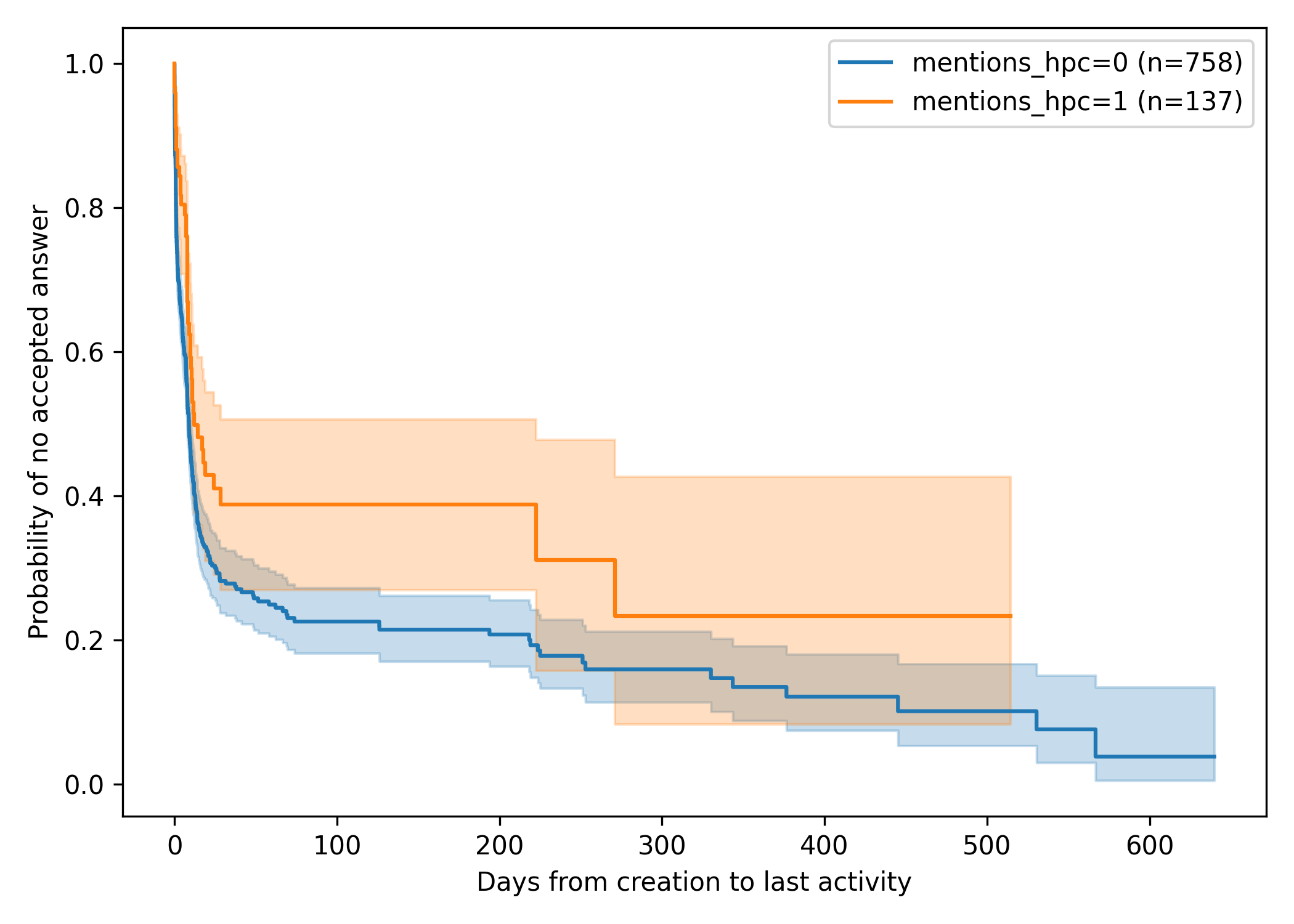}
        \caption{HPC mention}
        \label{fig:forum_km_hpc}
    \end{subfigure}

    \caption{Kaplan--Meier-style lifecycle curves for selected forum discussion features. The y-axis shows the probability that a discussion has not received an accepted answer over its observed activity span.}
    \label{fig:forum_km_selected}
\end{figure*}

\paragraph{Forum Discussion Lifecycle.}
We examined how selected discussion features relate to accepted-answer outcomes over the observed activity span of each forum discussion. The event was receiving an accepted answer, while discussions without accepted answers were treated as censored. For space, Figure~\ref{fig:forum_km_selected} reports three representative signals: code blocks, cloud mentions, and HPC mentions.

The curves show that discussions with code blocks moved more quickly toward accepted-answer outcomes, consistent with the bivariate results showing that concrete diagnostic evidence improves resolvability. In contrast, discussions mentioning cloud or HPC remained without accepted answers for longer, indicating that infrastructure-dependent support problems are harder to resolve conclusively. These patterns suggest that forum questions are more likely to reach resolution when they include actionable technical evidence, whereas cloud and HPC questions often depend on local configuration, credentials, schedulers, storage systems, or institutional execution policies.

\paragraph{Qualitative sampling for interpretation.}
To contextualize the quantitative forum results, we inspected representative discussions from four groups: accepted-answer discussions with high engagement, high-engagement discussions without accepted answers, unanswered discussions with no replies, and long-lifecycle discussions without accepted answers. This sampling was used to interpret the statistical patterns rather than to introduce a separate qualitative coding scheme.

Accepted-answer discussions with high engagement typically contained concrete technical details and active back-and-forth interaction. For example, \textit{How to wait until a process is complete?} received an accepted answer after 23 posts, 12 replies, 913 views, and an activity span of 12.96 days. Similarly, \textit{Timeout issue in nextflow process run on Azure batch} had 19 posts, 12 replies, 517 views, and 4 likes before reaching an accepted answer. These cases illustrate how detailed, active discussions can converge on a support outcome.

However, engagement alone did not guarantee resolution. Some high-engagement discussions remained without accepted answers, including \textit{Proper way to utilize $>$1 GPUs in a single machine?}, which had 14 posts, 9 replies, 509 views, and remained active for 530.82 days, and \textit{Is it possible to use Nextflow with the libmamba conda solver?}, which had 13 posts, 8 replies, 1,076 views, and 11 likes but no accepted answer. Unanswered discussions also attracted substantial visibility: \textit{MultiQC on fastp results} had no replies but 1,047 views, while \textit{Using conda environments in singularity containers with nextflow} had no replies but 627 views. Long-lifecycle unresolved discussions often involved development tooling, GPU use, HPC/cloud execution, or job-log recovery; for example, \textit{Remote-debug in IntelliJ is not stopping at breakpoints} remained active for 639.86 days without an accepted answer, and \textit{My jobs are being canceled and logs are unable to be recovered} remained unresolved for 514.50 days. Overall, the qualitative inspection supports the statistical findings: forum questions are more likely to reach accepted-answer resolution when they are bounded and technically diagnosable, while infrastructure-dependent or environment-specific questions may remain visible and useful but unresolved in formal accepted-answer terms.

\begin{center}
\fbox{
\begin{minipage}{0.96\linewidth}
\textbf{Summary of RQ3.}
\textbf{RQ3} shows that resolution in nf-core is associated with actionability, coordination, and infrastructure complexity. GitHub issues were more likely to close and closed faster when they showed maintainer attention and diagnostic clarity, such as assignees, comments, milestones, bug labels, error mentions, and version information. Pull requests were highly likely to merge overall, but successful integration was associated with contributor familiarity, checklists, fix/test framing, linked issues, requested reviewers, and well-aligned automation such as Renovate. In contrast, draft pull requests, development-branch targets, template-related changes, Dependabot updates, and bot-authored changes were less consistently integrated. Forum discussions were less formalized: accepted answers were more likely when discussions included code blocks, multiple comments, likes, and concrete technical evidence, while cloud, HPC, and workflow-semantics questions were harder to resolve. Overall, nf-core resolves artifacts most effectively when problems are concrete, technically diagnosable, and routed through clear coordination mechanisms.
\end{minipage}
}
\end{center}

\subsection{RQ4: Problem--Solution Flow Across Repository and Support Spaces}
\label{sec:rq4}

\subsubsection{Motivation}
\textbf{RQ1--RQ3} show what maintenance and support concerns appear in nf-core, how they differ across artifact types, and which factors are associated with resolution outcomes. However, these analyses do not show whether problems raised in one space are connected to solutions or follow-up work in another. This matters because maintenance and support in nf-core are distributed across GitHub issues, pull requests, and Seqera Community Forum discussions. A forum discussion may reveal a recurring user-facing execution problem, an issue may formalize that problem as a repository-level task, and a pull request may implement the corresponding fix or documentation update.

Understanding this flow is important for evaluating traceability in the ecosystem. Strong traceability helps maintainers connect reported problems to implementation work and durable improvements. Weak traceability, especially between user-support discussions and repository artifacts, may cause recurring support problems to remain isolated in forum threads rather than being converted into GitHub issues, pull requests, documentation updates, or template improvements. \textbf{RQ4} therefore examines how visible and traceable problem--solution flow is across repository-centered development spaces and community-centered support spaces.

\subsubsection{Approach}
We followed the problem--solution flow procedure described in Section~\ref{sec:problem_solution_flow}, but report the results in three layers. First, we analyzed \emph{explicit traceability} by extracting direct references, including GitHub issue and pull request URLs, forum URLs, GitHub \texttt{\#number} references, and closing-keyword references such as \textit{fixes}, \textit{closes}, and \textit{resolves}. These links provide the strongest evidence of visible cross-artifact relationships.

Second, we analyzed \emph{implicit issue--pull request relatedness} using semantic similarity after removing directly linked pairs. These pairs are treated as candidate semantic flows rather than confirmed problem--solution links because textual similarity does not prove that one artifact caused or directly informed another. We therefore interpret them as evidence of possible related maintenance work that is not explicitly linked.

Third, we compared \emph{technical-signal overlap} across issues, pull requests, and forum discussions. Signal overlap does not establish artifact-level traceability, but it shows whether similar technical concerns recur across repository and support spaces. Together, direct links, semantic-flow candidates, and technical-signal overlap allow us to distinguish strong explicit traceability from weaker thematic connections.

\subsubsection{Results of RQ4}
RQ4 shows that problem--solution flow in nf-core is strongly traceable within GitHub but weakly traceable between GitHub and the Seqera Community Forum. Pull requests frequently reference issues and other pull requests, creating a visible repository-internal pathway from reported problems to implementation work. In contrast, explicit links between forum discussions and GitHub artifacts are rare, even though forum discussions and GitHub artifacts share many technical concerns. This suggests that user-facing support knowledge often enters maintenance indirectly, through recurring technical themes, rather than through explicit cross-platform links.

\begin{table}[htbp]
\centering
\caption{Explicit direct-link flows across nf-core artifacts}
\label{tab:rq4_direct_flow}

\scriptsize
\setlength{\tabcolsep}{4pt}
\renewcommand{\arraystretch}{1.05}

\begin{tabularx}{0.85\columnwidth}{Xr}
\toprule
\textbf{Flow type} & \textbf{Links} \\
\midrule
Pull request $\rightarrow$ Issue & 7,599 \\
Pull request $\rightarrow$ Pull request & 6,021 \\
Issue $\rightarrow$ Issue & 1,886 \\
Issue $\rightarrow$ Pull request & 1,091 \\
Pull request $\rightarrow$ unresolved GitHub reference & 804 \\
Issue $\rightarrow$ unresolved GitHub reference & 99 \\
GitHub $\rightarrow$ Forum & 6 \\
Forum $\rightarrow$ GitHub & 5 \\
Forum $\rightarrow$ Forum & 1 \\
\bottomrule
\end{tabularx}

\caption*{\footnotesize \textit{Note.} Direct links include explicit GitHub/forum URLs, GitHub \texttt{\#number} references, and closing-keyword references.}

\end{table}

\paragraph{Explicit links are concentrated inside GitHub.}
Table~\ref{tab:rq4_direct_flow} summarizes explicit direct-link flows across nf-core artifacts. We identified 17,512 explicit cross-artifact references. The largest flow was from pull requests to issues, with 7,599 links. This included closing-keyword references, explicit issue URLs, and issue-number references. We also observed 6,021 pull request-to-pull request references, 1,886 issue-to-issue references, and 1,091 issue-to-pull request references.

These results indicate that GitHub provides the clearest traceable problem--solution pathway in nf-core. Although many detected references point from pull requests to issues, the underlying maintenance logic is problem-to-solution: issues describe reported problems, feature requests, or maintenance tasks, while pull requests provide implementation, review, or integration work. This pattern shows that nf-core contributors frequently connect related repository artifacts, especially when pull requests address, discuss, or supersede issue-level maintenance tasks.

\paragraph{Forum--GitHub traceability is limited.}

Explicit links between the Seqera Community Forum and GitHub artifacts were rare. We found only five forum-to-GitHub links and six GitHub-to-forum links, along with one forum-to-forum link (Table~\ref{tab:rq4_direct_flow}). This pattern indicates that user-facing support discussions are weakly connected to repository-centered maintenance through explicit references.

This weak traceability is important because forum discussions often contain maintenance-relevant knowledge. Repeated forum questions may reveal documentation gaps, confusing error messages, cloud or HPC configuration problems, container issues, or recurring pipeline-execution failures. However, without explicit links to GitHub issues, pull requests, or documentation updates, this knowledge may remain local to the support thread. As a result, forum support may help individual users but not always become durable repository-level improvement.

\begin{table}[htbp]
\centering
\caption{Implicit semantic flow after removing directly linked artifact pairs}
\label{tab:rq4_semantic_flow}

\scriptsize
\setlength{\tabcolsep}{3.5pt}
\renewcommand{\arraystretch}{1.05}

\begin{tabularx}{0.75\columnwidth}{Xrrr}
\toprule
\textbf{Flow type} & \textbf{Pairs} & \textbf{Median sim.} & \textbf{Median lag} \\
\midrule
Issue problem $\rightarrow$ Pull request solution & 5,985 & 0.359 & 18.27 \\
Pull request precedes issue & 1,137 & 0.343 & -7.88 \\
\bottomrule
\end{tabularx}

\caption*{\footnotesize \textit{Note.} Semantic-flow pairs are candidate relationships identified using TF--IDF cosine similarity after removing directly linked artifact pairs. Median lag is measured in days from source artifact creation to target artifact creation; negative values indicate that the pull request preceded the related issue.}

\end{table}

\paragraph{Semantic similarity reveals additional candidate issue--pull request flow.}
Because many related artifacts may not be explicitly linked, we also examined semantic similarity between GitHub issues and pull requests after removing directly linked pairs. Table~\ref{tab:rq4_semantic_flow} reports the resulting candidate relationships. We identified 5,985 candidate issue-to-pull-request flows, with a median cosine similarity of 0.359 and a median lag of 18.27 days. This suggests that related implementation work often appears within weeks after an issue is raised, even when the issue and pull request are not explicitly connected.

We also found 1,137 cases where a semantically related pull request preceded an issue, with a median lag of $-7.88$ days. These cases may reflect follow-up reports, regressions, documentation needs, or issues opened after related implementation work had already begun. Because these pairs are based on textual similarity and temporal ordering rather than explicit references, we interpret them as candidate implicit flows rather than confirmed problem--solution links.

\begin{table}[htbp]
\centering
\caption{Manual validation of sampled RQ4 linkage candidates}
\label{tab:rq4_manual_validation_all}

\scriptsize
\begin{tabular}{@{}p{0.24\columnwidth}rrrrr@{}}
\toprule
\textbf{Candidate type} & \textbf{Sample} & \textbf{Exact} & \textbf{Related} & \textbf{Weakly related} & \textbf{Unrelated} \\
\midrule
Direct link & 72 & 8 (11.1\%) & 47 (65.3\%) & 15 (20.8\%) & 2 (2.8\%) \\
Issue--pull request reference & 10 & 7 (70.0\%) & 3 (30.0\%) & 0 (0.0\%) & 0 (0.0\%) \\
Semantic similarity & 20 & 7 (35.0\%) & 8 (40.0\%) & 4 (20.0\%) & 1 (5.0\%) \\
\midrule
Overall & 102 & 22 (21.6\%) & 58 (56.9\%) & 19 (18.6\%) & 3 (2.9\%) \\
\bottomrule
\end{tabular}

\caption*{\footnotesize \textit{Note.} Exact pairs refer to the same concrete maintenance or support concern. Related pairs share a meaningful technical or maintenance concern but may not show direct problem--solution continuity. Weakly related pairs share limited vocabulary or broad context. Unrelated pairs do not provide meaningful evidence of maintenance flow.}

\end{table}

\paragraph{Manual validation of linkage candidates.}
Because direct references and semantic similarity can vary in evidential strength, we manually validated a stratified sample of over a hundred linkage candidates drawn from the RQ4 linkage pool. The sample included direct links, issue--pull request references, and semantic-similarity candidates across the observed flow types. Each sampled pair was classified as \emph{exact}, \emph{related}, \emph{weakly related}, or \emph{unrelated}. Exact pairs referred to the same problem, feature, bug, module, pull request, or implementation concern. Related pairs shared a meaningful maintenance or support concern but did not always show direct problem--solution continuity. Weakly related pairs shared limited vocabulary, repository context, or broad ecosystem terminology. Unrelated pairs did not show a meaningful maintenance connection.

Table~\ref{tab:rq4_manual_validation_all} reports the validation results. Overall, 22 of 102 sampled candidates were exact, 58 were related, 19 were weakly related, and 3 were unrelated. Thus, 80 of 102 sampled candidates were either exact or related. The validation also shows differences across evidence types. Direct links were often meaningful, but unresolved GitHub references and cross-platform links sometimes produced weak matches. Issue--pull request references were the strongest category, with all 10 sampled cases classified as exact or related. Semantic-similarity candidates were more mixed: 7 of 20 were exact, 8 were related, 4 were weakly related, and 1 was unrelated. This confirms that semantic similarity can identify useful implicit relationships, but it also produces false positives and should not be interpreted as definitive evidence of problem--solution flow.

\begin{table}[htbp]
\centering
\caption{Technical signal overlap across forum discussions, issues, and pull requests}
\label{tab:rq4_signal_overlap}

\footnotesize
\setlength{\tabcolsep}{2.6pt}
\renewcommand{\arraystretch}{1.05}

\begin{tabularx}{0.75\columnwidth}{Xrrr}
\toprule
\textbf{Signal} & \textbf{Forum} & \textbf{Issues} & \textbf{Pull requests} \\
\midrule
Execution error    & 52.85\% & 36.50\% & 52.80\% \\
Workflow execution & 83.69\% & 58.91\% & 61.66\% \\
Container/runtime  & 26.37\% & 16.30\% & 22.51\% \\
Cloud              & 27.04\% & 6.19\%  & 3.21\%  \\
HPC                & 18.77\% & 9.50\%  & 2.24\%  \\
MultiQC/reporting  & 23.35\% & 14.36\% & 7.08\%  \\
Testing/linting    & 52.96\% & 43.63\% & 65.98\% \\
Docs/training      & 16.09\% & 11.34\% & 54.12\% \\
Dependency update  & 23.13\% & 26.44\% & 64.07\% \\
Template sync      & 4.92\%  & 6.72\%  & 15.19\% \\
Module/subworkflow & 14.30\% & 27.72\% & 27.93\% \\
Configuration      & 48.60\% & 25.60\% & 31.49\% \\
Channel/dataflow   & 51.17\% & 39.14\% & 47.01\% \\
\bottomrule
\end{tabularx}

\caption*{\footnotesize \textit{Note.} Values are percentages of artifacts containing each signal. Signals are not mutually exclusive, so percentages within each artifact type may sum to more than 100\%.}

\end{table}

\paragraph{Technical signals recur across spaces even without explicit links.}

Technical-signal overlap provides a broader view of recurring maintenance and support concerns across artifact types. Table~\ref{tab:rq4_signal_overlap} shows that forum discussions were dominated by user-facing execution and infrastructure signals, including workflow execution (83.69\%), execution errors (52.85\%), configuration (48.60\%), cloud (27.04\%), and HPC (18.77\%). Pull requests showed a more implementation-oriented profile, with high levels of testing/linting (65.98\%), dependency updates (64.07\%), documentation/training (54.12\%), and template synchronization (15.19\%). Issues occupied an intermediate position, where execution problems, testing/linting, channel/dataflow behavior, and module/subworkflow concerns were formalized as repository-level maintenance tasks.

These overlaps suggest that similar technical concerns move across the ecosystem even when direct links are missing. For example, execution and configuration problems appear prominently in forum discussions as user-support needs, appear in issues as repository-level problem reports, and appear in pull requests as tests, fixes, documentation updates, dependency changes, or template improvements. However, because signal overlap is aggregate-level evidence, it should not be interpreted as proof that specific forum posts caused specific GitHub changes. Instead, it indicates that recurring support concerns and repository maintenance work share common technical domains.

\paragraph{Synthesis of problem--solution flow.}

Together, the three analyses show a layered pattern of traceability. Explicit links provide strong evidence that GitHub is the main traceable space where nf-core problems are connected to implementation work. Semantic similarity reveals additional candidate issue--pull request relationships that are not explicitly linked, suggesting that repository maintenance contains more related work than direct references alone capture. Technical-signal overlap shows that forum discussions, issues, and pull requests share recurring technical concerns, especially around execution, configuration, containers, cloud, HPC, reporting, testing, dependencies, and documentation.

However, the weakest layer is explicit forum--GitHub traceability. Forum discussions frequently contain support problems that overlap with GitHub maintenance concerns, but they are rarely linked to repository artifacts. This means that user-facing support knowledge often enters repository maintenance indirectly, if at all. Strengthening lightweight links between forum threads, GitHub issues, pull requests, and documentation updates could help convert recurring support problems into more durable ecosystem improvements.

\begin{center}
\fbox{
\begin{minipage}{0.96\linewidth}
\textbf{Summary of RQ4.}
\textbf{RQ4} shows that problem--solution flow in nf-core is strongly traceable within GitHub but weakly connected across GitHub and the Seqera Community Forum. Direct-link analysis identified dense repository-internal connections, especially pull request-to-issue references (7,599 links), showing that GitHub is the main space where reported problems are converted into implementation work. In contrast, explicit Forum--GitHub links were rare, with only five forum-to-GitHub and six GitHub-to-forum links, despite substantial technical-signal overlap across these spaces. Semantic similarity revealed additional implicit issue--pull request relationships, including 5,985 candidate issue-to-pull-request flows with a median lag of 18.27 days. Overall, nf-core has strong issue--pull request traceability, but user-facing support knowledge from forum discussions often enters repository maintenance indirectly through recurring technical concerns rather than explicit cross-platform links.
\end{minipage}
}
\end{center}

\section{Discussion}

Our findings show that sustaining nf-core requires more than workflow engines, standardized templates, and automated testing. Maintenance and support are distributed across several artifact types that play different roles in the ecosystem. GitHub issues surface problems and coordination needs, pull requests implement and validate changes, and forum discussions expose user-facing execution barriers that often depend on local infrastructure. This section discusses the implications of these findings for nf-core, for community-driven scientific pipeline ecosystems, and for empirical research on software maintenance in scientific computing.

\subsection{Maintenance and Support Are Distributed Across Artifact Types}

A central finding of this study is that nf-core maintenance is distributed across complementary socio-technical spaces. Issues, pull requests, and forum discussions do not simply provide different samples of the same activity. They capture different stages and forms of ecosystem work. Issues function as a repository-centered space for reporting problems, requesting features, tracking migrations, coordinating module work, and formalizing maintenance tasks. Pull requests represent the implementation and integration layer, where changes are reviewed, tested, standardized, and merged. Forum discussions reveal practical support problems that arise when users run workflows across local machines, HPC clusters, cloud platforms, containers, and managed execution services.

This division of work is important for understanding scientific pipeline sustainability. Scientific workflow research often emphasizes portability, reproducibility, and execution across heterogeneous infrastructure. Our results show that these properties are not maintained by workflow specifications alone. They require continuous work on modules, subworkflows, sample sheets, test datasets, templates, containers, dependency versions, execution profiles, documentation, and user support. In this sense, nf-core should be understood not only as a collection of reusable pipelines, but also as a maintained infrastructure whose sustainability depends on coordinated repository work and community support.

This finding also supports the value of a cross-artifact study design. An analysis limited to GitHub issues and pull requests would capture repository-centered maintenance but miss many operational barriers encountered by users. An analysis limited to forum discussions would reveal user difficulties but miss how durable repository-level changes are implemented. Studying these spaces together provides a more complete account of how scientific pipeline ecosystems are sustained.

\subsection{Resolution Depends on Actionability, Coordination, and Diagnostic Evidence}

Across issues, pull requests, and forum discussions, resolution outcomes were associated with actionability, coordination, and diagnostic evidence. Issues were more likely to close, and closed faster, when they had assignees, comments, milestones, bug labels, error mentions, and version information. Pull requests were more likely to merge when they included review-readiness signals such as checklists, fix/test framing, linked issues, requested reviewers, and contributor or maintainer authorship. Forum discussions were more likely to receive accepted answers when they included code blocks, concrete technical evidence, and sustained interaction.

These patterns should not be interpreted causally. Features such as assignees, comments, requested reviewers, replies, or likes often emerge during the handling of an artifact and may reflect maintainer attention, artifact priority, or community engagement. They are therefore process signals, not necessarily independent causes of resolution. Nevertheless, their consistent association with resolution across artifact types suggests that nf-core artifacts are easier to resolve when they provide enough context for maintainers or community members to understand the problem and act on it.

This has practical implications for issue templates, pull request checklists, and forum posting guidelines. These mechanisms are not merely administrative. They help transform ambiguous reports into actionable maintenance objects. For scientific pipelines, this is especially important because failures may involve tool versions, input files, reference datasets, profiles, containers, executors, scheduler policies, storage systems, and cloud credentials. Templates and guidelines that prompt users to provide commands, versions, profiles, logs, executor settings, container details, and minimal reproducible examples could improve diagnosability and reduce repeated clarification work.

\subsection{Infrastructure-Specific Support Remains a Persistent Boundary}

The forum results highlight a persistent boundary in scientific workflow support: many difficult problems arise from the interaction between workflows and execution environments rather than from pipeline code alone. Cloud, HPC, container, configuration, and scheduler-related discussions were prominent in the forum and were less consistently resolved through accepted answers. These questions often depend on local information that maintainers cannot easily reproduce, such as institutional scheduler policies, storage permissions, private registries, cloud credentials, queue limits, memory constraints, and executor-specific behavior.

This suggests that documentation focused only on pipeline parameters is insufficient. Users also need infrastructure-oriented guidance, including tested execution profiles, examples for common HPC schedulers, cloud deployment recipes, storage and permission checklists, container troubleshooting guides, and guidance on collecting diagnostic logs. Forum discussions can help identify where such guidance is missing. Even when a forum thread does not correspond to a pipeline defect, it may reveal a recurring usability or infrastructure pain point that can be addressed through better documentation, examples, templates, or profile validation.

The finding also suggests that accepted-answer status should be interpreted cautiously. Some unresolved forum discussions may still be useful to later users because they contain logs, partial explanations, workarounds, or infrastructure-specific context. Thus, the absence of an accepted answer does not necessarily mean the discussion has no value; rather, it indicates that the forum did not reach a formally marked resolution.

\subsection{Automation Helps When It Aligns with Review Workflows}

Automation is central to nf-core maintenance, but our results show that automated maintenance is not uniformly successful. Renovate pull requests had very high merge rates, while Dependabot and other bot-authored pull requests were less consistently integrated and were more often closed without merge. This difference suggests that automation is most useful when its output is scoped, predictable, compatible with nf-core validation workflows, and easy for maintainers to review.

This finding has broader implications for scientific software ecosystems that rely on automated dependency updates, template propagation, and continuous integration. Automation should not be evaluated only by the number of pull requests it creates. Its value depends on whether it produces reviewable and context-aware changes. Automated updates that are noisy, obsolete, poorly scoped, or difficult to validate may increase maintainer workload rather than reduce it. Ecosystems such as nf-core may benefit from automation policies that distinguish routine low-risk updates from higher-risk changes, provide clearer evidence for automated updates, and route automation-generated pull requests to appropriate maintainers or validation workflows.

\subsection{Cross-Artifact Traceability Is a Missing Layer of Support Infrastructure}

\textbf{RQ4} shows that nf-core has strong repository-internal traceability but weak explicit traceability between GitHub and the Seqera Community Forum. Pull request-to-issue links provide a visible pathway from reported problems to implementation work inside GitHub. In contrast, explicit links between forum discussions and GitHub artifacts were rare, despite substantial technical-signal overlap across forum discussions, issues, and pull requests.

This creates an important maintenance gap. Forum discussions often contain maintenance-relevant knowledge about recurring execution failures, documentation gaps, confusing errors, cloud and HPC configuration, container behavior, and pipeline usability. However, when these discussions are not linked to GitHub issues, pull requests, or documentation updates, the knowledge may remain local to the support thread. In such cases, forum support may help individual users without necessarily becoming durable ecosystem improvement.

Semantic similarity analysis suggests that related maintenance work may occur even without explicit links, but these pairs should be interpreted only as candidate implicit relationships. Textual similarity and temporal ordering do not prove that one artifact caused or informed another. The stronger and more actionable finding is therefore the lack of visible forum--GitHub traceability. Lightweight linking practices could help close this gap. For example, recurring forum discussions could be periodically triaged into documentation issues, infrastructure-support issues, or template-improvement tasks. Forum posts could encourage users or maintainers to link related GitHub issues, and GitHub issue templates could include prompts for related forum discussions. Such mechanisms would make support knowledge more discoverable and easier to convert into repository-level maintenance.

\subsection{Implications for nf-core and Similar Scientific Workflow Ecosystems}

Although this study focuses on nf-core, the findings are relevant to other scientific workflow ecosystems that combine reusable workflows, shared components, community support, and heterogeneous execution infrastructure. Ecosystems such as Galaxy, Snakemake, Pegasus, and related community-maintained workflow platforms face similar sustainability challenges: workflows must remain executable as dependencies, containers, reference data, schedulers, cloud services, and user environments change.

The results suggest three broader lessons. First, community support spaces should be treated as maintenance-relevant evidence, not only as help channels. Support discussions can reveal recurring problems before they become formal issues. Second, resolution mechanisms should be designed to improve actionability by encouraging structured diagnostic evidence and clear routing. Third, cross-artifact traceability should be considered part of ecosystem infrastructure. Without links between support discussions and repository-level changes, recurring user problems may remain visible but unresolved at the ecosystem level.

\subsection{Actionable Recommendations}

Based on these findings, we identify five actionable recommendations for nf-core and similar community-driven scientific pipeline ecosystems.

\begin{table}[htbp]
\centering
\caption{Actionable recommendations derived from the cross-artifact analysis.}
\label{tab:discussion_recommendations}

\footnotesize
\setlength{\tabcolsep}{4pt}
\renewcommand{\arraystretch}{1.08}

\begin{tabularx}{0.85\columnwidth}{
>{\raggedright\arraybackslash}p{0.15\columnwidth}
>{\raggedright\arraybackslash}X
}
\toprule
\textbf{Area} & \textbf{Recommendation} \\
\midrule
Issue reporting & Strengthen issue templates with prompts for pipeline version, command, profile, executor, container runtime, logs, relevant files, and minimal reproducible examples. \\
Pull request review & Preserve checklist-based review readiness, but distinguish routine updates from coordination-intensive changes requiring domain-specific review. \\
Automation & Evaluate automated pull requests by mergeability, scope, validation evidence, and maintainer burden, not only by update frequency. \\
Forum support & Add structured forum prompts for cloud, HPC, container, and executor-related questions, including scheduler, storage, credential, profile, and log details. \\
Traceability & Encourage lightweight links between forum discussions, GitHub issues, pull requests, and documentation updates when support threads reveal recurring problems. \\
\bottomrule
\end{tabularx}

\end{table}

These recommendations follow directly from the observed patterns. Artifacts with clearer diagnostic information were more often resolved; review-ready pull requests were more consistently integrated; infrastructure-specific questions were harder to answer; and forum knowledge was weakly linked to repository maintenance. Improving structure, routing, and traceability can therefore help convert recurring support needs into durable ecosystem improvements.

Overall, this study shows that maintaining reusable scientific pipelines is a continuous coordination problem. Workflow engines, templates, automated tests, and containers provide essential technical foundations, but they do not eliminate the need for ongoing maintenance and user support. nf-core remains sustainable because contributors and users collectively report problems, review changes, update dependencies, maintain templates, troubleshoot infrastructure, and share knowledge. The main challenge is to make this distributed work more actionable and traceable. Stronger diagnostic templates, review-aware automation, infrastructure-specific guidance, and lightweight links between support and repository artifacts can help scientific pipeline ecosystems turn recurring user problems into maintainable, reusable, and durable improvements.

\section{Threats to Validity}\label{threats-to-validity}
Threats to validity are factors that may affect the accuracy, reliability, or generalizability of study findings. They concern how research conclusions may diverge from reality \cite{bean2007qualitative}. In this study, we acknowledge the following potential threats to validity.

\subsection{Construct Validity}
A threat to construct validity concerns whether our measures accurately capture maintenance, support, resolution, and problem--solution flow \cite{wohlin2012experimentation}. We used issue closure, pull request merge or closed-without-merge status, and forum accepted-answer status as observable resolution outcomes. These measures follow the conventions of each artifact type, but they may not capture all forms of resolution. For example, a closed issue may not fully solve the underlying problem, a merged pull request may address only part of a larger maintenance task, and a forum discussion may be useful even without an accepted answer.

Textual and technical-signal features also introduce measurement risk. Signals such as cloud, HPC, containers, configuration, testing, and dependency updates were detected using keyword-based rules, which may miss relevant cases or include false positives. Similarly, BERTopic clusters summarize recurring themes but do not perfectly separate artifacts that contain multiple concerns. We mitigated these risks by using domain-informed signal definitions, modeling each artifact type separately, tuning topic-model parameters, and manually inspecting representative artifacts before assigning topic labels.

\subsection{External Validity}
Our findings are based on nf-core, a mature and highly structured ecosystem of standardized Nextflow pipelines. The exact topic distributions, resolution rates, and traceability patterns may not generalize to smaller projects, less formal workflow communities, or ecosystems with different governance and review practices. However, many challenges observed in nf-core, including dependency updates, container failures, cloud and HPC execution, input validation, documentation gaps, and user support, are common across scientific workflow ecosystems. Thus, while the numeric results are nf-core-specific, the broader implications may apply to ecosystems such as Galaxy, Snakemake, Pegasus, and related community-maintained workflow platforms.

\subsection{Internal Validity}
This study is observational, so the reported associations should not be interpreted as causal effects. Features such as assignees, comments, milestones, requested reviewers, checklists, or code blocks may be associated with resolution because they improve actionability, but they may also reflect unobserved factors such as artifact priority, maintainer availability, contributor expertise, or pipeline maturity. For example, assigned issues may close faster because assignment improves accountability, or because maintainers assign issues that are already more actionable.

Temporal interpretation also requires caution. Some maintenance and support work may occur outside the observed artifacts, such as private communication, local debugging, or informal coordination. For forum discussions, accepted-answer timestamps were unavailable, so we used activity span as a lifecycle proxy rather than exact time-to-answer. For RQ4, semantic-flow pairs indicate textual relatedness and plausible temporal ordering, but they do not prove that one artifact directly caused or informed another. We therefore interpret them as candidate implicit flows, not definitive problem--solution links.

\subsection{Conclusion Validity}
Large repository-mining datasets can produce statistically significant results even when effects are small. To reduce this threat, we reported odds ratios, hazard ratios, median differences, and effect sizes where appropriate, and applied Benjamini--Hochberg correction for multiple comparisons. We also interpreted findings based on consistency across descriptive statistics, bivariate tests, regression models, lifecycle curves, and qualitative inspection.

Some models required feature pruning or regularization because the data contained sparse categories, correlated predictors, and skewed distributions. Regularized model coefficients should therefore be read as adjusted directional associations rather than formal causal estimates. We avoid overinterpreting individual coefficients and focus on patterns that recur across analyses.

\section{Conclusion}
\label{sec:conclusion}

This study examined maintenance and support in nf-core through a cross-platform analysis of 15,760 GitHub issues, 35,411 GitHub pull requests, and 895 Seqera Community Forum discussions. The findings show that nf-core maintenance is distributed across complementary spaces: issues capture problem reporting and coordination, pull requests capture implementation and integration, and forum discussions capture user-facing execution support.

Across these spaces, resolution was associated with actionability, coordination, and diagnostic evidence. Issues closed more effectively when they included maintainer attention and concrete technical details; pull requests merged more consistently when they were review-ready, traceable, and supported by tests or checklists; and forum discussions were more likely to receive accepted answers when they included code, interaction, and sufficient diagnostic context. However, infrastructure-specific problems involving cloud, HPC, containers, and workflow semantics remained harder to resolve.

The problem--solution flow analysis showed strong traceability within GitHub, especially through pull request-to-issue links, but weak explicit linkage between forum discussions and repository artifacts. This suggests that user-facing support knowledge often informs maintenance indirectly rather than through visible cross-platform connections. Strengthening links between forum discussions, GitHub issues, pull requests, and documentation updates could help convert recurring support problems into durable ecosystem improvements.

Overall, sustaining community-driven scientific pipeline ecosystems requires not only workflow engines, templates, containers, and tests, but also coordinated maintenance practices, actionable support processes, and stronger traceability across development and user-support spaces. Future work can extend this study by comparing nf-core with other scientific workflow ecosystems and by evaluating traceability mechanisms that connect recurring forum support problems to GitHub issues, pull requests, and documentation updates.

\section*{Acknowledgments}
This research is supported in part by the Natural Sciences and Engineering Research Council of Canada (NSERC), and by the industry-stream NSERC CREATE in Software Analytics Research (SOAR).

\section*{Data Availability Statements}
Our replication package can be found in our online appendix \citep{anonymous2026nfcore}.

\section*{Conflict of Interest}
We have no conflict of interest.


\bibliography{sn-bibliography}

\end{document}